\documentclass[aps,pra,twocolumn,nopacs,lettersize,superscriptaddress,floatfix,longbibliography]{revtex4-1}
\usepackage{array}
\usepackage{rotating}
\usepackage{multirow}
\usepackage{placeins}
\usepackage{xcolor}
\usepackage{hyperref}
\usepackage{lipsum} 
\usepackage{soul}
\hypersetup{
    colorlinks=true,
    linkcolor=blue,
    citecolor=blue,
    citebordercolor=white,
    linkbordercolor=white,
}
\usepackage[margin=0.65in]{geometry}
\usepackage{graphicx}
\graphicspath{ {} }
\usepackage{float}
\usepackage{amsmath,amssymb,amsfonts}
\usepackage{mathrsfs}

\usepackage{cleveref}

\usepackage{natbib}
\usepackage{tabularx}
\newcolumntype{Y}{>{\centering\arraybackslash}X}

\usepackage[]{titlesec}

\newcommand{\eqnreft}[1]{{Eq.~(\ref{#1})}}

\newcommand{\eqnsreft}[2]{{Eqs.~(\ref{#1}) and (\ref{#2})}}

\newcommand{\rr}{\mathbf{r}}
\newcommand{\uu}{\mathbf{u}}
\newcommand{\dd}{\mathrm{d}}

\begin{document}

\title{Turbulent relaxation to equilibrium in a two-dimensional quantum vortex gas}

\author{Matthew T. Reeves}
\email{m.reeves@uq.edu.au}
\affiliation{Australian Research Council Centre of Excellence in Future Low-Energy Electronics Technologies, School of Mathematics and Physics, University of Queensland, St Lucia, QLD 4072, Australia.}
\author{Kwan Goddard-Lee}
\affiliation{Australian Research Council Centre of Excellence for Engineered Quantum Systems, School of Mathematics and Physics, University of Queensland, St. Lucia, QLD 4072, Australia.}
\author{Guillaume Gauthier}
\affiliation{Australian Research Council Centre of Excellence for Engineered Quantum Systems, School of Mathematics and Physics, University of Queensland, St. Lucia, QLD 4072, Australia.}
\author{Oliver R. Stockdale}
\affiliation{Australian Research Council Centre of Excellence in Future Low-Energy Electronics Technologies, School of Mathematics and Physics, University of Queensland, St Lucia, QLD 4072, Australia.}
\author{Hayder Salman}
\affiliation{School of Mathematics, University of East Anglia, Norwich, NR4 7TJ, United Kingdom.}
\author{Timothy Edmonds}
\affiliation{Australian Research Council Centre of Excellence in Future Low-Energy Electronics Technologies, School of Mathematics and Physics, University of Queensland, St Lucia, QLD 4072, Australia.}
\author{Xiaoquan Yu}
\affiliation{Graduate School of  China Academy of Engineering Physics, Beijing 100193, China}
\affiliation{Department of Physics, Centre for Quantum Science, and Dodd-Walls Centre for Photonic and Quantum Technologies, University of Otago, Dunedin, New Zealand.}
\author{Ashton S. Bradley}
\affiliation{Department of Physics, Centre for Quantum Science, and Dodd-Walls Centre for Photonic and Quantum Technologies, University of Otago, Dunedin, New Zealand.}
\author{Mark Baker}
\affiliation{Australian Research Council Centre of Excellence for Engineered Quantum Systems, School of Mathematics and Physics, University of Queensland, St. Lucia, QLD 4072, Australia.}
\author{Halina Rubinsztein-Dunlop}
\affiliation{Australian Research Council Centre of Excellence for Engineered Quantum Systems, School of Mathematics and Physics, University of Queensland, St. Lucia, QLD 4072, Australia.}
\author{Matthew J. Davis}
\affiliation{Australian Research Council Centre of Excellence in Future Low-Energy Electronics Technologies, School of Mathematics and Physics, University of Queensland, St Lucia, QLD 4072, Australia.}
\author{Tyler W. Neely}
\email{t.neely@uq.edu.au}
\affiliation{Australian Research Council Centre of Excellence for Engineered Quantum Systems, School of Mathematics and Physics, University of Queensland, St. Lucia, QLD 4072, Australia.}

\date{\today}

\begin{abstract}

 We experimentally study emergence of microcanonical equilibrium states in the turbulent relaxation dynamics of a two-dimensional chiral vortex gas.  Same-sign vortices are injected into a quasi-two-dimensional disk-shaped atomic Bose-Einstein condensate using a range of mechanical stirring  protocols.  The resulting long-time vortex distributions are found to be in excellent agreement with the mean-field Poisson-Boltzmann equation for the system describing the microcanonical ensemble at fixed energy $\mathcal{H}$ and angular momentum $\mathcal{M}$. The equilibrium states are characterized by the corresponding thermodynamic variables of inverse temperature $\hat \beta$ and rotation frequency $\hat \omega$.  
We are able to realize equilibria spanning the full phase diagram of the vortex gas, including on-axis states near zero-temperature, infinite temperature, and negative absolute temperatures.  At sufficiently high energies the system exhibits a symmetry-breaking transition, resulting in an off-axis equilibrium phase at negative absolute temperature that no longer shares the symmetry of the container.  We introduce a point-vortex model with phenomenological damping and noise that is able to quantitatively reproduce the equilibration dynamics. 

\end{abstract}

\maketitle

\section{Introduction} 

Turbulence continues to stand as one of the most challenging problems in physics despite several centuries of study. Most phenomena occurring in the turbulent motion of fluids are strongly nonequilibrium in nature, making the problem highly intractable for theoretical treatment. The chaotic fluid motion ultimately requires a probabilistic description, yet one of the most powerful probabalistic tools available --- the maximum entropy principle of statistical mechanics --- is effectively rendered useless;  turbulent flows generally  defy a description in terms of equilibrium statistical mechanics, due to their strong dissipation of energy and consequent lack of detailed balance~\cite{Thalabard_2015,frisch1995turbulence,batchelor1953theory}. 

A notable exception occurs in the case of quasi-two-dimensional flows, where, due to the suppression of vortex stretching, energy is conserved in the limit of large Reynolds number~\cite{kraichnan1967inertial,batchelor1969}. In such flows, large and long-lived  isolated vortices tend to spontaneously form out of the turbulent background. Examples are regularly seen in a range of systems including electron plasmas~\cite{sarid_breaking_2004,rodgers2009hydrodynamic}, soap films~\cite{kellay2017hydrodynamics}, stratified fluid layers~\cite{hansen1998two}, and planetary atmospheres~\cite{adriani2018clusters,bouchet2002emergence} --- Jupiter's Great Red Spot, which has persisted for over 350 years, is perhaps the most famous example.
The prevalence and long-lived nature of these structures suggests they are an aspect of turbulence to which equilibrium statistical mechanics could be successfully applied.

The idea to apply statistical mechanics to turbulent flows originated with the seminal work  Onsager~\cite{onsager1949statistical}, who investigated the statistical mechanics of a system of point vortices in a perfect (i.e., inviscid) fluid. In this simple Hamiltonian model, the vortices are treated as a kind of ``gas", whose particles interact via long-range interactions.  The equilibria of this model are indeed typically dominated by one or two large clusters of vortices, reflecting what is typical of real fluids. While the point-vortex approach could not be quantitatively applied to real fluids (which have continuous vorticity distributions), Onsager's maximum entropy approach was, naturally, a highly appealing prospect; a significant body of work in the following decades aimed to bridge the gap between the discrete and continuous vorticity distributions~\cite{joyce_montgomery_1973,edwards1974negative,Kraichnan:1975ku,miller1990statistical,robert1991maximum,robert_sommeria_1991}, in the hope to connect the maximum entropy approach to real fluids (for a summary of theoretical developments, see, e.g., \cite{maestrini2019entropy}). 

Unfortunately however, although equilibrium theories have proven to be successful in some cases~\cite{denoix1992,matthaeus1991,rodgers2009hydrodynamic,bouchet2002emergence}, they also fail in many situations. It has been argued that ``statistical  approaches have not been proven yet to offer a more than qualitative framework for the interpretation of experimental observation"~\cite{tabeling2002two}.   Some examples explicitly avoid assuming global entropy maximization~\cite{Sarid2004}, while  others require abandoning entropy altogether~\cite{driscoll1990,carnevale1992rates,marteau1995equilibrium,huang1994relaxation,Patker2018}. One major culprit, it seems, is the ergodicity assumption; two-dimensional turbulent systems often do not exhibit sufficiently vigorous (ergodic) mixing to justify the search for global equilibrium states~\cite{tabeling2002two}. Indeed, it is now known that a general property of long-range interacting systems is that they are unable to thermalize when they contain a large number of degrees of freedom~\cite{levin2014nonequilibrium} 
---  precisely the situation which occurs in large Reynolds number flows~\cite{batchelor1953theory,batchelor1969}. A second major complication arises due  to contributions in the boundary layer, which introduces crucial changes to the flow near the container walls for any nonvanishing value of viscocity~\cite{brands1999statistical, Li1996decaying,vanHeijst_2006}.

Superfluid atomic gases confined in uniform box traps~\cite{gaunt2013bose,chomaz2015emergence,gauthier2016direct,mukherjee2017homogeneous,tajik2019designing,navon2021quantum} have recently emerged as a new platform to test fundamental theories of turbulence and vortex dynamics in a highly tunable system~\cite{Navon:2016cb,navon2019synthetic,glidden2021bidirectional,gauthier2019giant,johnstone2019evolution,stockdale2019universal,kwon2021sound}. A natural question which arises is whether the maximum entropy approach may more accurately describe coherent vortices in a superfluid, due to several advantages these systems offer. First, in superfluids the viscosity is identically zero, as is assumed in the maximum entropy theories discussed above. Second, in thin-layer superfluids the vorticity is genuinely pointlike in nature, and the condensate wave function constrains the vorticity to be quantized with value $\Gamma = \pm h/m$, where $h$ is Planck's constant and $m$ is the mass of a superfluid particle. In fact, provided the vortex cores are small, the vortex dynamics are governed precisely by the Hamiltonian point-vortex system originally considered by Onsager~\cite{fetter1966vortices,groszek2018motion}. These systems therefore offer the unique prospect of experimentally testing the maximum entropy approach in a system which is genuinely inviscid, and contains a relatively small number of degrees of freedom (determined by the vortex number, $N$), where simulations suggest that the ergodicity assumption may hold~\cite{dritschel2015ergodicity,esler2015universal,salman2016long}.  Significant experimental progress in this direction has been made in two recent works~\cite{gauthier2019giant,johnstone2019evolution} (one by some of the present authors~\cite{gauthier2019giant}), which observe signatures consistent with maximum entropy vortex distributions.  However, these experiments both suffer from  key limitations: (i) The temperature of the  vortex distributions could only be inferred from \textit{a priori} assumptions of equilibrium, and (ii) the relaxation to equilibrium is not tested for a wide range of nonequilibrium initial conditions. Without having tested these aspects it cannot yet be said whether the maximum entropy approach proves useful for describing two-dimensional turbulent flows in superfluids.

 In this work we demonstrate that the maximum entropy approach quantitatively agrees with experiment over a wide range of parameters. We experimentally consider a chiral (single-sign circulation) vortex gas confined to a disk geometry, which we realize in an ultracold atomic Bose-Einstein condensate confined with a fully configurable optical potential.  In contrast to the previous experiments~\cite{gauthier2019giant,johnstone2019evolution} which consider neutral vortex gas systems~\cite{joyce_montgomery_1973,yu2016theory}, the chiral system exhibits nontrivial (i.e., spatially nonuniform) equilibria over the entire the phase diagram~\cite{smith_phase-transition_1989}, 
 facilitating a comparison between experiment and theory over the full phase diagram of the vortex gas.  Furthermore, as all vortices have the same sign, vortex-antivortex annihilation is  completely suppressed in the bulk of the superfluid. By suppressing this nonequilibrium process, the interpretation of the system in terms of the microcanonical ensemble is simplified considerably.
 
 Using controllable optical potentials  to stir the superfluid, we are able to initialize vortex distributions of $N\sim 10$ -- $16$ vortices with essentially arbitrary initial values of energy and angular momentum, which are the two control parameters determining the equilibrium states. We find that, despite some residual dissipation, the vortex gas lives long enough to reach equilibrium. We first initialize the system directly into a near-equilibrium state, and demonstrate that it remains near equilibrium, undergoing a gradual cooling.  Then, by initializing the vortices in nonequilibrium configurations with different energy and angular momenta, we demonstrate relaxation to a range of equilibrium distributions predicted by the microcanonical ensemble. Finally, we introduce a point-vortex model with  damping and noise that is able to quantitatively reproduce the equilibration dynamics. 

The outline of this paper is as follows. In Sec.~\ref{sec:PVM} we outline the point-vortex system and summarize the known results from statistical mechanics and the mean-field phase diagram of the chiral vortex gas. In Sec.~\ref{sec:Expt} we present our experimental results and compare the observed vorticity distributions with the equilibrium predictions presented in Sec.~\ref{sec:PVM}. Our results on the dynamics of the vortex gas are presented in Sec.~\ref{sec:vortexDynamics}, which we show can be quantitatively described by a point-vortex model supplemented by friction and noise. Section~\ref{sec:conclusions} presents conclusions and outlook. 
\begin{figure*}
    \centering
    \includegraphics[width=0.95\textwidth]{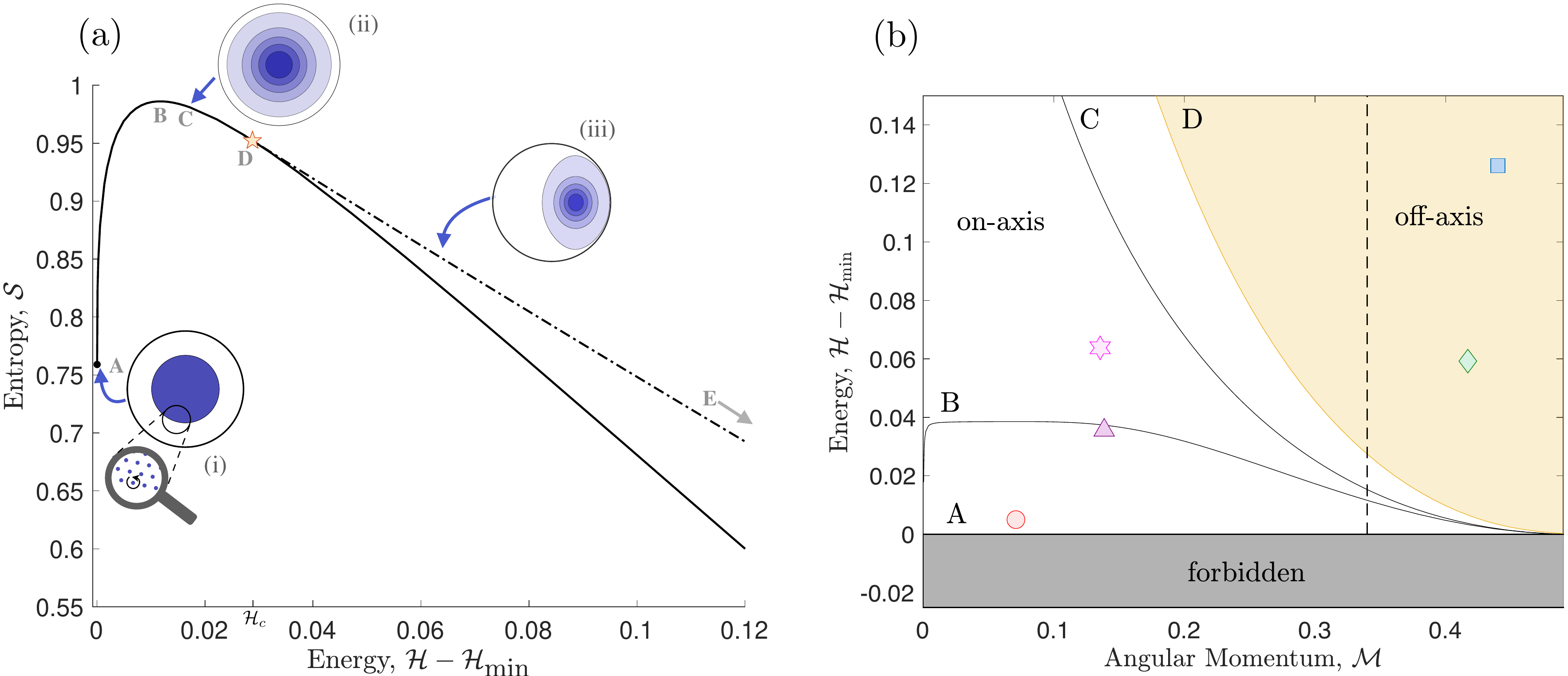}
    \caption{(a)  Entropy and equilibrium states of a chiral vortex gas in a disk at fixed angular momentum, obtained by numerically solving the Poisson-Boltzmann equation, Eq.~(\ref{eqn:meanfield}), for states with $\mathcal{M}= 0.34$ and subtracting off the minimum energy $\mathcal{H}_{\rm min} = \tfrac{1}{4} - \tfrac{1}{2}\log(2\mathcal{M})$ (see Table~\ref{tab:MFsolutions}). The solid curve shows states where cylindrical symmetry is enforced; at low energy these states are the maximum entropy solutions (i,ii; solid curve).
   At high energies the maximum entropy states break the rotational symmetry and sit off-axis (iii; dash-dot curve).  Insets show example contour plots of the average vortex density. Labels \textbf{A}, \textbf{B}, \textbf{C}, \textbf{D}, and \textbf{E} mark significant equilibrium states. 
    \textbf{A} Rankine; \textbf{B} Gaussian; \textbf{C} Riccati; \textbf{D} Off-Axis; \textbf{E} supercondensate
    (see text, Sec.~\ref{sec:StatMechSection}). 
     (b) Full equilibrium phase diagram of the chiral vortex gas in the $\mathcal{H}$-$\mathcal{M}$ plane. Lines show boundaries of the states \textbf{A} -- \textbf{D}, relative to the minimum energy of the Rankine (\textbf{A}) state.  Markers show the best-fit  values for the five different experiments shown in Sec.~\ref{sec:Expt}: blue square (I); red circle (II); pink star  (III); purple triangle (IV); green diamond (V).  The vertical dashed line indicates $\mathcal{M}=0.34$ as in (a).
}

    \label{fig:Overview}
\end{figure*}


\section{Chiral vortex gas in a disk}
\label{sec:PVM}

Before presenting our experimental results, we first provide context by briefly introducing the model of a chiral vortex gas in a disk, and we review the known equilibrium results obtained from statistical mechanics.

\subsection{point-vortex model}
\label{sec:PointVortexModel}

We consider a two-dimensional fluid containing a chiral vortex gas of $N$ point vortices with quantized circulations $ \Gamma = +h/m$, where $h$ is Planck's constant and $m$ is the mass of a fluid particle.  The fluid is assumed to be incompressible and inviscid, with a uniform (areal) density $\rho_0$, and is confined to a disk of radius $R$. Hereafter we may set $R=1$ without loss of generality. In addition to the vortex number $N$,  the kinetic energy and  angular momentum of the fluid are conserved. The kinetic energy of the fluid
can be expressed in terms of the vortex locations $\rr_j$ as~\cite{newton2013nvortexproblem,smith1990nonaxisymmetric}
\begin{equation}
        H =    -  \sum_{j\neq k} \textrm{ln} \left | 	  \rr_j -\rr_k \right| + \sum_{j,k}   \ln \left| r_j(\rr_j-\bar{\rr}_k) \right|   \label{eqn:PVHamiltonian}.
    \end{equation}

Here $H$ is expressed in units of the energy $E_0= \rho_0\Gamma^2/4 \pi$.  Notice that, although the energy is entirely kinetic, the Hamiltonian resembles the interaction energy in a ``gas" of charged particles in two dimensions;  the first term describes the Coulomb-like interaction between vortices, while second term describes the interaction between vortices and image vortices, which have circulation $-\Gamma$ and are located outside the disk at position $\bar{\rr}_j =  \rr_j / r_j^2$ where $r_j = |\rr_j|$. The fictitious image vortices enforce the condition that the fluid may not flow through the boundary, i.e., $\mathbf{u} \cdot \hat{\rr} |_{r=1} =0$, where $\hat \rr$ is the radial unit vector. The vortex dynamics governed by $H$ are given by 
 \begin{align}
 \dot x_j &= \partial H /\partial y_j, & \dot y_j &= -\partial H /\partial x_j.
 \label{eqn:PVdynamics}
 \end{align}
It can be seen from Eq.~(\ref{eqn:PVdynamics}) that the $x$ and $y$ coordinates of the vortices are canonically conjugate variables. This unusual feature of this Hamiltonian system has a profound effect on the statistical mechanics, as will be discussed in the next section. 
 The angular momentum is $L = \rho_0 \int \mathrm{d}^2\rr\; \rr \times \uu(\rr) =   \tfrac{1}{2} \rho_0 \Gamma  (N - M)$, where 
\begin{equation}
M = {\sum_j}  |\rr_j|^2.
\label{eqn:PVAngMom}
\end{equation}
The angular momentum hence constrains the mean-square radius of the vortex distribution \footnote{$M$ is often referred to as the angular momentum~\cite{newton2013nvortexproblem,aref1979motion}.  The distinction is not so important when applying the point-vortex model to the dynamics of a classical Euler fluid (where the vortex number is conserved), but is for a superfluid as vortex annihilation can occur at the boundary; here $L$ varies continuously as a vortex leaves at the boundary, whereas $M$ changes discontinuously.}.

\subsection{Statistical mechanics}
\label{sec:StatMech}

For the Hamiltonian system described by \eqnreft{eqn:PVHamiltonian}, at sufficiently large $N$ one may hope to invoke the ergodicity hypothesis to determine the long-time behavior of the system from the tools of statistical mechanics. For the vortex gas dynamically evolving at fixed energy [Eq.~(\ref{eqn:PVHamiltonian})] and angular momentum [Eq.~(\ref{eqn:PVAngMom})], the system is described by the microcanonical ensemble
\begin{equation}
\delta S - \beta (\delta H - \omega \delta M) = 0,
\label{eqn:microensemble}
\end{equation}
where $S$ is the entropy, and
\begin{align}
\beta &= \frac{\partial S}{\partial H}\bigg|_M, & \omega = \frac{1}{\beta}\frac{\partial S}{\partial M}\bigg|_H. 
\label{eqn:betaandomega}
\end{align}
 Here $\beta$ is the inverse temperature and $\beta \omega$ is a thermodynamic potential for the angular momentum. The quantity $\omega$ may be interpreted as a  rotation frequency~\footnote{Note however that for a given solution $\omega$ does not necessarily correspond to any particular physical rotation rate in the system; see Ref.~\cite{smith1990nonaxisymmetric}.}. 
    
A remarkable property of \eqnsreft{eqn:PVHamiltonian}{eqn:PVdynamics} is that the canonical coordinates are determined only by the circulations and the physical-space coordinates of the vortices. As first appreciated by Onsager~\cite{onsager1949statistical}, this property has profound effects on the statistical mechanics of the system;  if the physical space is bounded by a container of area $A$, it follows that the total accessible phase space volume is bounded:

\begin{equation}
 \int d^2\rr_1 \dots d^2\rr_N = A^N.
\label{eqn:phasespace}
\end{equation} 

It follows directly from this property~\cite{onsager1949statistical} that entropy reaches a maximum at a finite value of the energy [see Fig.~\ref{fig:Overview}(a), point B]. Above this energy, the entropy decreases with increasing energy, and by Eq.~(\ref{eqn:betaandomega}) these equilibria are thus characterized by negative absolute temperatures. The bounded phase space property starkly contrasts with most systems, for which the phase space is unbounded and the entropy monotonically increases with energy.   Notice crucially that  the unusual Hamiltonian structure occurs because the vortices are massless objects; there is no term in the Hamiltonian of the form $\frac{1}{2} m_v v^2$ (for hypothetical vortex mass $m_v$). The appearance of such a term would  break the bounded phase space condition \eqnreft{eqn:phasespace}. Finally, it should be noted that as this system supports negative temperatures,  ensemble equivalence does not hold in general, and the canonical ensemble is therefore not appropriate~\cite{levin2014nonequilibrium,smith1990nonaxisymmetric}.

  \begin{table*}
\renewcommand{\arraystretch}{1.5}
    \centering
   \begin{tabularx}{\textwidth}{c | YYYYY}
   \hline\hline
  Solution & \textbf{A:}  Rankine & \textbf{B:}  Gaussian & \textbf{C:}  Riccati & \textbf{D:}  off-Axis & \textbf{E:}  supercondensate \\

      Density     &\includegraphics[width = 0.15\columnwidth]{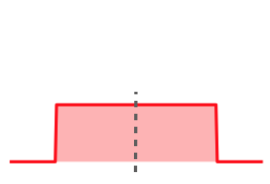} & \includegraphics[width = 0.15\columnwidth]{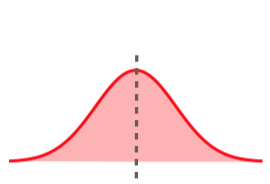} & \includegraphics[width = 0.15\columnwidth]{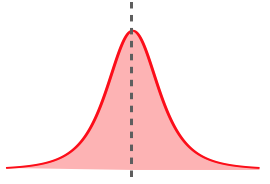} & \includegraphics[width = 0.15\columnwidth]{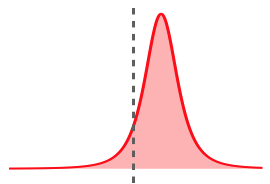} &\includegraphics[width = 0.17\columnwidth]{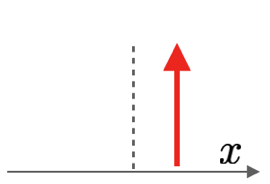} \\
    \hline
         $\hat \beta$  &   $+\infty$ & $0^{\pm}$ & $f( \hat \beta)=\mathcal{M}$ & --- & -2 \\ 
                  $\hat \omega$  & $(2\mathcal{M})^{-1}$ & $\pm\infty$ & 0 & 1 & $(1-\mathcal{M})^{-1}$ \\ 
                           $\mathcal{H}$  & $\tfrac{1}{4} - \tfrac{1}{2}\log(2 \mathcal{M})$ & $g( \hat \beta \hat \omega)$  & $\hspace{-0.2cm}-\tfrac{2}{ \hat \beta^2}\left[\ln(1 + \tfrac{ \hat \beta}{2}) - \tfrac{ \hat \beta}{2} \right]$ & --- & $+\infty$ \\
                                    $n(\rr)/n_0$  & $\Theta(\sqrt{2\mathcal{M}} -r)$ & $e^{- \hat \beta \hat \omega r^2}$ & $\hspace{0.4cm}(1 - \pi \hat \beta n_0 r^2/2 )^{-2}$ & --- & $\delta^{(2)}(\rr - \boldsymbol{\mathcal{D}})$ \\

    \hline\hline
\end{tabularx}
    \caption{Summary of analytical, maximum entropy solutions to~\eqnreft{eqn:meanfield} at fixed $\mathcal{M}$ as obtained by Smith and O'Neil~\cite{smith1990nonaxisymmetric}.  The points corresponding to solutions \textbf{A--E} are also labeled in Fig.~\ref{fig:Overview}. \textbf{A:} Uniform (Rankine) distribution. \textbf{B:} Gaussian. As $\hat \beta\rightarrow0$,  $\hat \beta \hat \omega$ remains finite, and is  determined by the transcendental equation $\mathcal{M} = (1-e^{\hat \beta \hat \omega})^{-1} + (\hat \beta \hat \omega)^{-1}$. The energy is $g(\alpha) = \tfrac{1}{8}e^\alpha \mathrm{csch}^2(\tfrac{\alpha}{2})(\gamma + \mathrm{Ei}(-2 \alpha) - 2\mathrm{Ei} (-\alpha) + \ln(\alpha/2) )$ where $\gamma$ is Euler's constant and $\mathrm{Ei}(x )$ is the exponential integral function. \textbf{C:} Riccati.  At  $\hat\omega = 0$,  \eqnreft{eqn:meanfield} reduces to a Riccati equation with an exact solution. Here $f(\hat \beta) = (1+\tfrac{2}{\hat \beta})[ 1 -\tfrac{2}{\hat \beta}\ln (1+ \tfrac{\hat \beta}{2}) ]$. \textbf{D}: off-axis, marks the bifurcation point where $\hat \omega(\mathcal{H}_c,\mathcal{M}) = 1$. At higher energies the on-axis states are no longer stable. \textbf{E:} supercondensate, where the density distribution collapses to a point.  Fields left blank cannot be expressed in closed form or as a transcendental equation, and must be evaluated numerically.  $\Theta(r)$: Heaviside step function; $\delta(\rr)$: Dirac delta function. The schematic density profiles show a slice along the $x$-axis, assuming the dipole moment points along $+\hat{\mathbf{x}}$.}
    \label{tab:MFsolutions}
\end{table*}

\subsection{Mean-field theory of the vortex gas}
\label{sec:StatMechSection}

To describe our experiment, we consider a mean-field approach to the vortex equilibria~\cite{salman2016long}, as first developed by Joyce and Montgomery~\cite{joyce_montgomery_1973}. In the mean-field theory the point-vortex distribution is replaced with a coarse-grained field $\sum_i\delta({\mathbf{r}-\mathbf{r}_i})/N \rightarrow n(\rr)$; the entropy is given by 
\begin{equation}
\mathcal{S}[n(\mathbf{r})] = - \int d^2\rr \; n(\rr) \ln n(\rr).
\label{eqn:entropy}
\end{equation}
The energy and angular momentum are rescaled to remove explicit dependence on $N$ via $\mathcal{H}\equiv H/N^2$, $\mathcal{M}\equiv M/N$, and their
rescaled conjugate variables are the inverse temperature $\hat{\beta} \equiv \beta E_0N$ and rotation frequency $\hat{\omega}\equiv \omega R^2/(E_0 N)$.
In terms of $n(\rr)$,
\begin{align} 
\mathcal{H} &= \tfrac{1}{2} \int d^2\rr\; n(\rr) \phi(\rr), & \mathcal{M} &= \int d^2\rr\; r^2 n(\rr), 
\end{align}
where $\phi(\rr)$ is the stream function that satisfies the Poisson equation $\nabla^2 \phi(\rr) = -4 \pi n(\rr)$. The equilibrium vortex distributions maximize $\mathcal{S}$ subject to the constraints of fixed energy and angular momentum and can be shown to satisfy the Poisson-Boltzmann equation
\begin{align} 
 n(\rr) &=n_0 \exp\{- \hat \beta [\phi(\rr) + \hat {\omega} r^2]\},
\label{eqn:meanfield}
\end{align}
where the density prefactor $n_0$  is determined by the normalization condition, which we are free to choose as  $\int d^2\mathbf{r}\; n(\rr) = 1$. 

\subsection{Phase diagram}
\label{sec:PhaseDiagram}
The mean-field phase diagram of the chiral vortex gas in a disk geometry is laid out in the  work of Smith and O'Neil~\cite{smith1990nonaxisymmetric}. They show that the system exhibits a symmetry breaking at high energy due to a competition between the energy, which requires vortices to be in close proximity and far from the container walls [\eqnreft{eqn:PVHamiltonian}], and the angular momentum, which fixes the mean-square radius [\eqnreft{eqn:PVAngMom}]. At low energy, equilibria share the underlying rotational symmetry of the container, whereas the high-energy equilibria are non-axisymmetric states which break this symmetry.  An overview of the system is shown in Fig.~\ref{fig:Overview}. In Fig.~\ref{fig:Overview}(a) we show an example of the entropy versus energy for the system found from numerically solving \eqnreft{eqn:meanfield} at a fixed angular momentum $\mathcal{M}=0.34$ (for numerical details see Appendix~\ref{sec:AppMF}).

 Analytical solutions can be obtained to \eqnreft{eqn:meanfield} for a few special cases~\cite{smith1990nonaxisymmetric}; these solutions will provide useful reference points to compare against our experiment, and are hence summarized in Table~\ref{tab:MFsolutions}. We also label these solutions as  \textbf{A}--\textbf{E} in Fig.~\ref{fig:Overview}(a). For all energies below point \textbf{D} the equilibria are axisymmetric, and depend only on the radial variable~$r$. For a given $\mathcal{M}$, the lowest-energy solution \textbf{A} is a uniform (Rankine) vortex that extends to radius $r=\sqrt{2\mathcal{M}}$, rotating rigidly at frequency $\hat \omega_R = (2\mathcal{M})^{-1}$ . Increasing the energy rounds out the edge of the density profile until  \textbf{B}, where $\hat \beta\rightarrow 0^{\pm}$ and $\hat \omega\rightarrow \pm \infty$, in such a way that $\hat \beta \hat \omega$ remains finite and the density becomes Gaussian $n(r) = n_0 e^{-\hat \beta \hat \omega r^2}$. At energies higher than point \textbf{B}, $\hat \beta$ and $\hat \omega$ become negative, but the solutions remain axisymmetric; the solution becomes more strongly peaked at the origin to increase the energy, while developing longer tails to satisfy the angular momentum constraint.  At  \textbf{C}, where $\hat \omega =0$, \eqnreft{eqn:meanfield} reduces to a Ricati equation with the exact solution shown in Table~\ref{tab:MFsolutions}, with the vortex distribution taking the form of a squared Lorentzian, $n(r) = n_0/(1-\pi \hat \beta n_0 r^2/2)^2$. 
 
 The onset of the off-axis phase occurs at the bifurcation point \textbf{D}, where $\hat{\omega}(\mathcal{M},\mathcal{H}_c)=1$. For energies above $\mathcal{H}_c$ the on-axis states are no longer stable; the off-axis states have the highest entropy and are hence the relevant solutions for thermal equilibrium~ [see Fig.~\ref{fig:Overview}(a)]. Within mean-field theory, it is typical to use the growth of the macroscopic dipole moment 
\begin{equation}
  \boldsymbol{ \mathcal{{D}}} = \int d^2\rr \; \rr\, n(\rr),
\end{equation}
to mark the transition to the off-axis phase. It can be rigorously treated within perturbation theory near the bifurcation point, and can be shown to grow as $|  \boldsymbol{ \mathcal{{D}}}| \sim |\mathcal{H}-\mathcal{H}_c|^{1/2} $ for $ \mathcal{H} \gtrsim \mathcal{H}_c$~\cite{smith_phase-transition_1989,yu2016theory}. It must be noted however, that the square-root growth is easily accessible only in the limit of very large $N$~\cite{yu2016theory}; for small $N$,  as is relevant here, the dipole moment is 
\begin{equation}
D = \sum_j \rr _j,
\label{eqn:dipoleFiniteN}
\end{equation}
which exhibits a noise floor $ |D| \sim M/\sqrt{N}$~\cite{smith1990nonaxisymmetric,yu2016theory} in the on-axis phase \footnote{This behaviour follows from the fact that below the off-axis transition the vortex positions are approximately normally distributed about the origin, with  the width of the normal distribution determined by the angular momentum.}. This noise floor washes out the square-root growth with energy in the mean value of $|D|$~\cite{yu2016theory}.

Finally, \textbf{E} marks the so-called ``supercondensation" limit $\mathcal{H}\rightarrow \infty$~\cite{Kraichnan:1975ku}, where the density distribution collapses to a point. Here $ \hat \beta$ tends to a universal value which is independent of the container geometry, $\hat \beta_s = -2 $ and $\hat \omega$ approaches $\hat \omega_s =  (1-\mathcal{M})^{-1}$, which corresponds to the orbit frequency of a single point vortex located off axis at $ |\boldsymbol{\mathcal{D}}_s| = \sqrt{\mathcal{M}}$.

Figure~\ref{fig:Overview}(b) shows the full phase diagram for the chiral vortex gas on a disk as a function of energy $\mathcal{H}$ and angular momentum $\mathcal{M}$.  Points \textbf{A}--\textbf{D} extend to be lines in this two-dimensional plane.  The colored symbols in Fig.~\ref{fig:Overview}(b) indicate the vortex gas equilibria that we observe in our experiment, which we present in the next section.

\begin{figure*}
    \centering
    \includegraphics[width=\textwidth]{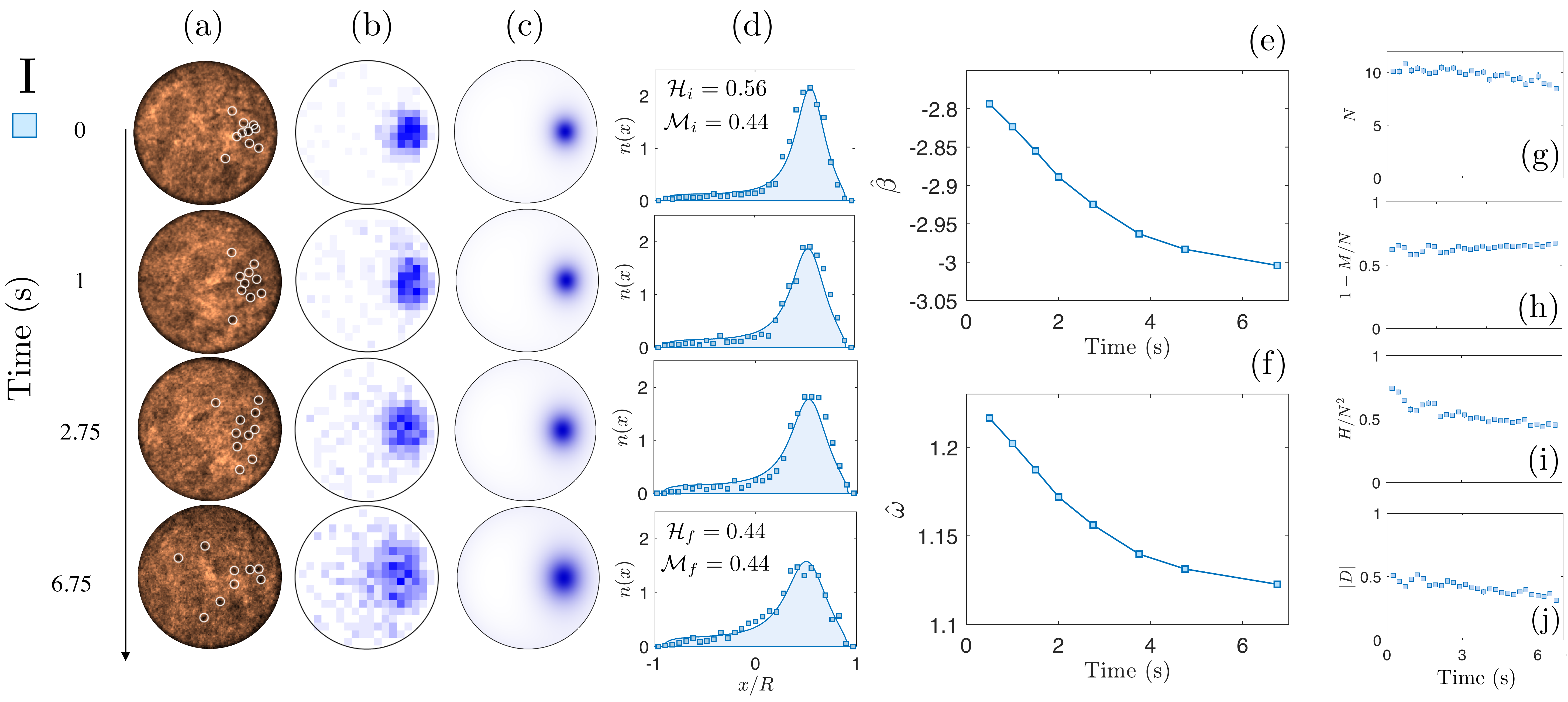}
  \caption{\textbf{Experiment I}: injection and evolution of a near-equilibrium state. (a) Examples of measured vortex distributions against hold time. The white circles indicated detected vortices. (b) Vortex position histograms, with samples aligned along the $x$-axis. (c) best-fit distributions from solving the mean-field Poisson-Boltzmann equation. (d) Comparison of experimental data against mean-field for the column-integrated vortex density $n(x)$; all fits have the same value of angular momentum, $\mathcal{M}= 0.44$. (e) and (f) show mean-field best-fit values for the inverse temperature $\hat \beta$ and rotation frequency $\hat \omega$ respectively. (g)-(j) show macroscopic measures of the vortex distribution vs. time, as calculated from the experimentally-measured vortex positions. (g) Vortex number, $N$. (h) Angular momentum (per vortex) $1 - M/N$. (i) Energy (per vortex squared) $H/N^2$. (j) Dipole moment $|\mathbf{D}|$. }
    \label{fig:StirringSchematic}
\end{figure*}

\section{Experiment}
\label{sec:Expt}
The vortex gas system may be realized in an oblate atomic Bose-Einstein condensate which is  near zero temperature and trapped by a hard-walled confining potential~\cite{groszek2018motion}. We briefly summarize our experimental setup here, with full details provided in Appendix~\ref{app:Expt}. The experimental system consists of approximately $2 \times 10^6$ $^{87}$Rb atoms confined in a gravity-compensated optical potential. The potential results from the combination of an oblate red-detuned optical trap (harmonic trapping frequencies $\{\omega_x,\omega_y,\omega_z\} \sim 2 \pi \times \{1.8,1.6,108\}$~Hz), with the blue-detuned optical potential produced from direct imaging of a digital micromirror device (DMD), of depth approximately $5 \mu$~\cite{gauthier2016direct,gauthier2019giant}, where $\mu$ is the chemical potential. The DMD projection provides nearly hard-walled circular confinement, here configured to produce a disk-shaped trap. The result is a nearly uniform condensate with a horizontal radius of $50~\mu$m, vertical Thomas-Fermi radius of $6~\mu$m, healing length of approximately $\xi \sim 500$~nm~\citep{gauthier2019giant}, and condensate fraction of approximately $80\%$. Neglecting the residual harmonic confinement from the red-detuned trap, the radial potential is of the form $V(r)\propto (r/R_0)^\alpha$, for $r\leq R_0 = 50~\mu$m; numerically estimating the projection resolution of approximately $650$~nm full-width-half-maximum~\cite{gauthier2016direct} results in a steep-walled trap with $\alpha\sim 30$. The $1/e$ lifetime of the condensate atom number is approximately 21~s.

The DMD also offers real-time dynamical control of the optical potential, which we use to inject vortices into the superfluid through a variety of stirring protocols.  We use a combination of paddle-shaped stirring potentials~\cite{gauthier2019giant} and circular pinning potentials~\cite{samson2012generating}, which allow a high degree of control over the initial vortex positions. The stirring protocols also minimize the creation of other (undesirable) excitations, such as sound waves, which would cause unwanted heating of the condensate. The precise details of the stirring mechanisms are not central to our analysis; rather, what is important are only the initial vortex positions $\{\rr_i\}$, as these completely specify the initial condition, determining both the energy $H$ and the angular momentum $M$ [see Sec.~\ref{sec:PointVortexModel}
].  However, for completeness the stirring protocols used are detailed in  Appendix~\ref{app:Expt}, along with numerical simulations schematically illustrating the stirring processes. To measure the vortex positions, images are captured utilizing dark-ground Faraday imaging~\cite{bradley1997bose} after a short (3--5)~ms time of flight that expands the vortex cores to improve visibility. The vortex positions are determined from images of the condensate density using a blob-detection algorithm~\cite{rakonjac2016measuring}.

 \begin{figure*}[t!]
     \centering
     \includegraphics[width=.95\textwidth]{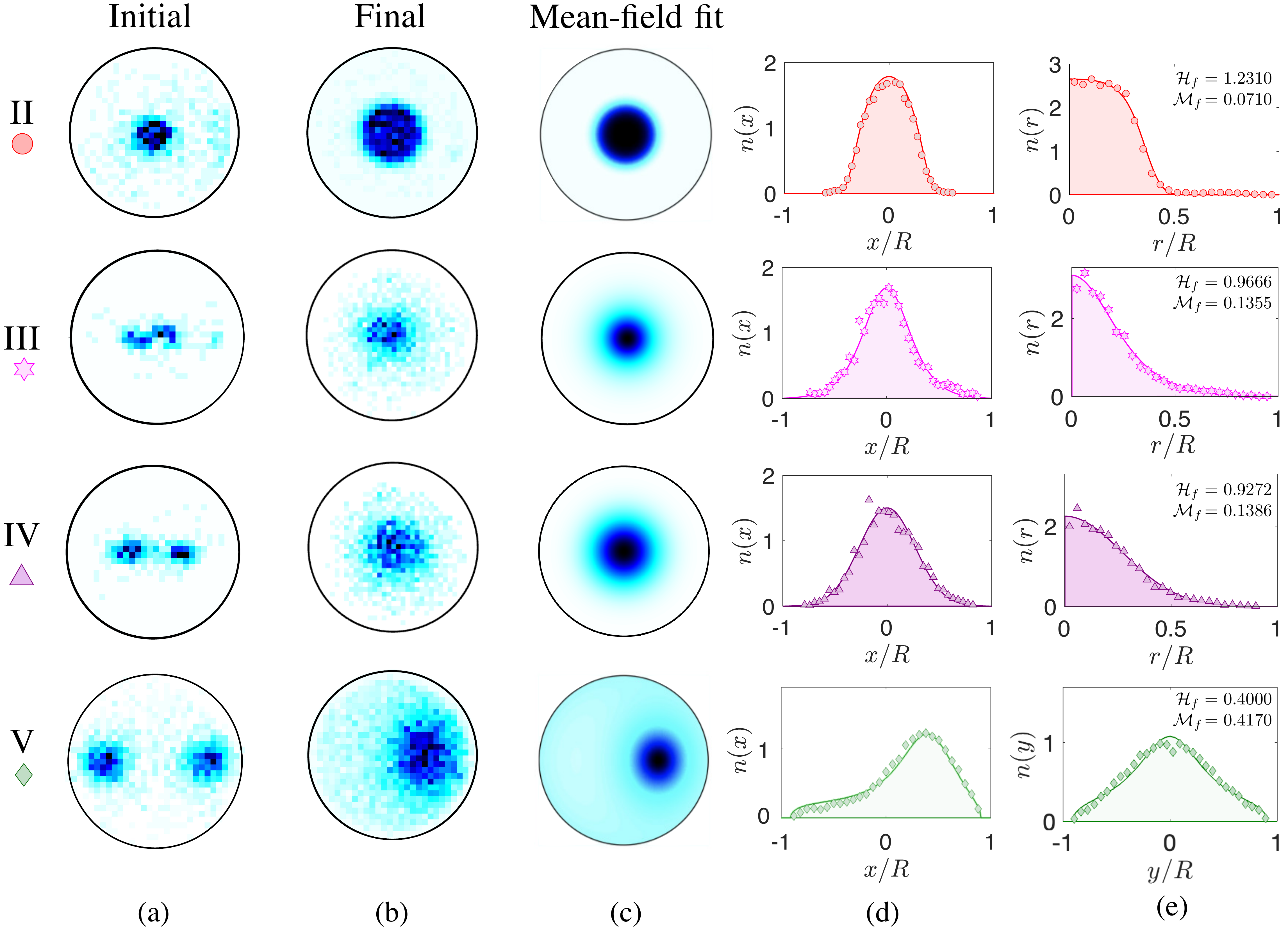}
     \caption{\textbf{Experiments II--V:} comparison of experimental vortex distributions with solutions of the Poisson-Boltzmann equation.  (a) and (b) show the initial and final experimental vortex density distributions respectively. (c) Best-fit solutions from solving the Poisson-Boltzmann equation. Note that in the off-axis and bimodal distributions  the samples have been oriented along the $x$-axis. (d, e): Integrated 1D density profiles $\textstyle {n(x) = \int dy\; n(x,y)} $ and $n(r) = \int d\phi\; n(\rr)$ comparing experimental results (markers) with the mean-field solutions (shaded curves). Note for for the off-axis state we show $\textstyle {n(y) = \int dx\; n(x,y)} $ instead of $n(r)$.}
     \label{fig:histograms_compare}
 \end{figure*}

\subsection{Injection and cooling of a near-equilibrium state}

\textbf{Experiment I:} Our first experiment, shown in Fig.~\ref{fig:StirringSchematic}, considers a  scenario similar to that in Ref.~\cite{gauthier2019giant} --- we inject a vortex distribution which closely resembles an equilibrium state, and track its subsequent evolution. By dragging a paddle-shaped barrier through one edge of the condensate [see Appendix~\ref{app:Expt}], a single off-axis cluster, concentrated  near $r/R \sim 0.5$, is injected [Fig.~\ref{fig:StirringSchematic}(a), top]. This initial condition closely resembles an equilibrium state in the negative temperature, off-axis phase [Fig.~\ref{fig:Overview}(iii)].   The other panels in Fig.~\ref{fig:StirringSchematic}(a) show examples of the measured vortex distribution at different times, and Fig.~\ref{fig:StirringSchematic}(b) shows vortex histograms gathered from approximately 40 samples at each time.  Note that in the actual dynamics, the vortex cluster orbits within the trap; the distributions are oriented such that the dipole moment points along the $x$-axis.  In Fig.~\ref{fig:StirringSchematic}(b) it is clear that the cluster remains off axis for the entire duration of the experiment, and slowly expands with time. 

We compare the observed vortex density distributions with the predictions of the Poisson-Boltzmann equation [Eq.~(\ref{eqn:meanfield})] by minimising the least-squares error to the column-integrated vortex densities 
\begin{align}
n(x) &= \int \dd y\, n(x,y), & n(y) &= \int \dd x\, n(x,y),
\end{align}
 using $\hat \beta$ and $\hat \omega$ as fitting parameters (equivalently fitting $\mathcal{H}$ and $\mathcal{M}$).  Figure~\ref{fig:StirringSchematic}(c) shows the best-fit distributions $n(\rr)$ for the corresponding times shown in Fig.~\ref{fig:StirringSchematic}(b), and Fig.~\ref{fig:StirringSchematic}(d) compares the experimental data against the mean-field solution for the column-integrated density $n(x)$. The fits match the data well, indicating that the system is in equilibrium at negative temperature, as was indirectly inferred in Ref.~\cite{gauthier2019giant}. As time progresses the vortex gas gradually loses energy and become more diffuse while maintaining an approximately fixed angular momentum $\mathcal{M} \approx 0.44$ [Fig.~\ref{fig:StirringSchematic}(d)]. 
 
 As shown in Figs.~\ref{fig:StirringSchematic}(e) and (f), this leads to a gradual decrease in both the inverse temperature $\hat \beta$ and the rotation frequency $\hat \omega$. Notably $\hat \beta$ decays as the energy decreases (i.e. as time increases), demonstrating that the microcanonical specific heat $C \propto - \partial \mathcal{H} / \partial{\hat \beta}$ is \emph{negative} within the off-axis phase, breaking ensemble equivalence~\cite{smith1990nonaxisymmetric}.  
 Despite the slow decay, the state at the end of the experiment is still deep within the off-axis phase with $\hat \omega > 1$ throughout the entire experiment [cf. Table~\ref{tab:MFsolutions}]. The location of the final equilibrium state within the mean-field phase diagram is shown in Fig.~\ref{fig:Overview}(b).

 Corroborating evidence for equilibrium is shown in Figs.~\ref{fig:StirringSchematic}(g)--(i), where we show the time evolution of the nominally conserved quantities: the vortex number $N$, the angular momentum $M$ [Eq.~(\ref{eqn:PVAngMom})], and the energy $H$ [Eq.~(\ref{eqn:PVHamiltonian})], as calculated directly from the experimentally measured vortex positions. Also shown in Fig.~\ref{fig:StirringSchematic}(j) is the dipole moment $|D|$ [Eq.~(\ref{eqn:dipoleFiniteN})] , which is not a dynamically conserved quantity, but its value is constant provided the system is in thermal equilibrium. 

Consistent with our equilibrium mean-field analysis, we find the quantities decay only gradually throughout the experiment. We find that the vortex number [Fig.~\ref{fig:StirringSchematic}(g)] and angular momentum [Fig.~\ref{fig:StirringSchematic}(h)] are quite robust to the presence of dissipation arising from the thermal cloud, with both quantities deviating $\lesssim$ 15\% from their initial values. We note that the robustness of angular momentum to dissipation in a circular container is also noted in viscous fluids~\cite{Li1996decaying}.  The energy is slightly less robust to the dissipation, although the energy decay slows with time and is quite gradual at late times (e.g., for $t>4$ s, $H$ decays only by roughly $10\%$). The dipole moment is the most significantly affected by the dissipation, (decaying roughly 50\% over the duration of the experiment). Nonetheless it remains large throughout the experiment (well above the noise floor $\sim~0.15$ for $M \sim 0.4$, see Sec~\ref{sec:PhaseDiagram}), consistent with a symmetry broken, off-axis phase. 

Finally, we point out that the slow evolution of the experimental system through a series of seemingly microcanonical equilibrium states implies that there is a separation
of the time scales associated with inter-vortex interactions in relation to those associated with the dissipative dynamics due to the presence of a thermal cloud.
Indeed, when we discuss the equilibrium state of any real system, we are implicitly assuming a separation of time scales between
the system we are considering, and its interaction with its surroundings.  True equilibrium is of course an approximation state for any real system --- our results suggest that the dissipation due to the thermal cloud occurs on a much slower time scale that the equilibration of the vortices due to their self-interaction.  We discuss this point in more detail in Sec.~\ref{sec:SPVM} where we present experimental results for the nonequilibrium dynamics of the vortex gas, and simulate the dynamics of the system using a phenomenological stochastic point-vortex model.

\subsection{Nonequilibrium relaxation to maxiumum entropy states}


Having established that the vortex gas is able to attain equilibrium, we now test whether equilibrium can be achieved through turbulent relaxation from nonequilibrium initial conditions. We considered a range of nonequilibrium initial conditions in experiments II -- V, which are shown in Fig.~\ref{fig:histograms_compare}. In experiment II, we test the decay of a multiquantum vortex of circulation $\Gamma = N h/m$, with $N\sim 12$, which is highly unstable and decays into $N$ singly quantized vortices. Experiments III, IV, and V probe the relaxation of two separated vortex clusters each containing $N = 6$ -- $8$ vortices, initially separated by varying distances $d$. The two-cluster distributions contain more spatial structure than a single cluster, and thus have a lower entropy. The maximum entropy principle thus predicts that, under dynamical evolution, the two clusters will merge into a single cluster to maximize the entropy. The initial conditions in experiments II -- V were chosen target regions of the phase diagram within the neighborhoods of the solutions \textbf{A}-\textbf{D} outlined in Table~\ref{tab:MFsolutions}, and are informed by modeling with a dissipative point-vortex model which is presented in the next section (Section~\ref{sec:vortexDynamics}).  The best-fit vortex density distributions are compared against the experimental data in  Fig.~\ref{fig:histograms_compare}, and their positions within the mean-field phase diagram are presented in Fig.~\ref{fig:Overview}(b). 

\textbf{Experiment II:} We find that the decay of the multiquantum vortex produces an on-axis vortex distribution which clearly exhibits a flattop shape with slight rounding at $r/R \approx 0.3$, seen in the 2D distribution $n(\rr)$ [Fig.~\ref{fig:histograms_compare}(II,b)] and the radial density $n(r)$ [Fig.~\ref{fig:histograms_compare}(II,e)]. Similarly, the column-integrated density $n(x)$ exhibits a nearly inverse-parabolic profile [Fig.~\ref{fig:histograms_compare}(II,d)], as expected for the column integration of a constant density. Consistent with these observations, for this state we find $ \hat \beta = +10.348$ and $ \hat \omega = +8.043$, indicating this state a near-minimum-energy Rankine state (\textbf{A}), for which $\hat\beta \rightarrow \infty$ and $\hat \omega = 1/2\mathcal{M} \approx 7.05$ for the best-fit value of angular momentum $\mathcal{M} = 0.071$ [cf.  Table~\ref{tab:MFsolutions}]. Indeed, in the phase diagram [Fig.~\ref{fig:Overview}(b)], it can be seen that the solution is close to the minimum allowed energy determined by the Rankine vortex.

\textbf{Experiment III:} Here we create two clusters initially separated by a distance $d/R = 0.5$ [Fig.~\ref{fig:histograms_compare}(III)(a)]. The clusters are observed to rapidly merge (within $\sim$ 500 ms). After a hold time of $1$~s, the distribution relaxes to an on-axis state at negative absolute temperature with $\hat \beta = -1.07 $ and $\hat \omega = -3.01$. This state lies between the Gaussian (\textbf{B}) and Riccati (\textbf{C}) states within the on-axis phase, as shown in Fig.~\ref{fig:Overview}(b). The state is qualitatively similar to the Riccati solution \textbf{C} --- as can be seen in  Figs.~\ref{fig:histograms_compare}(d) and ~\ref{fig:histograms_compare}(e). The mean-field solution has noticeably longer tails than a Gaussian distribution. 

\textbf{Experiment IV:}  Increasing the initial cluster separation to $d/R = 0.666$[Fig.~\ref{fig:histograms_compare}(IV)(a))], we find the clusters merge after approximately $t \sim 1$ s.  After a hold time of $4$~s, we find the distribution relaxes into a Gaussian-like state. The best-fit from the Poisson-Boltzmann equation gives values $ \hat \beta= +0.088$ and $ \hat{\beta} \hat{\omega} = 7.5$, indicating this state is very close to the infinite-temperature Gaussian state (\textbf{B}) [see Table~\ref{tab:MFsolutions} and Fig.~\ref{fig:Overview}(b)]. A pure Gaussian fit to the data is graphically almost indistinguishable, and yields very similar parameters ($\hat \beta = 0$ and $\hat \beta \omega = 7.07$).

\textbf{Experiment V:}  Upon further increasing the cluster distance to $d/R = 0.7$, after $t \sim 3.75$  s we find the two clusters merge into a single off axis cluster  [Fig.~\ref{fig:histograms_compare}(V)(a) and (b)]. The distribution is more diffuse than in experiment I  and peaks closer to the origin, near $r/R\sim 0.3$). For this state we find $\hat \beta = -3.093$ and $ \hat \omega = +1.052$ indicating that this state, while also at negative temperature and in the symmetry-broken phase, is closer to the transition boundary $\hat \omega = 1$ than the final state in experiment I [see Fig.~\ref{fig:Overview}(b)]].

\begin{figure}
    \centering
    \includegraphics[width=0.75\columnwidth]{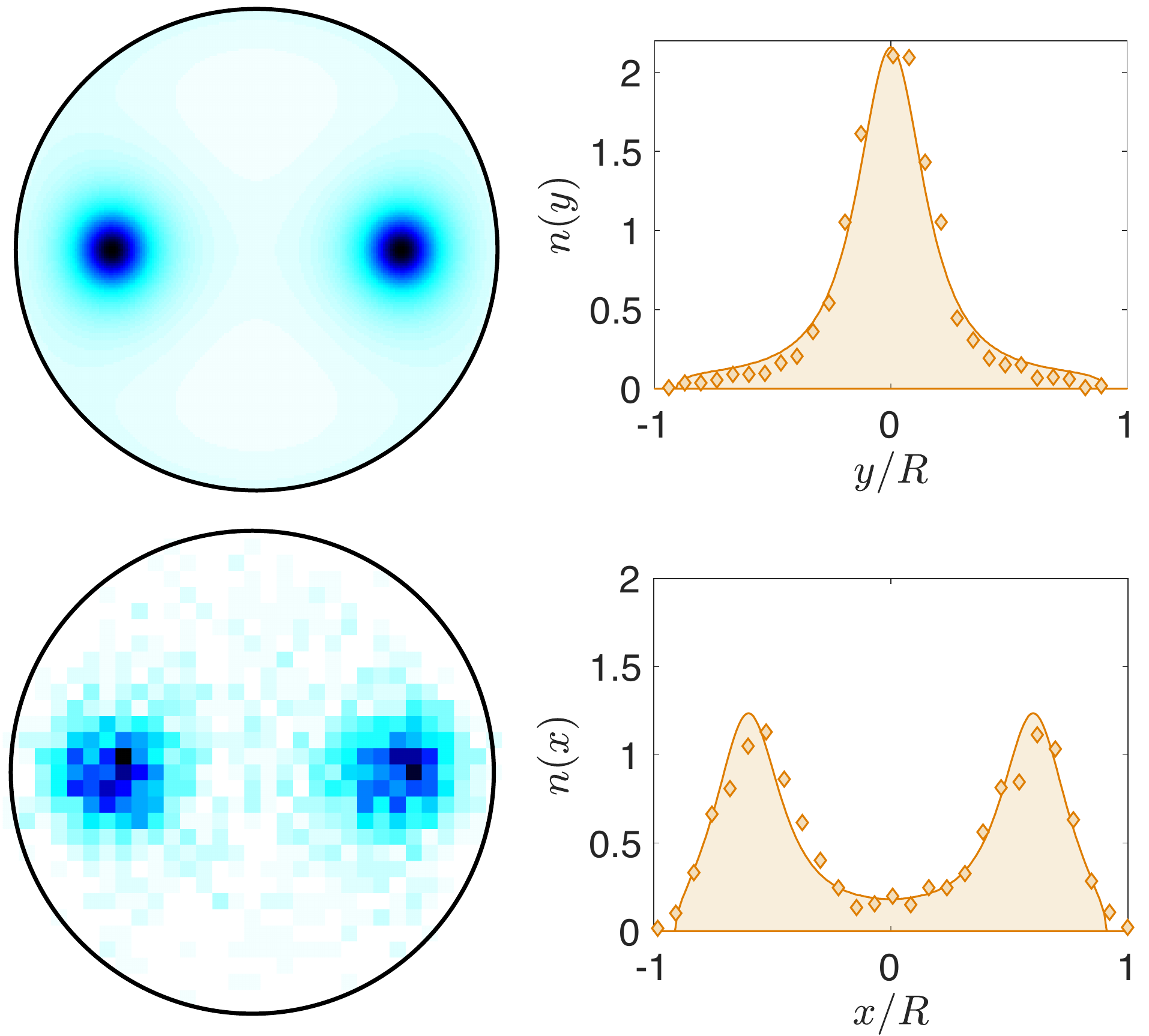}
    \caption{\textbf{Experiment V:} comparison of the initial state with a bimodal solution of the Poisson-Boltzmann equation; note the solution is a (nonequilibrium) local entropy maximum, unlike the (equilibrium) global entropy maximum solutions shown in Fig.~\ref{fig:Overview} and Fig.~\ref{fig:histograms_compare}.}
    \label{fig:bimodal}
\end{figure}

Remarkably, in experiment V we find that we are also able to characterize the initial, \emph{nonequilibrium} vortex distribution via a solution of the Poisson-Boltzmann equation.  Motivated by the linearized analysis of Chavanis and Sommeria~\cite{chavanis1996classification}, we trace out an additional branch of solutions corresponding to a bimodal vorticity distribution.  This branch corresponds to solutions to the fully nonlinear problem not  previously reported by O'Neil and Smith \cite{smith1990nonaxisymmetric}; unlike the on-axis and off-axis branches (Fig.~\ref{fig:Overview}), this bimodal branch is not a state of maximum entropy at any energy. We find it
appears to exist only at high energies where the specific heat capacities are negative. In Fig.~\ref{fig:bimodal}, we compare the early-time ($t<1$~s) vortex distribution in  experiment V against this solution branch. The experimental data are clearly well described by this bimodal branch; this allows us to explicitly calculate the entropy production for this experiment. For the best-fit values $\mathcal{M}=0.45$ and $\mathcal{H}=0.4$ we find the entropy is $\mathcal{S}=0.574$. Upon relaxing to the single-cluster distribution presented in Fig.~\ref{fig:histograms_compare}(V,b), the best fit values are $\mathcal{M}=0.417$ and $\mathcal{H}=0.4$ and the corresponding entropy is $S=0.9282$. This demonstrates that the entropy exhibits a marked increase even though both the angular momentum and energy remain essentially constant.
 
 \section{Vortex gas dynamics}
 \label{sec:vortexDynamics}

\label{sec:SPVM}
So far we have demonstrated that our experiment evolves several initial vortex distributions toward a state that is consistent with microcanonical equilibrium. However, there is some small amount of dissipation present in the system due to the thermal cloud.  This is demonstrated in the results of experiment I presented in Fig.~\ref{fig:StirringSchematic}(e)--\ref{fig:StirringSchematic}(j), which show a rather slow evolution of the thermodynamic variables $\beta$ and $\omega$ due to this dissipation.   As each variable changes by only around 10\% over the course of 6 s of dynamics, the experimental system clearly approximates the conservative dynamics~Eq.~(\ref{eqn:PVdynamics}) over reasonably long timescales.   In this section we attempt to provide a more quantitative understanding of the dissipation.

\begin{figure}[t!]
     \centering
     \includegraphics[width=\columnwidth]{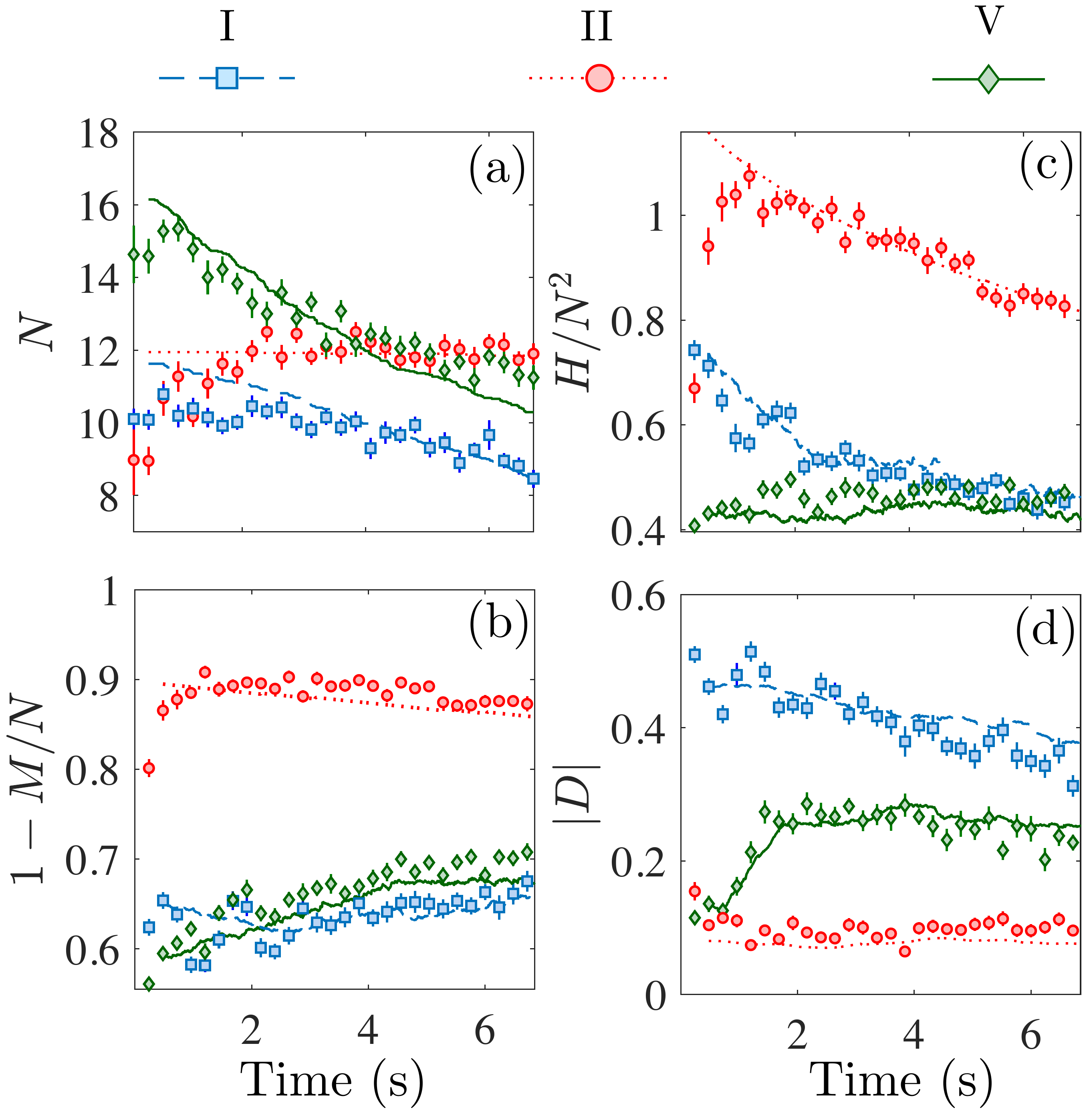}
     \caption{Evolution of the macroscopic properties of the vortex gas following injection. The markers show experimental measurements of (a) the vortex number $N$, (b) energy $H/N^2$, (c) angular momentum $(1-M/N)$ , and (d) dipole moment $|D|$ , as functions of time, for the three different experiments investigated.  Lines show simulation results from the stochastic point-vortex model [\eqnreft{eqn:StochasticPV}], using the experimental data as inputs for the initial conditions (see text).
 } 
     \label{fig:TimeData}
 \end{figure}

To probe the effect of the dissipation, for three of our experiments (I,II and V), we take full series of time data to probe the vortex gas evolution ($7$ s of evolution, $250$ms intervals, $40$ samples at each interval). The data are shown in Figs.~\ref{fig:TimeData}(a)--\ref{fig:TimeData}(d), where we show data for the vortex number $N$, the angular momentum $M$, and the energy $H$ [Sec.~\ref{sec:PVM}] and the dipole moment $|D|$ respectively, [the data shown for experiment I are the same as those shown in Fig.~\ref{fig:StirringSchematic}(g)--\ref{fig:StirringSchematic}(j); they are reproduced here for convenience].

 We find that the dynamical evolution of the quantities shown in Figs.~\ref{fig:TimeData}(a)--\ref{fig:TimeData}(d) can be reproduced by a modified point vortex model~\cite{carnevale1992rates,weiss1993temporal} which incorporates additional dynamical aspects specific to superfluids~\cite{simula2014emergence,billam2015spectral}. Our modified point-vortex model is supplemented by mutual friction and Brownian motion, taking the form
 \begin{equation}
      \mathrm{d} \rr_j = [ \mathbf{v}_j - \gamma \hat{\mathbf{z}} \times  \mathbf{v}_j]\, \mathrm{d}t + \sqrt{2 \eta}\,\mathrm{d} \mathbf{W}_j,
      \label{eqn:StochasticPV}
 \end{equation}
 where $\gamma$ is the mutual friction coefficient, and $\eta$ is the vortex diffusion rate. The noises $\mathrm{d} \mathbf{W}_j = (\mathrm{d} {W}_j^x,\mathrm{d} {W}_j^y)$ are independent complex Gaussian random variables with $\langle \mathrm{d}W_j^\alpha(t) \mathrm{d}W^\beta_k(t) \rangle = \delta_{jk}\delta_{\alpha \beta} \mathrm{d}t$, and where all other correlations vanish. The vortex velocities $\mathbf{v}_j$ are given by the Hamiltonian evolution $\dot x_j$ and $\dot y_j$ as expressed in \eqnreft{eqn:PVdynamics}. Vortex annihilation at the boundary is also included by removing vortices that come within a distance of one healing length $\xi$ to the boundary.

The lines in Fig.~\ref{fig:TimeData} are the results obtained from numerically simulating~\eqnreft{eqn:StochasticPV}. The experimentally measured vortex positions at $t=0.25$~s are used as the initial conditions, and we averaged over all approximately 40 experimental runs for each stirring protocol.  A small number of additional vortices are added at random locations in each trajectory to account for undercounting at early times. The undercounting is most evident for experiment II [Fig.~\ref{fig:TimeData}(a), circles], where the vortices are initially so densely packed that they cannot be individually resolved [cf.  Figs.~\ref{fig:histograms_compare}(II,a) and Fig.~\ref{fig::vortex_injection_simulation}] and the detected vortex number hence gradually increases until $t\sim2$~s~\footnote{The positions of the additional vortices were determined by using the experimental histograms [cf.~Fig.~\ref{fig:StirringSchematic}] as probability distributions for rejection sampling. For experiments I and V, one additional vortex was added to each experimental run. For experiment II, between 2 and 3 vortices were added with equal probability.}. 

The model yields good quantitative agreement with the experimental observations for all three experiments, and captures the salient features of the evolution quite well. For example, the model captures the rapid decay in $H$ observed for $t<2$s in experiment I, and the slower decay for $t>2$s. Similarly, in experiment V, the model accurately captures the growth of the dipole moment with time;  $|D|$  is initially small due to the symmetric cluster injection [cf. Fig.~\ref{fig:histograms_compare}(V,a)], but jumps suddenly to $|D|\sim 0.3$ at $t\sim 1$ -- $2$~s as the clusters merge into a single off-axis cluster (by contrast, notice in experiment II that $|D|\sim 0.1$ throughout the evolution, consistent with the finite $N$ noise floor~[Sec.~\ref{sec:PhaseDiagram}]). 

For the simulation results shown in Fig.~\ref{fig:TimeData} the magnitude of the mutual friction coefficient is identical for all three cases, $\gamma = 2 \times 10 ^{-3}$. Curiously however, the on axis and off axis scenarios require different values of noise to capture the trends in the data. In experiment II, where the cluster is on axis, the decay is described purely by mutual friction, with $\eta = 0$, whereas in experiments I and V, where the cluster is off axis, significant diffusion is required, with $\eta = 3.5 \times 10^{-2}$. The noise is found to be crucial to reproducing the trends of the experimental data for the off-axis states; in particular, mutual friction alone yields essentially no decay of the dipole moment for reasonable values of $\gamma$, and also cannot reproduce the slight increase in the angular momentum per vortex observed in Fig.~\ref{fig:TimeData}(c).

\begin{figure}[!t]
\includegraphics[width = \columnwidth]{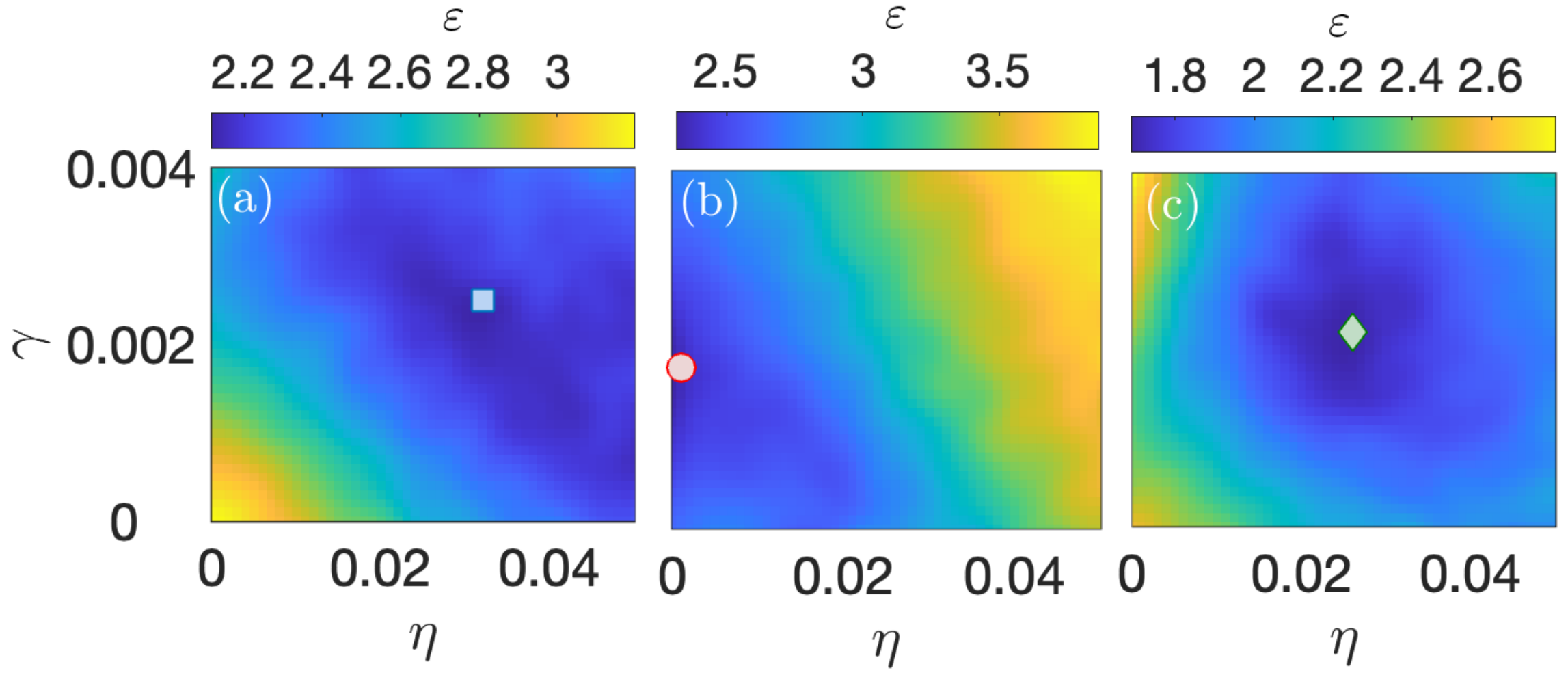}
\caption{Errors for the fitting of the stochastic point-vortex model shown obtained by minimizing \eqnreft{eqn:time_err}. (a) experiment I, (b) experiment II, and experiment V. Markers indicate the best-fit parameters.}
\label{fig:tdata_err}
\end{figure}

 A more comprehensive fitting analysis, presented in Fig.~\ref{fig:tdata_err}, confirms this picture. Figure~\ref{fig:tdata_err} shows estimated optimal values for  $\gamma$ and  $\eta$ obtained by minimizing the sum-of-squares error $\varepsilon$ between the theoretical and experimental values of $N$, $H$, $M$ and $|D|$, i.e., by minimizing
 \begin{equation}
 \label{eqn:time_err}
\varepsilon(\gamma, \eta) =   \sum_{t_k \in \mathbb{A}} \sum_{f \in \mathbb{B}} \left (\frac{f_{\rm{exp}}[t_k] - f_{\rm{th}}[t_k] }{ f_{\rm{exp}}[t_k] } \right)^2,
\end{equation}
where $\mathbb{A} = \{t_i, t_{i+1}, \dots, t_f\}$ and $\mathbb{B} = \{N,H,M,|D|\}$. We use $[t_i,t_f] = [2,6.75]$s, with the  initial time $t_i$  chosen such that the experimental data could be used as the initial conditions, without needing to compensate for undercounting at early times. It is clear from Fig.~\ref{fig:tdata_err}  that the optimal value for $\gamma$ is similar in all three cases. However, the on-axis case requires no noise while the off-axis cases require  $\eta \sim 10^{-2}$. The exact numerical values should be interpreted with some caution, as these fluctuate with the choice of initial conditions $t_i$ (roughly a factor of $2$ for different choices of $t_i$). Nonetheless, the on-axis case (experiment II) consistently yields $\eta \sim 0$, whereas the two off-axis cases (experiments I and V) yield $\eta \sim 10^{-2}$.

The mutual friction term involving $\gamma$ may be rigorously derived~\cite{tornkvist1997vortex} from the damped Gross-Pitaevskii equation familiar from $c$-field methods~\cite{blakie2005projected,Gardiner_2002,gardiner2003stochastic,rooney2012stochastic}, where $\gamma$ describes the damping rate of the condensate due to Bose-enhanced collisions with a stationary thermal cloud~\cite{blakie2008dynamics}. While we add the Brownian motion term as a phenomenological parameter, we note that  thermally driven density and phase fluctuations of the condensate, which are neglected within a pure point-vortex treatment, will also contribute to the vortex dynamics~\cite{tornkvist1997vortex,groszek2018motion}. 
 
 Our findings in this section show that Brownian motion is essential to capturing the thermal decay observed in global measures of the vortex distribution. This finding appears to be contrary to the standard paradigm of mutual friction  (e.g., see ~\cite{kim2016role,sachkou2019coherent}).  While one might anticipate a fluctuation term based on the usual fluctuation-dissipation arguments, it should be noted that these arguments cannot be applied in a straightforward manner here; typically the competition between dissipation and noise serves to produce a steady thermal distribution, whereas here both terms are effectively dissipative --- the equilibrium is one in which no vortices are present at all. In fact, by analogy with classical vortex methods for viscous fluids~\cite{chorin2013vorticity}, the noise can be interpreted as  an effective viscosity. The noise is found to be necessary only for the off-axis states, which are at negative absolute temperature. A possible cause is that trap noise due to diffraction in the optical potentials may be more important at the trap edge. A more intriguing possibility is that the noise might be due to the negative-temperature vortex system being  coupled to a positive-temperature phonon bath. 
 
 \section{Conclusions and outlook}
 \label{sec:conclusions}
We have experimentally studied equilibrium and nonequilibrium states of a chiral vortex gas confined within a disk-shaped atomic Bose-Einstein condensate.  The steady-state distributions have been shown to be in quantitative agreement with the predictions of the microcanonical ensemble, at both positive and negative temperatures. While there is a small amount of dissipation present in the system, this is slow compared to the mixing timescale of the conservative vortex dynamics; this leads to the system slowly evolving through different microcanonical equilibria as time progresses. We have also quantified the sources of dissipation, showing that the dynamics of the quantum vortices are quantitatively described by point-vortex dynamics supplemented by thermal friction and Brownian motion.

We have found that the mean-field predictions for the vortex density describe our system remarkably well, despite the system containing only $N\sim 12$ vortices, suggesting beyond-mean-field corrections may be negligible in much of the vortex gas phase diagram. While this might seem surprising in a low-dimensional system, we note that the long-range interactions mean that the vortices effectively have many nearest neighbors, supporting the use of a mean-field approach. 

Note, however, the mean-field description cannot describe physics at the scales of the intervortex distance. This can become important, for example, at low energies, where the vortex distribution approaches the Rankine vortex [\textbf{A}, Table~\ref{tab:MFsolutions}]. Recently, Bogatskiy and Weigmann~\cite{bogatskiy2019edge} have shown the Rankine vortex is expected to exhibit quantum Hall analog physics, including density oscillations on scales comparable to the intervortex distance, and quantized edge solitons. Our results in experiment II (also Ref~\cite{stockdale2019universal}) show that the system is nearing the Rankine vortex regime, suggesting such physics is within reach experimentally.

Our results present atomic gas superfluids as complementary to existing platforms for the fundamental study of two-dimensional turbulence. The mean-field Poisson-Boltzmann equation, although originally developed for classical fluids, cannot be straightforwardly  applied to the continuous vorticity distributions of classical fluids as it is derived from the point-vortex approximation~\cite{miller1990statistical,robert1991maximum,robert_sommeria_1991}.  This is because the point-vortex approximation does not respect the conservation of other important quantities such as the peak vorticity and the so-called `Casimirs' that are preserved by the full Euler equation~\cite{miller1990statistical,robert_sommeria_1991}. By contrast, in our system the point-vortex approximation is an excellent description; for larger $N$ and lower dissipation, the Poisson-Boltzmann equation could in principle be applied with no fitted parameters. Another complementary aspect of our system is control over dimensionality; in two-dimensional classical fluid flows it is often difficult to determine the influence of the third dimension~\cite{tabeling2012turbulence}.
While our particular system is quasi-2D, we note that quantum systems permit the complete freezing out of the third dimension to the zero-point motion~\cite{hadzibabic2006berezinskii}.

As Hamiltonian systems with long-range interactions are  known to exhibit divergent thermalization times in the large-particle limit~\cite{levin2014nonequilibrium,Patker2018}, it would be interesting to experimentally probe systems with a larger number of vortices $N$
to test whether superfluids suffer the same issue. Here more complex routes to equilibrium (involving many cluster mergers) could be tested, to determine whether local entropy maxima such as vortex crystals~\cite{fine1995,jin1998,adriani2018clusters}, or nonequilibrium steady states such as core-halo states~\cite{Patker2018} emerge in superfluid systems.  The 2D quantum vortex gas presents a rather unique system to study turbulent phenomena; here, unlike in a classical, viscous fluid, the number of active degrees of freedom can be varied independently of the dissipation mechanism. In a viscous fluid both are controlled by the Reynolds number, $\mathrm{Re}$. In 2D the number of active degrees of freedom scales as $N\sim \mathrm{Re}$, while the energy dissipation rate scales as $\varepsilon \sim \mathrm{Re}^{-1}$~\cite{batchelor1969}). In contrast, for the quantum vortex gas the dissipation rate from mutual friction $\gamma$ is \emph{independent} of the vortex number $N$.

Future work on vortex matter may also provide new insights into the role of vortex mass in quantum fluids. Theoretical mass estimates range from $0$ to $\infty$~\cite{simula2018vortex}, while others argue the corrections do not affect the dynamics or are undefined~\cite{thouless2007vortex,lucas2014sound}. Crucially, recall  that the absence of vortex mass in \eqnreft{eqn:PVHamiltonian} is the property that results in the bounded phase space~\eqnreft{eqn:phasespace} and hence the existence of negative-temperature states. For large enough mass, the mass-dependent cyclotron-like motion~\cite{driscoll1990,shukla2020Magnus,richaud2020vortices} would alter the equilibrium profiles and eventually destroy the negative-temperature states~\cite{smith1990nonaxisymmetric}; our observations in agreement with massless vortices suggest that any vortex mass corrections are likely to be small.

The continued study of vortex matter will also benefit the growing interest in technological applications of atomtronic devices~\cite{eckel2016contact,burchianti2018connecting, gauthier2019quantitative}, and renewed interest in nanomechanical  resonators  with  superfluid helium~\cite{sachkou2019coherent,harris2016laser,souris2017ultralow,he2020strong,varga2020observation}.
These experiments suggest an enhanced understanding of superfluid turbulence will be required.
Equilibrium theories of vortex matter may prove a useful tool in predicting the end states of turbulence in such applications 
provided fluid flows are sufficiently two dimensional
--- particularly in helium, where vortices cannot be directly imaged. As some systems are known to exhibit effective equilibria under steady driving and loss~\cite{FossFeig2017}, it would be interesting to consider the possibility of emergent equilibria in driven superfluid systems, such as the superfluid helium Helmholtz resonator recently studied in Ref.~\cite{varga2020observation}.

\section*{Acknowledgements}
We thank E. Kozik, N. Proukakis, and T. Simula for useful discussions.  M.T.R. and T.W.N. thank the Institute for Nuclear Theory (INT) at the University of Washington for its kind hospitality and stimulating research environment at the INT-19-1a workshop, during which some of this research was undertaken. This research was supported in part by the INT's U.S. Department of Energy Grant No. DE-FG02- 00ER41132.  This research was supported by the Australian Research Council (ARC) Centre of Excellence for Engineered Quantum Systems (EQUS, CE170100009), and ARC Discovery Projects Grant DP160102085. This research was also partially supported by the ARC Centre of Excellence in Future Low-Energy Electronics Technologies (FLEET, project number CE170100039) and funded by the Australian Government. G.G. acknowledges the support of an Australian Government Research and Training Program Scholarship. XY acknowledges the support from NSAF with grant No. U1930403 and NSFC with grant No. 12175215. T.W.N. acknowledges the support of an ARC Future Fellowship FT190100306. Computing support was provided by the Getafix cluster at the University of Queensland. 

\section{Experimental procedure}
\label{app:Expt}
\subsection*{Initial state preparation}
\begin{figure*}
\centering
\includegraphics[angle=0,width =1.5\columnwidth]{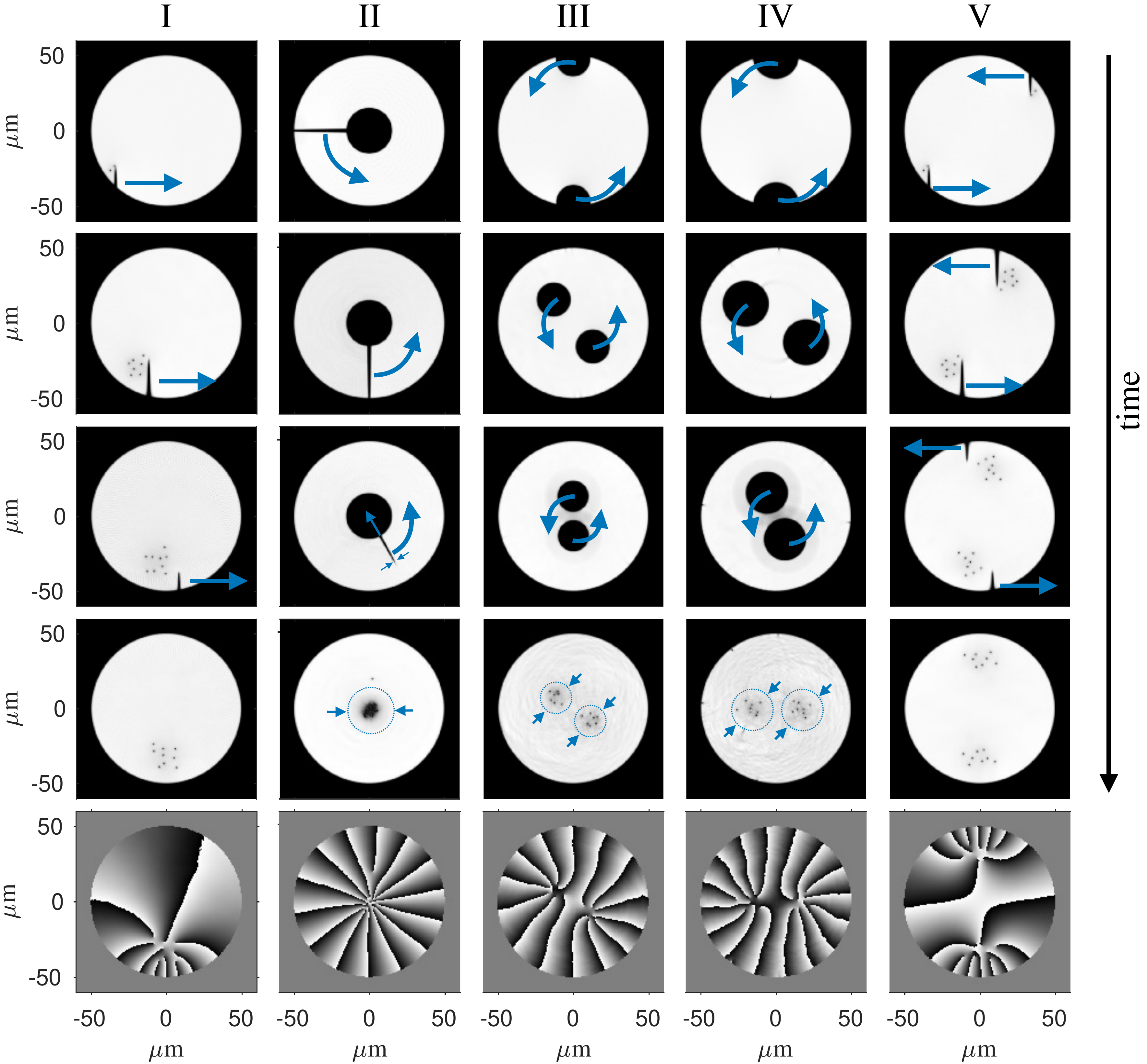}
\caption{Damped GPE simulations of vortex injection methods as described in the main text. Columns, left to right: \textbf{experiments I-V}.  Time increases moving down the columns but is not the same for the images across each row due to the different stirring schemes. Instead, frames from the simulation are chosen to schematically illustrate the stirring process.  The blue arrows indicate the motion of the time-dependent stirring potentials. The second-from-bottom and bottom rows show the condensate density and phase respectively, after all stirring potentials are ramped off. The phase profiles demonstrate that all vortices generated are of the same sign.  Note that \textbf{experiments II-IV} initially result in a giant (multiquantum) vortices at the pinning beam locations, from which singly quantized vortices eventually emerge with increasing hold time. This illustrates the difficulty with initially identifying the vortex positions at early times as discussed in Sec.~\ref{sec:SPVM} (see also Ref.~\cite{stockdale2019universal}).  Movies of the five stirring sequences can be found in Supplemental Material~\cite{SM}.}
\label{fig::vortex_injection_simulation}
\end{figure*}

A key ingredient for observing the predicted equilibrium vortex gas states, and the relaxation of nonequilibrium configurations, are experimental techniques that facilitate the injection of vortices with nearly arbitrary angular momentum and energy. The topological nature of single-quantized vortices means that they must be introduced from the boundary of the condensate. Since the DMD provides dynamic control of the potential, vortices can introduced into the condensate by ``paddle" barriers that intersect the condensate edge and are stirred through the superfluid~\cite{gauthier2019giant,johnstone2019evolution}. The number of vortices in a cluster, and the cluster location, can be controlled by the speed and location of the paddle. A series of frames from GPE simulations of the vortex injection protocols are shown in Fig.~\ref{fig::vortex_injection_simulation}, schematically illustrating the stirring schemes. 

\textbf{Experiment I:} For injecting a single cluster, the experimental sequence follows: After transfer to the final potential, and a one-second equilibration period, the elliptically shaped paddle, with a major and minor axis of $50~\mu$m and $2~\mu$m, respectively, is swept through the Bose-Einstein condensate (BEC) at constant velocity; the paddle intersects the edge of the circular trap at its midpoint. The paddle sweep is defined by a set of 250 frames and the paddles sweep at a constant $150~\mu \textrm{m}~\textrm{s}^{-1}$ velocity (approximately $0.1 c $, where $c \sim 1290~\mu \textrm{m}~\textrm{s}^{-1}$). After crossing the halfway point of their translation, the paddle is linearly ramped to zero intensity by reducing the major and minor axis widths to zero DMD pixels. The resulting vortex cluster is shown in the left column in Fig.~\ref{fig:StirringSchematic}. 

\textbf{Experiment II:} For creating a vortex cluster on the axis of the disk trap, a different approach is used, as placement of the cluster immediately on axis via an external paddle is found to have poor repeatability. Following procedures for producing persistent currents in ring-trapped BECs~\cite{eckel2014hysteresis,wright2013driving}, we first initialize the BEC in an annular trap, resulting from ramping on of an additional central barrier with $R_0 = 15~\mu$m to the circular trap over 200 ms. Simultaneously, an elliptical stirring barrier that crosses the annulus is added, resulting in a split ring.  The elliptical paddle has a major and minor axis of $85~\mu$m and $2~\mu$m respectively. Over a time of 400 ms, the stirring paddle is linearly accelerated at $~980~\mu\textrm{m}\,\textrm{s}^{-2}$. While still moving the paddle at the final velocity, it is then ramped off  by reducing both the barrier width and length over 100~ms. After a 400-ms period of equilibration in the annular trap, the central barrier is then removed over 200~ms by linearly reducing its radius to zero. This results in a high-energy cluster of approximately 12 vortices localized to the trap center. During the retraction of the paddle, sometimes one or two extraneous vortices of the same sign are produced, however, this is later shown to have little effect on the dynamics of the central cluster. Energy damping results in the cluster gradually spreading, and after $t\sim 2$~s the vortices can be easily resolved.

\textbf{Experiment III, IV:} To realize the Gaussian and near-Riccati equilibrium states, a third stirring technique was used, based on the methods of Ref.~\cite{wilson2021generation}. Two pinning beams are spiraled into the condensate until separated by a distance $d$, resulting in two multicharge vortices. On removing the pins, the multicharge vortices break up into two distinct clusters of approximately 7 vortices each, before merging into a single cluster over the course of the dynamics. The stirring process occurs as follows. The beams are initially located on the edge of the condensate and rotate at a constant frequency, while the radial position of the beams is linearly ramped from $R$ to $d/2$ over $550$~ms, resulting in spiral paths. For experiment III, the stirring beams have a diameter of $22~\mu$m, rotated at $1.6$~Hz, and the final spacing is $d=25~\mu$m. On reaching their final locations, the radii of the pinning potentials are linearly ramped to zero over $50~$ms. For experiment IV, the stirring beams have a diameter of 30~$\mu$m, rotated at $1.4$~Hz, and the final spacing is $d=33.3~\mu$m. 

\textbf{Experiment V:} This scenario uses the same technique as experiment I; by using two paddles with the same parameters, but  propagating in opposite directions, two same-sign vortex clusters can be realized.

\subsection*{Experimental data collection and vortex imaging} The sensitivity of the vortex configurations to density gradients requires fine control of the magnetic levitation field to ensure a uniform BEC density. This was achieved by applying small magnetic correction gradients and ensuring that the central vortex cluster created in experiment II remains centered in the BEC over the initial 500~ms of evolution. By periodically repeating this calibration procedure, we find that the experiment exhibits slow, small-scale drifts in the density balance, which settle after an initial period of approximately 8 h of running. The data presented in the main text are thus collected in a continuous period subsequent to this warm-up.

High-resolution images of the BEC and vortex cores are obtained after a short, 3-ms time of flight that allows the vortex cores to expand and become visible. The radial distribution of the condensate is essentially unchanged from this expansion. Dark-ground Faraday imaging~\cite{bradley1997bose} is used, with light detuned by 220~MHz from the $^{87}$Rb $|F=1\rangle\rightarrow|F'=2\rangle$ transition, in the 80-G magnetic field resulting from the magnetic compensation of gravity. The vortex positions are then obtained using a Gaussian fitting algorithm~\citep{rakonjac2016measuring,gauthier2019giant}. The Faraday imaging light is sufficiently closely detuned that the images of the BEC are destructive. 

For experiments  I, II, and V, up to 40 images are collected for each hold time in 250-ms increments up to the maximum time of 6.75~s. This amount of data generates reliable statistics for vortex distributions as a function of time. The resulting histograms for the several hold times for experiment I are shown in Fig.~\ref{fig:StirringSchematic}. 

The results of  Sec.~\ref{sec:vortexDynamics} confirm that our system is well described by a simple point-vortex model, and that the dissipative losses are sufficiently weak at late times that the system can be treated as being in approximate microcanonical equilibrium for the time period 3.25 s -- 6.75~s. The histograms shown in Fig.~\ref{fig:histograms_compare} for experiments II and V thus show the average quasiequilibrium vortex density histograms for the $\simeq550$ images collected for $t>3.25$~s. 

In contrast, we do not record time series data for experiments III and IV.  Instead, sets of images are collected for the initial state and at a single, sufficiently long  hold time such that quasiequilibrium has been reached. For experiment III, 120 images are collected at a 4-s hold time, and for experiment IV, 361 images were collected at a 1.5-s hold time.

\subsection*{GPE modeling of stirring protocols}

We simulated the initial vortex injection using a phenomenologically damped GPE (dGPE) model. Working in length units of the healing length $\xi = \hbar/\sqrt{\rho_0 g}$ and time units defined by the chemical potential $\tau = \hbar/\mu$, we simulate the following equation for $\Psi(\rr,t)$
\begin{equation}
 i \partial_t \Psi =   (1-i \gamma) \left( -\tfrac{\nabla^2}{2} + V + |\Psi|^2 - 1 \right)\Psi,
\end{equation}
where $V(\rr,t)$ is the external potential and $\gamma$ is a phenomenological damping factor of $2\times 10^{-3}$ which matches the thermal friction coefficient observed in the experiment. Note that the Brownian motion term is not included in this model, as these effects are not significant to the stirring dynamics. The circular trapping and pinning potentials and the elliptical stirring potentials are modeled as hard-wall potentials of the general form
\begin{multline}
    V(\rr) = \\ V_0\tanh \left[d \sqrt{x'^2+y'^2}\left(1- \tfrac{ab}{\sqrt{(a x')^2+(b y')^2}}\right) \right],
\end{multline}
where $V_0$ is the strength of the potential, $a$ and $b$ are the semimajor and semiminor axis respectively, $x' = x - x_0$ and $y' = y - y_0$ to allow for translation or $x' = R \cos(\theta) - R\sin(\theta)$ and $y' = R\sin(\theta) + R\cos(\theta)$ to allow for rotations by angle $\theta$, and $d$ controls the steepness of the potential. 
 Figure~\ref{fig::vortex_injection_simulation} shows the resulting condensate density snapshots from simulating the stirring protocols with similar parameters to those performed in the experiment. All vortices injected were of the same sign. Movies of dGPE simulations of the vortex injection methods are provided in the Supplemental Material~\cite{SM}.

\section{Solution of Poisson-Boltzmann equation}
\label{sec:AppMF}

To solve the Poisson equation given by
\begin{align}
\nabla^2 \phi(\rr) = -4 \pi n(\rr) \, ,
\end{align}
subject to the boundary conditions $\phi(r=1,\varphi)=0$ and the constraints of prescribed angular momentum (angular impulse), energy, and normalisation given by
 \begin{align}
     \mathcal{M} = \int \dd^2\rr\; r^2 n , & &
     \mathcal{H} = \tfrac{1}{2} \int \dd^2\rr\; n \phi , & &
     \mathcal{C} = \int \dd^2\rr\; n \, ,
 \end{align}
we adapt the method described in Ref.~\cite{turkington1996statistical}. First, we represent the stream function $\phi$ in space using a Fourier series decomposition in terms of the azimuthal angle so that
\begin{align}
\phi(r,\varphi) = \sum_m f_m(r) e^{i m \varphi} \, .
\end{align}
The above equation then reduces to
\begin{align}
\frac{1}{r} \frac{\dd}{\dd r} \left( r \frac{\dd f_m}{\dd r}\right) - \frac{m^2}{r^2} f_m =  -4\pi \int_0^{2\pi} n(r,\varphi) e^{-i m \varphi} \dd \varphi \, .
\end{align}
To satisfy the boundary condition, we require that $f_m(r=1)=0, \, \forall \, m$, and $f_m(r=0)=0, \, \forall \, m \ne 0$. For $m=0$, we set $f_0'(r=0)=0$ where a prime denotes differentiation with respect to the radius $r$. This radial equation is then discretized using a centered finite-differencing scheme. The resulting discretized equation can be inverted to recover the streamfunction from the vortex density field.

To find a self-consistent solution of the Poisson-Boltzmann equation subject to the constraints given, we define the functional 
\begin{align}
F[n] = S[n]-\hat{\alpha} \big( \mathcal{C}[n]-1 \big)&-\hat{\beta} \big( \mathcal{H}[n]-H_0 \big)  \nonumber \\
&- \hat{\Omega} \big( \mathcal{M}[n] - M_0 \big) \, ,
\end{align}
where $\hat{\Omega}=\hat{\beta} \hat{\omega}$. We want to find $n$ such that $F[n]$ is stationary so that $\delta F = F[n+\delta n]-F[n]=0$. This requires that
\begin{align}
\delta S[n] = \hat{\alpha} \delta \mathcal{C}[n] + \hat{\beta} \delta \mathcal{H}[n] + \hat{\Omega} \delta \mathcal{M}[n] \, ,
\end{align}
which implies that
\begin{align}
n(\rr) = \exp \left( -1-\hat{\alpha} -\hat{\beta} \phi(\rr) - \hat{\Omega} r^2 \right) \,.
\end{align}
It follows that we can rewrite the three constraints in the form
\begin{align}
-\left( \frac{\partial}{\partial \hat{\alpha}} , \frac{\partial}{\partial \hat{\beta}} , \frac{\partial}{\partial \hat{\Omega}} \right)
\int n(\rr) \dd^2 \rr = \left( 1 , 2H_0 , M_0 \right) \, .
\end{align}
We note that the constraints $\mathcal{C}$ and $\mathcal{M}$ are both linear in the vortex density $n$. However, the constraint for the energy is nonlinear since it can expressed in terms of the Green's function $G(\rr,\rr')$ of the Laplacian operator as
\begin{align}
\mathcal{H} = \frac{1}{2} \int n \phi \; \dd^2 \rr = \frac{1}{2} \iint n(\rr) G(\rr,\rr') n(\rr') \, \dd^2 \rr \, \dd^2 \rr' \, ,
\end{align}
where $G(\rr,\rr')$ is the 2D Green's function for a point vortex confined within a circular disk~\cite{newton2013nvortexproblem}. Linearizing the energy about the state $n^k$, we obtain
\begin{align}
\mathcal{H}[n^{k+1}] &\simeq \mathcal{H}[n^{k}] + \iint n(\rr') G(\rr,\rr') \delta n(\rr) \dd^2 \rr \dd^2 \rr' \, , \nonumber \\
&= \mathcal{H}[n^{k}] + \int \frac{\delta \mathcal{H}[n^k]}{\delta n} \big( n^{k+1}(\rr)-n^k(\rr) \big) \dd^2 \rr \, , \nonumber \\
&= \mathcal{H}[n^{k}] + \int \phi^k(\rr) \big( n^{k+1}(\rr)-n^k(\rr) \big) \dd^2 \rr \, .
\end{align}
The linearized constraint on energy is then given by 
\begin{align}
\mathcal{H}[n^{k}] + \int \frac{\delta \mathcal{H}[n^k]}{\delta n} \big( n^{k+1}(\rr)-n^k(\rr) \big) \dd^2 \rr = H_0 \, . 
\end{align}
We can now rewrite the three constraints in the linearized form
\begin{multline}
\mathbf{F} = \left( C_0 , H_0+\mathcal{H}^k , M_0 \right) \\ + \left( \frac{\partial}{\partial \hat{\alpha}} , \frac{\partial}{\partial \hat{\beta}} , \frac{\partial}{\partial \hat{\Omega}} \right) \int n^k(\rr) \dd^2 \rr  = \mathbf{0} ,
\end{multline}
where $n^{k+1}(\rr) = \exp \left( -1-\hat{\alpha}-\hat{\beta} \phi^k(\rr) - \hat{\Omega} r^2 \right)$ and $\nabla^2 \phi^k(\rr) = -4 \pi n^k(\rr)$. The advantage of casting the constraints in this form is that we can now proceed to update the system of Lagrange multipliers $(\hat{\alpha},\hat{\beta},\hat{\Omega})$ by using a Newton-Raphson iteration scheme. Alternatively, given that the constraints are now expressed in terms of a gradient operator, a gradient descent algorithm can now be used to give
\begin{align}
\big( \hat{\alpha}^{l+1},\hat{\beta}^{l+1},\hat{\Omega}^{l+1} \big) = \big( \hat{\alpha}^{l},\hat{\beta}^{l},\hat{\Omega}^{l} \big) - s \mathbf{F}[n^k;\hat{\alpha}^{l},\hat{\beta}^{l},\hat{\Omega}^{l} ] \, ,
\end{align}
where $s$ is a relaxation parameter $0 < s \le 1$. In practice, we use a gradient descent algorithm to obtain a good estimate of $n$ followed by a Newton-Raphson scheme in order to accelerate convergence. The Newton-Raphson scheme requires the evaluation of the Hessian of $\mathbf{F}[n^k]$ that we denote by $\boldsymbol{\mathsf{H}}$ . The updates can then be evaluated as
\begin{align}
\big( \hat{\alpha}^{l+1},\hat{\beta}^{l+1},\hat{\Omega}^{l+1} \big) = \big( \hat{\alpha}^{l},\hat{\beta}^{l},\hat{\Omega}^{l} \big) - s \boldsymbol{\mathsf{H}}^{-1} \, \mathbf{F} \, . 
\end{align}
The iterations are stopped once the values for $\hat{\alpha}$, $\hat{\beta}$, and $\hat{\Omega}$ converge to within a tolerance of $10^{-9}$.

To obtain the branch corresponding to axisymmetric (centered) flows, we solve the above by setting the coefficients of all modes corresponding to $m \ne 0$ to zero. To find the symmetric and off-centered maximum entropy solutions, we typically started with an exact solution such as the Gaussian profile corresponding to $\hat{\beta} = 0$ and then trace out the branches by varying the energy for fixed $\mathcal{M}$ and $\mathcal{C}$.


\begin{thebibliography}{98}%
\makeatletter
\providecommand \@ifxundefined [1]{%
 \@ifx{#1\undefined}
}%
\providecommand \@ifnum [1]{%
 \ifnum #1\expandafter \@firstoftwo
 \else \expandafter \@secondoftwo
 \fi
}%
\providecommand \@ifx [1]{%
 \ifx #1\expandafter \@firstoftwo
 \else \expandafter \@secondoftwo
 \fi
}%
\providecommand \natexlab [1]{#1}%
\providecommand \enquote  [1]{``#1''}%
\providecommand \bibnamefont  [1]{#1}%
\providecommand \bibfnamefont [1]{#1}%
\providecommand \citenamefont [1]{#1}%
\providecommand \href@noop [0]{\@secondoftwo}%
\providecommand \href [0]{\begingroup \@sanitize@url \@href}%
\providecommand \@href[1]{\@@startlink{#1}\@@href}%
\providecommand \@@href[1]{\endgroup#1\@@endlink}%
\providecommand \@sanitize@url [0]{\catcode `\\12\catcode `\$12\catcode
  `\&12\catcode `\#12\catcode `\^12\catcode `\_12\catcode `\%12\relax}%
\providecommand \@@startlink[1]{}%
\providecommand \@@endlink[0]{}%
\providecommand \url  [0]{\begingroup\@sanitize@url \@url }%
\providecommand \@url [1]{\endgroup\@href {#1}{\urlprefix }}%
\providecommand \urlprefix  [0]{URL }%
\providecommand \Eprint [0]{\href }%
\providecommand \doibase [0]{http://dx.doi.org/}%
\providecommand \selectlanguage [0]{\@gobble}%
\providecommand \bibinfo  [0]{\@secondoftwo}%
\providecommand \bibfield  [0]{\@secondoftwo}%
\providecommand \translation [1]{[#1]}%
\providecommand \BibitemOpen [0]{}%
\providecommand \bibitemStop [0]{}%
\providecommand \bibitemNoStop [0]{.\EOS\space}%
\providecommand \EOS [0]{\spacefactor3000\relax}%
\providecommand \BibitemShut  [1]{\csname bibitem#1\endcsname}%
\let\auto@bib@innerbib\@empty
\bibitem [{\citenamefont {Thalabard}\ \emph {et~al.}(2015)\citenamefont
  {Thalabard}, \citenamefont {Saint-Michel}, \citenamefont {Herbert},
  \citenamefont {Daviaud},\ and\ \citenamefont {Dubrulle}}]{Thalabard_2015}%
  \BibitemOpen
  \bibfield  {author} {\bibinfo {author} {\bibfnamefont {Simon}\ \bibnamefont
  {Thalabard}}, \bibinfo {author} {\bibfnamefont {Brice}\ \bibnamefont
  {Saint-Michel}}, \bibinfo {author} {\bibfnamefont {Eric}\ \bibnamefont
  {Herbert}}, \bibinfo {author} {\bibfnamefont {Fran{\c{c}}ois}\ \bibnamefont
  {Daviaud}}, \ and\ \bibinfo {author} {\bibfnamefont {B{\'{e}}reng{\`{e}}re}\
  \bibnamefont {Dubrulle}},\ }\bibfield  {title} {\enquote {\bibinfo {title} {A
  statistical mechanics framework for the large-scale structure of turbulent
  von {K}{\'{a}}rm{\'{a}}n flows},}\ }\href {\doibase
  10.1088/1367-2630/17/6/063006} {\bibfield  {journal} {\bibinfo  {journal}
  {New Journal of Physics}\ }\textbf {\bibinfo {volume} {17}},\ \bibinfo
  {pages} {063006} (\bibinfo {year} {2015})}\BibitemShut {NoStop}%
\bibitem [{\citenamefont {Frisch}(1995)}]{frisch1995turbulence}%
  \BibitemOpen
  \bibfield  {author} {\bibinfo {author} {\bibfnamefont {Uriel}\ \bibnamefont
  {Frisch}},\ }\href@noop {} {\emph {\bibinfo {title} {Turbulence: the legacy
  of AN Kolmogorov}}}\ (\bibinfo  {publisher} {Cambridge university press},\
  \bibinfo {year} {1995})\BibitemShut {NoStop}%
\bibitem [{\citenamefont {Batchelor}(1953)}]{batchelor1953theory}%
  \BibitemOpen
  \bibfield  {author} {\bibinfo {author} {\bibfnamefont {George~Keith}\
  \bibnamefont {Batchelor}},\ }\href@noop {} {\emph {\bibinfo {title} {The
  theory of homogeneous turbulence}}}\ (\bibinfo  {publisher} {Cambridge
  university press},\ \bibinfo {year} {1953})\BibitemShut {NoStop}%
\bibitem [{\citenamefont {Kraichnan}(1967)}]{kraichnan1967inertial}%
  \BibitemOpen
  \bibfield  {author} {\bibinfo {author} {\bibfnamefont {Robert~H}\
  \bibnamefont {Kraichnan}},\ }\bibfield  {title} {\enquote {\bibinfo {title}
  {Inertial ranges in two-dimensional turbulence},}\ }\href {\doibase
  10.1063/1.1762301} {\bibfield  {journal} {\bibinfo  {journal} {The Physics of
  Fluids}\ }\textbf {\bibinfo {volume} {10}},\ \bibinfo {pages} {1417--1423}
  (\bibinfo {year} {1967})}\BibitemShut {NoStop}%
\bibitem [{\citenamefont {Batchelor}(1969)}]{batchelor1969}%
  \BibitemOpen
  \bibfield  {author} {\bibinfo {author} {\bibfnamefont {G.~K.}\ \bibnamefont
  {Batchelor}},\ }\bibfield  {title} {\enquote {\bibinfo {title} {Computation
  of the energy spectrum in homogeneous two‐dimensional turbulence},}\ }\href
  {\doibase 10.1063/1.1692443} {\bibfield  {journal} {\bibinfo  {journal} {The
  Physics of Fluids}\ }\textbf {\bibinfo {volume} {12}},\ \bibinfo {pages}
  {II--233--II--239} (\bibinfo {year} {1969})}\BibitemShut {NoStop}%
\bibitem [{\citenamefont {Sarid}\ \emph
  {et~al.}(2004{\natexlab{a}})\citenamefont {Sarid}, \citenamefont
  {Teodorescu}, \citenamefont {Marcus},\ and\ \citenamefont
  {Fajans}}]{sarid_breaking_2004}%
  \BibitemOpen
  \bibfield  {author} {\bibinfo {author} {\bibfnamefont {Eli}\ \bibnamefont
  {Sarid}}, \bibinfo {author} {\bibfnamefont {Catalin}\ \bibnamefont
  {Teodorescu}}, \bibinfo {author} {\bibfnamefont {Philip~S.}\ \bibnamefont
  {Marcus}}, \ and\ \bibinfo {author} {\bibfnamefont {Joel}\ \bibnamefont
  {Fajans}},\ }\bibfield  {title} {\enquote {\bibinfo {title} {Breaking of
  {{Rotational Symmetry}} in {{Cylindrically Bounded 2D Electron Plasmas}} and
  {{2D Fluids}}},}\ }\href {\doibase 10.1103/PhysRevLett.93.215002} {\bibfield
  {journal} {\bibinfo  {journal} {Physical Review Letters}\ }\textbf {\bibinfo
  {volume} {93}},\ \bibinfo {pages} {215002} (\bibinfo {year}
  {2004}{\natexlab{a}})}\BibitemShut {NoStop}%
\bibitem [{\citenamefont {Rodgers}\ \emph {et~al.}(2009)\citenamefont
  {Rodgers}, \citenamefont {Servidio}, \citenamefont {Matthaeus}, \citenamefont
  {Montgomery}, \citenamefont {Mitchell},\ and\ \citenamefont
  {Aziz}}]{rodgers2009hydrodynamic}%
  \BibitemOpen
  \bibfield  {author} {\bibinfo {author} {\bibfnamefont {D.~J.}\ \bibnamefont
  {Rodgers}}, \bibinfo {author} {\bibfnamefont {S.}~\bibnamefont {Servidio}},
  \bibinfo {author} {\bibfnamefont {W.~H.}\ \bibnamefont {Matthaeus}}, \bibinfo
  {author} {\bibfnamefont {D.~C.}\ \bibnamefont {Montgomery}}, \bibinfo
  {author} {\bibfnamefont {T.~B.}\ \bibnamefont {Mitchell}}, \ and\ \bibinfo
  {author} {\bibfnamefont {T.}~\bibnamefont {Aziz}},\ }\bibfield  {title}
  {\enquote {\bibinfo {title} {Hydrodynamic relaxation of an electron plasma to
  a near-maximum entropy state},}\ }\href {\doibase
  10.1103/PhysRevLett.102.244501} {\bibfield  {journal} {\bibinfo  {journal}
  {Physical Review Letters}\ }\textbf {\bibinfo {volume} {102}},\ \bibinfo
  {pages} {244501} (\bibinfo {year} {2009})}\BibitemShut {NoStop}%
\bibitem [{\citenamefont {Kellay}(2017)}]{kellay2017hydrodynamics}%
  \BibitemOpen
  \bibfield  {author} {\bibinfo {author} {\bibfnamefont {H.}~\bibnamefont
  {Kellay}},\ }\bibfield  {title} {\enquote {\bibinfo {title} {Hydrodynamics
  experiments with soap films and soap bubbles: A short review of recent
  experiments},}\ }\href {\doibase 10.1063/1.4986003} {\bibfield  {journal}
  {\bibinfo  {journal} {Physics of Fluids}\ }\textbf {\bibinfo {volume} {29}},\
  \bibinfo {pages} {111113} (\bibinfo {year} {2017})}\BibitemShut {NoStop}%
\bibitem [{\citenamefont {Hansen}\ \emph {et~al.}(1998)\citenamefont {Hansen},
  \citenamefont {Marteau},\ and\ \citenamefont {Tabeling}}]{hansen1998two}%
  \BibitemOpen
  \bibfield  {author} {\bibinfo {author} {\bibfnamefont {A.~E.}\ \bibnamefont
  {Hansen}}, \bibinfo {author} {\bibfnamefont {D.}~\bibnamefont {Marteau}}, \
  and\ \bibinfo {author} {\bibfnamefont {P.}~\bibnamefont {Tabeling}},\
  }\bibfield  {title} {\enquote {\bibinfo {title} {Two-dimensional turbulence
  and dispersion in a freely decaying system},}\ }\href {\doibase
  10.1103/PhysRevE.58.7261} {\bibfield  {journal} {\bibinfo  {journal} {Phys.
  Rev. E}\ }\textbf {\bibinfo {volume} {58}},\ \bibinfo {pages} {7261--7271}
  (\bibinfo {year} {1998})}\BibitemShut {NoStop}%
\bibitem [{\citenamefont {Adriani}\ \emph {et~al.}(2018)\citenamefont
  {Adriani}, \citenamefont {Mura}, \citenamefont {Orton}, \citenamefont
  {Hansen}, \citenamefont {Altieri}, \citenamefont {Moriconi}, \citenamefont
  {Rogers}, \citenamefont {Eichst{\"a}dt}, \citenamefont {Momary},
  \citenamefont {Ingersoll} \emph {et~al.}}]{adriani2018clusters}%
  \BibitemOpen
  \bibfield  {author} {\bibinfo {author} {\bibfnamefont {Alberto}\ \bibnamefont
  {Adriani}}, \bibinfo {author} {\bibfnamefont {A}~\bibnamefont {Mura}},
  \bibinfo {author} {\bibfnamefont {G}~\bibnamefont {Orton}}, \bibinfo {author}
  {\bibfnamefont {C}~\bibnamefont {Hansen}}, \bibinfo {author} {\bibfnamefont
  {F}~\bibnamefont {Altieri}}, \bibinfo {author} {\bibfnamefont {M.~L.}\
  \bibnamefont {Moriconi}}, \bibinfo {author} {\bibfnamefont {J}~\bibnamefont
  {Rogers}}, \bibinfo {author} {\bibfnamefont {G}~\bibnamefont
  {Eichst{\"a}dt}}, \bibinfo {author} {\bibfnamefont {T}~\bibnamefont
  {Momary}}, \bibinfo {author} {\bibfnamefont {Andrew~P}\ \bibnamefont
  {Ingersoll}},  \emph {et~al.},\ }\bibfield  {title} {\enquote {\bibinfo
  {title} {Clusters of cyclones encircling {J}upiter's poles},}\ }\href
  {https://doi.org/10.1038/nature25491} {\bibfield  {journal} {\bibinfo
  {journal} {Nature}\ }\textbf {\bibinfo {volume} {555}},\ \bibinfo {pages}
  {216} (\bibinfo {year} {2018})}\BibitemShut {NoStop}%
\bibitem [{\citenamefont {Bouchet}\ and\ \citenamefont
  {Sommeria}(2002)}]{bouchet2002emergence}%
  \BibitemOpen
  \bibfield  {author} {\bibinfo {author} {\bibfnamefont {F.}~\bibnamefont
  {Bouchet}}\ and\ \bibinfo {author} {\bibfnamefont {J.}~\bibnamefont
  {Sommeria}},\ }\bibfield  {title} {\enquote {\bibinfo {title} {Emergence of
  intense jets and {J}upiter's {G}reat {R}ed {S}pot as maximum-entropy
  structures},}\ }\href {\doibase 10.1017/S0022112002008789} {\bibfield
  {journal} {\bibinfo  {journal} {Journal of Fluid Mechanics}\ }\textbf
  {\bibinfo {volume} {464}},\ \bibinfo {pages} {165–207} (\bibinfo {year}
  {2002})}\BibitemShut {NoStop}%
\bibitem [{\citenamefont {Onsager}(1949)}]{onsager1949statistical}%
  \BibitemOpen
  \bibfield  {author} {\bibinfo {author} {\bibfnamefont {Lars}\ \bibnamefont
  {Onsager}},\ }\bibfield  {title} {\enquote {\bibinfo {title} {Statistical
  hydrodynamics},}\ }\href {\doibase 10.1007/BF02780991} {\bibfield  {journal}
  {\bibinfo  {journal} {Il Nuovo Cimento (1943-1954)}\ }\textbf {\bibinfo
  {volume} {6}},\ \bibinfo {pages} {279--287} (\bibinfo {year}
  {1949})}\BibitemShut {NoStop}%
\bibitem [{\citenamefont {Joyce}\ and\ \citenamefont
  {Montgomery}(1973)}]{joyce_montgomery_1973}%
  \BibitemOpen
  \bibfield  {author} {\bibinfo {author} {\bibfnamefont {Glenn}\ \bibnamefont
  {Joyce}}\ and\ \bibinfo {author} {\bibfnamefont {David}\ \bibnamefont
  {Montgomery}},\ }\bibfield  {title} {\enquote {\bibinfo {title} {Negative
  temperature states for the two-dimensional guiding-centre plasma},}\ }\href
  {\doibase 10.1017/S0022377800007686} {\bibfield  {journal} {\bibinfo
  {journal} {Journal of Plasma Physics}\ }\textbf {\bibinfo {volume} {10}},\
  \bibinfo {pages} {107--121} (\bibinfo {year} {1973})}\BibitemShut {NoStop}%
\bibitem [{\citenamefont {Edwards}\ and\ \citenamefont
  {Taylor}(1974)}]{edwards1974negative}%
  \BibitemOpen
  \bibfield  {author} {\bibinfo {author} {\bibfnamefont {S.~F.}\ \bibnamefont
  {Edwards}}\ and\ \bibinfo {author} {\bibfnamefont {J.~B.}\ \bibnamefont
  {Taylor}},\ }\bibfield  {title} {\enquote {\bibinfo {title} {Negative
  temperature states of two-dimensional plasmas and vortex fluids},}\ }\href
  {http://www.jstor.org/stable/78450} {\bibfield  {journal} {\bibinfo
  {journal} {Proceedings of the Royal Society of London. Series A, Mathematical
  and Physical Sciences}\ }\textbf {\bibinfo {volume} {336}},\ \bibinfo {pages}
  {257--271} (\bibinfo {year} {1974})}\BibitemShut {NoStop}%
\bibitem [{\citenamefont {Kraichnan}(1975)}]{Kraichnan:1975ku}%
  \BibitemOpen
  \bibfield  {author} {\bibinfo {author} {\bibfnamefont {R.~H.}\ \bibnamefont
  {Kraichnan}},\ }\bibfield  {title} {\enquote {\bibinfo {title} {{Statistical
  dynamics of two-dimensional flow}},}\ }\href {\doibase
  https://doi.org/10.1017/S0022112075000225} {\bibfield  {journal} {\bibinfo
  {journal} {J. Fluid Mech.}\ }\textbf {\bibinfo {volume} {67}},\ \bibinfo
  {pages} {155--175} (\bibinfo {year} {1975})}\BibitemShut {NoStop}%
\bibitem [{\citenamefont {Miller}(1990)}]{miller1990statistical}%
  \BibitemOpen
  \bibfield  {author} {\bibinfo {author} {\bibfnamefont {Jonathan}\
  \bibnamefont {Miller}},\ }\bibfield  {title} {\enquote {\bibinfo {title}
  {Statistical mechanics of {E}uler equations in two dimensions},}\ }\href
  {\doibase 10.1103/PhysRevLett.65.2137} {\bibfield  {journal} {\bibinfo
  {journal} {Physical Review Letters}\ }\textbf {\bibinfo {volume} {65}},\
  \bibinfo {pages} {2137--2140} (\bibinfo {year} {1990})}\BibitemShut {NoStop}%
\bibitem [{\citenamefont {Robert}(1991)}]{robert1991maximum}%
  \BibitemOpen
  \bibfield  {author} {\bibinfo {author} {\bibfnamefont {Raoul}\ \bibnamefont
  {Robert}},\ }\bibfield  {title} {\enquote {\bibinfo {title} {A
  maximum-entropy principle for two-dimensional perfect fluid dynamics},}\
  }\href {https://link.springer.com/article/10.1007/BF01053743} {\bibfield
  {journal} {\bibinfo  {journal} {Journal of Statistical Physics}\ }\textbf
  {\bibinfo {volume} {65}},\ \bibinfo {pages} {531--553} (\bibinfo {year}
  {1991})}\BibitemShut {NoStop}%
\bibitem [{\citenamefont {Robert}\ and\ \citenamefont
  {Sommeria}(1991)}]{robert_sommeria_1991}%
  \BibitemOpen
  \bibfield  {author} {\bibinfo {author} {\bibfnamefont {R.}~\bibnamefont
  {Robert}}\ and\ \bibinfo {author} {\bibfnamefont {J.}~\bibnamefont
  {Sommeria}},\ }\bibfield  {title} {\enquote {\bibinfo {title} {Statistical
  equilibrium states for two-dimensional flows},}\ }\href {\doibase
  10.1017/S0022112091003038} {\bibfield  {journal} {\bibinfo  {journal}
  {Journal of Fluid Mechanics}\ }\textbf {\bibinfo {volume} {229}},\ \bibinfo
  {pages} {291--310} (\bibinfo {year} {1991})}\BibitemShut {NoStop}%
\bibitem [{\citenamefont {Maestrini}\ and\ \citenamefont
  {Salman}(2019)}]{maestrini2019entropy}%
  \BibitemOpen
  \bibfield  {author} {\bibinfo {author} {\bibfnamefont {Davide}\ \bibnamefont
  {Maestrini}}\ and\ \bibinfo {author} {\bibfnamefont {Hayder}\ \bibnamefont
  {Salman}},\ }\bibfield  {title} {\enquote {\bibinfo {title} {Entropy of
  negative temperature states for a point vortex gas},}\ }\href@noop {}
  {\bibfield  {journal} {\bibinfo  {journal} {Journal of Statistical Physics}\
  }\textbf {\bibinfo {volume} {176}},\ \bibinfo {pages} {981--1008} (\bibinfo
  {year} {2019})}\BibitemShut {NoStop}%
\bibitem [{\citenamefont {Denoix}(1992)}]{denoix1992}%
  \BibitemOpen
  \bibfield  {author} {\bibinfo {author} {\bibfnamefont {Marie-Alice}\
  \bibnamefont {Denoix}},\ }\emph {\bibinfo {title} {{Experimental study of
  stable structures in two-dimensional turbulence - Comparison with a
  statistical mechanical theory}}},\ \href
  {https://hal.archives-ouvertes.fr/tel-01340648} {\bibinfo {type} {Thesis}},\
  \bibinfo  {school} {{Institut National Polytechnique de Grenoble}} (\bibinfo
  {year} {1992})\BibitemShut {NoStop}%
\bibitem [{\citenamefont {Matthaeus}\ \emph {et~al.}(1991)\citenamefont
  {Matthaeus}, \citenamefont {Stribling}, \citenamefont {Martinez},
  \citenamefont {Oughton},\ and\ \citenamefont {Montgomery}}]{matthaeus1991}%
  \BibitemOpen
  \bibfield  {author} {\bibinfo {author} {\bibfnamefont {W.H.}\ \bibnamefont
  {Matthaeus}}, \bibinfo {author} {\bibfnamefont {W.T.}\ \bibnamefont
  {Stribling}}, \bibinfo {author} {\bibfnamefont {D.}~\bibnamefont {Martinez}},
  \bibinfo {author} {\bibfnamefont {S.}~\bibnamefont {Oughton}}, \ and\
  \bibinfo {author} {\bibfnamefont {D.}~\bibnamefont {Montgomery}},\ }\bibfield
   {title} {\enquote {\bibinfo {title} {Decaying, two-dimensional,
  {N}avier-{S}tokes turbulence at very long times},}\ }\href {\doibase
  https://doi.org/10.1016/0167-2789(91)90259-C} {\bibfield  {journal} {\bibinfo
   {journal} {Physica D: Nonlinear Phenomena}\ }\textbf {\bibinfo {volume}
  {51}},\ \bibinfo {pages} {531 -- 538} (\bibinfo {year} {1991})}\BibitemShut
  {NoStop}%
\bibitem [{\citenamefont {Tabeling}(2002)}]{tabeling2002two}%
  \BibitemOpen
  \bibfield  {author} {\bibinfo {author} {\bibfnamefont {Patrick}\ \bibnamefont
  {Tabeling}},\ }\bibfield  {title} {\enquote {\bibinfo {title}
  {Two-dimensional turbulence: a physicist approach},}\ }\href {\doibase
  https://doi.org/10.1016/S0370-1573(01)00064-3} {\bibfield  {journal}
  {\bibinfo  {journal} {Physics Reports}\ }\textbf {\bibinfo {volume} {362}},\
  \bibinfo {pages} {1--62} (\bibinfo {year} {2002})}\BibitemShut {NoStop}%
\bibitem [{\citenamefont {Sarid}\ \emph
  {et~al.}(2004{\natexlab{b}})\citenamefont {Sarid}, \citenamefont
  {Teodorescu}, \citenamefont {Marcus},\ and\ \citenamefont
  {Fajans}}]{Sarid2004}%
  \BibitemOpen
  \bibfield  {author} {\bibinfo {author} {\bibfnamefont {Eli}\ \bibnamefont
  {Sarid}}, \bibinfo {author} {\bibfnamefont {Catalin}\ \bibnamefont
  {Teodorescu}}, \bibinfo {author} {\bibfnamefont {Philip~S.}\ \bibnamefont
  {Marcus}}, \ and\ \bibinfo {author} {\bibfnamefont {Joel}\ \bibnamefont
  {Fajans}},\ }\bibfield  {title} {\enquote {\bibinfo {title} {Breaking of
  rotational symmetry in cylindrically bounded {2D} electron plasmas and {2D}
  fluids},}\ }\href {\doibase 10.1103/PhysRevLett.93.215002} {\bibfield
  {journal} {\bibinfo  {journal} {Physical Review Letters}\ }\textbf {\bibinfo
  {volume} {93}},\ \bibinfo {pages} {215002} (\bibinfo {year}
  {2004}{\natexlab{b}})}\BibitemShut {NoStop}%
\bibitem [{\citenamefont {Driscoll}\ and\ \citenamefont
  {Fine}(1990)}]{driscoll1990}%
  \BibitemOpen
  \bibfield  {author} {\bibinfo {author} {\bibfnamefont {C.~F.}\ \bibnamefont
  {Driscoll}}\ and\ \bibinfo {author} {\bibfnamefont {K.~S.}\ \bibnamefont
  {Fine}},\ }\bibfield  {title} {\enquote {\bibinfo {title} {Experiments on
  vortex dynamics in pure electron plasmas},}\ }\href {\doibase
  10.1063/1.859556} {\bibfield  {journal} {\bibinfo  {journal} {Physics of
  Fluids B: Plasma Physics}\ }\textbf {\bibinfo {volume} {2}},\ \bibinfo
  {pages} {1359--1366} (\bibinfo {year} {1990})}\BibitemShut {NoStop}%
\bibitem [{\citenamefont {Carnevale}\ \emph {et~al.}(1992)\citenamefont
  {Carnevale}, \citenamefont {McWilliams}, \citenamefont {Pomeau},
  \citenamefont {Weiss},\ and\ \citenamefont {Young}}]{carnevale1992rates}%
  \BibitemOpen
  \bibfield  {author} {\bibinfo {author} {\bibfnamefont {G.~F.}\ \bibnamefont
  {Carnevale}}, \bibinfo {author} {\bibfnamefont {J.~C.}\ \bibnamefont
  {McWilliams}}, \bibinfo {author} {\bibfnamefont {Y.}~\bibnamefont {Pomeau}},
  \bibinfo {author} {\bibfnamefont {J.~B.}\ \bibnamefont {Weiss}}, \ and\
  \bibinfo {author} {\bibfnamefont {W.~R.}\ \bibnamefont {Young}},\ }\bibfield
  {title} {\enquote {\bibinfo {title} {Rates, pathways, and end states of
  nonlinear evolution in decaying two-dimensional turbulence: Scaling theory
  versus selective decay},}\ }\href {\doibase 10.1063/1.858251} {\bibfield
  {journal} {\bibinfo  {journal} {Physics of Fluids A: Fluid Dynamics}\
  }\textbf {\bibinfo {volume} {4}},\ \bibinfo {pages} {1314--1316} (\bibinfo
  {year} {1992})}\BibitemShut {NoStop}%
\bibitem [{\citenamefont {Marteau}\ \emph {et~al.}(1995)\citenamefont
  {Marteau}, \citenamefont {Cardoso},\ and\ \citenamefont
  {Tabeling}}]{marteau1995equilibrium}%
  \BibitemOpen
  \bibfield  {author} {\bibinfo {author} {\bibfnamefont {D.}~\bibnamefont
  {Marteau}}, \bibinfo {author} {\bibfnamefont {O.}~\bibnamefont {Cardoso}}, \
  and\ \bibinfo {author} {\bibfnamefont {P.}~\bibnamefont {Tabeling}},\
  }\bibfield  {title} {\enquote {\bibinfo {title} {Equilibrium states of
  two-dimensional turbulence: An experimental study},}\ }\href {\doibase
  10.1103/PhysRevE.51.5124} {\bibfield  {journal} {\bibinfo  {journal}
  {Physical Review E}\ }\textbf {\bibinfo {volume} {51}},\ \bibinfo {pages}
  {5124--5127} (\bibinfo {year} {1995})}\BibitemShut {NoStop}%
\bibitem [{\citenamefont {Huang}\ and\ \citenamefont
  {Driscoll}(1994)}]{huang1994relaxation}%
  \BibitemOpen
  \bibfield  {author} {\bibinfo {author} {\bibfnamefont {X.-P.}\ \bibnamefont
  {Huang}}\ and\ \bibinfo {author} {\bibfnamefont {C.~F.}\ \bibnamefont
  {Driscoll}},\ }\bibfield  {title} {\enquote {\bibinfo {title} {Relaxation of
  {2D} turbulence to a metaequilibrium near the minimum enstrophy state},}\
  }\href {\doibase 10.1103/PhysRevLett.72.2187} {\bibfield  {journal} {\bibinfo
   {journal} {Physical Review Letters}\ }\textbf {\bibinfo {volume} {72}},\
  \bibinfo {pages} {2187--2190} (\bibinfo {year} {1994})}\BibitemShut {NoStop}%
\bibitem [{\citenamefont {Pakter}\ and\ \citenamefont
  {Levin}(2018)}]{Patker2018}%
  \BibitemOpen
  \bibfield  {author} {\bibinfo {author} {\bibfnamefont {Renato}\ \bibnamefont
  {Pakter}}\ and\ \bibinfo {author} {\bibfnamefont {Yan}\ \bibnamefont
  {Levin}},\ }\bibfield  {title} {\enquote {\bibinfo {title} {Nonequilibrium
  statistical mechanics of two-dimensional vortices},}\ }\href {\doibase
  10.1103/PhysRevLett.121.020602} {\bibfield  {journal} {\bibinfo  {journal}
  {Physical Review Letters}\ }\textbf {\bibinfo {volume} {121}},\ \bibinfo
  {pages} {020602} (\bibinfo {year} {2018})}\BibitemShut {NoStop}%
\bibitem [{\citenamefont {Levin}\ \emph {et~al.}(2014)\citenamefont {Levin},
  \citenamefont {Pakter}, \citenamefont {Rizzato}, \citenamefont {Teles},\ and\
  \citenamefont {Benetti}}]{levin2014nonequilibrium}%
  \BibitemOpen
  \bibfield  {author} {\bibinfo {author} {\bibfnamefont {Yan}\ \bibnamefont
  {Levin}}, \bibinfo {author} {\bibfnamefont {Renato}\ \bibnamefont {Pakter}},
  \bibinfo {author} {\bibfnamefont {Felipe~B}\ \bibnamefont {Rizzato}},
  \bibinfo {author} {\bibfnamefont {Tarc{\'\i}sio~N}\ \bibnamefont {Teles}}, \
  and\ \bibinfo {author} {\bibfnamefont {Fernanda~PC}\ \bibnamefont
  {Benetti}},\ }\bibfield  {title} {\enquote {\bibinfo {title} {Nonequilibrium
  statistical mechanics of systems with long-range interactions},}\ }\href
  {\doibase https://doi.org/10.1016/j.physrep.2013.10.001} {\bibfield
  {journal} {\bibinfo  {journal} {Physics Reports}\ }\textbf {\bibinfo {volume}
  {535}},\ \bibinfo {pages} {1--60} (\bibinfo {year} {2014})}\BibitemShut
  {NoStop}%
\bibitem [{\citenamefont {Brands}\ \emph {et~al.}(1999)\citenamefont {Brands},
  \citenamefont {Maassen},\ and\ \citenamefont
  {Clercx}}]{brands1999statistical}%
  \BibitemOpen
  \bibfield  {author} {\bibinfo {author} {\bibfnamefont {H.}~\bibnamefont
  {Brands}}, \bibinfo {author} {\bibfnamefont {S.~R.}\ \bibnamefont {Maassen}},
  \ and\ \bibinfo {author} {\bibfnamefont {H.~J.~H.}\ \bibnamefont {Clercx}},\
  }\bibfield  {title} {\enquote {\bibinfo {title} {Statistical-mechanical
  predictions and {N}avier-{S}tokes dynamics of two-dimensional flows on a
  bounded domain},}\ }\href {\doibase 10.1103/PhysRevE.60.2864} {\bibfield
  {journal} {\bibinfo  {journal} {Phys. Rev. E}\ }\textbf {\bibinfo {volume}
  {60}},\ \bibinfo {pages} {2864--2874} (\bibinfo {year} {1999})}\BibitemShut
  {NoStop}%
\bibitem [{\citenamefont {Li}\ and\ \citenamefont
  {Montgomery}(1996)}]{Li1996decaying}%
  \BibitemOpen
  \bibfield  {author} {\bibinfo {author} {\bibfnamefont {Shuojun}\ \bibnamefont
  {Li}}\ and\ \bibinfo {author} {\bibfnamefont {David}\ \bibnamefont
  {Montgomery}},\ }\bibfield  {title} {\enquote {\bibinfo {title} {Decaying
  two-dimensional turbulence with rigid walls},}\ }\href {\doibase
  https://doi.org/10.1016/0375-9601(96)00401-X} {\bibfield  {journal} {\bibinfo
   {journal} {Physics Letters A}\ }\textbf {\bibinfo {volume} {218}},\ \bibinfo
  {pages} {281--291} (\bibinfo {year} {1996})}\BibitemShut {NoStop}%
\bibitem [{\citenamefont {Van~Heijst}\ \emph {et~al.}(2006)\citenamefont
  {Van~Heijst}, \citenamefont {Clercx},\ and\ \citenamefont
  {Molenaar}}]{vanHeijst_2006}%
  \BibitemOpen
  \bibfield  {author} {\bibinfo {author} {\bibfnamefont {G.~J.~F.}\
  \bibnamefont {Van~Heijst}}, \bibinfo {author} {\bibfnamefont {H.~J.~H.}\
  \bibnamefont {Clercx}}, \ and\ \bibinfo {author} {\bibfnamefont
  {D.}~\bibnamefont {Molenaar}},\ }\bibfield  {title} {\enquote {\bibinfo
  {title} {The effects of solid boundaries on confined two-dimensional
  turbulence},}\ }\href {\doibase 10.1017/S002211200600886X} {\bibfield
  {journal} {\bibinfo  {journal} {Journal of Fluid Mechanics}\ }\textbf
  {\bibinfo {volume} {554}},\ \bibinfo {pages} {411–431} (\bibinfo {year}
  {2006})}\BibitemShut {NoStop}%
\bibitem [{\citenamefont {Gaunt}\ \emph {et~al.}(2013)\citenamefont {Gaunt},
  \citenamefont {Schmidutz}, \citenamefont {Gotlibovych}, \citenamefont
  {Smith},\ and\ \citenamefont {Hadzibabic}}]{gaunt2013bose}%
  \BibitemOpen
  \bibfield  {author} {\bibinfo {author} {\bibfnamefont {Alexander~L.}\
  \bibnamefont {Gaunt}}, \bibinfo {author} {\bibfnamefont {Tobias~F.}\
  \bibnamefont {Schmidutz}}, \bibinfo {author} {\bibfnamefont {Igor}\
  \bibnamefont {Gotlibovych}}, \bibinfo {author} {\bibfnamefont {Robert~P.}\
  \bibnamefont {Smith}}, \ and\ \bibinfo {author} {\bibfnamefont {Zoran}\
  \bibnamefont {Hadzibabic}},\ }\bibfield  {title} {\enquote {\bibinfo {title}
  {{B}ose-{E}instein {C}ondensation of {A}toms in a {U}niform {P}otential},}\
  }\href {\doibase 10.1103/PhysRevLett.110.200406} {\bibfield  {journal}
  {\bibinfo  {journal} {Phys. Rev. Lett.}\ }\textbf {\bibinfo {volume} {110}},\
  \bibinfo {pages} {200406} (\bibinfo {year} {2013})}\BibitemShut {NoStop}%
\bibitem [{\citenamefont {Chomaz}\ \emph {et~al.}(2015)\citenamefont {Chomaz},
  \citenamefont {Corman}, \citenamefont {Bienaim{\'e}}, \citenamefont
  {Desbuquois}, \citenamefont {Weitenberg}, \citenamefont {Nascimbene},
  \citenamefont {Beugnon},\ and\ \citenamefont
  {Dalibard}}]{chomaz2015emergence}%
  \BibitemOpen
  \bibfield  {author} {\bibinfo {author} {\bibfnamefont {Lauriane}\
  \bibnamefont {Chomaz}}, \bibinfo {author} {\bibfnamefont {Laura}\
  \bibnamefont {Corman}}, \bibinfo {author} {\bibfnamefont {Tom}\ \bibnamefont
  {Bienaim{\'e}}}, \bibinfo {author} {\bibfnamefont {R{\'e}mi}\ \bibnamefont
  {Desbuquois}}, \bibinfo {author} {\bibfnamefont {Christof}\ \bibnamefont
  {Weitenberg}}, \bibinfo {author} {\bibfnamefont {Sylvain}\ \bibnamefont
  {Nascimbene}}, \bibinfo {author} {\bibfnamefont {J{\'e}r{\^o}me}\
  \bibnamefont {Beugnon}}, \ and\ \bibinfo {author} {\bibfnamefont {Jean}\
  \bibnamefont {Dalibard}},\ }\bibfield  {title} {\enquote {\bibinfo {title}
  {Emergence of coherence via transverse condensation in a uniform
  quasi-two-dimensional {B}ose gas},}\ }\href {\doibase
  https://doi.org/10.1038/ncomms7162} {\bibfield  {journal} {\bibinfo
  {journal} {Nature Communications}\ }\textbf {\bibinfo {volume} {6}},\
  \bibinfo {pages} {1--10} (\bibinfo {year} {2015})}\BibitemShut {NoStop}%
\bibitem [{\citenamefont {Gauthier}\ \emph {et~al.}(2016)\citenamefont
  {Gauthier}, \citenamefont {Lenton}, \citenamefont {Parry}, \citenamefont
  {Baker}, \citenamefont {Davis}, \citenamefont {Rubinsztein-Dunlop},\ and\
  \citenamefont {Neely}}]{gauthier2016direct}%
  \BibitemOpen
  \bibfield  {author} {\bibinfo {author} {\bibfnamefont {G.}~\bibnamefont
  {Gauthier}}, \bibinfo {author} {\bibfnamefont {I.}~\bibnamefont {Lenton}},
  \bibinfo {author} {\bibfnamefont {N.~McKay}\ \bibnamefont {Parry}}, \bibinfo
  {author} {\bibfnamefont {M.}~\bibnamefont {Baker}}, \bibinfo {author}
  {\bibfnamefont {M.~J.}\ \bibnamefont {Davis}}, \bibinfo {author}
  {\bibfnamefont {H.}~\bibnamefont {Rubinsztein-Dunlop}}, \ and\ \bibinfo
  {author} {\bibfnamefont {T.~W.}\ \bibnamefont {Neely}},\ }\bibfield  {title}
  {\enquote {\bibinfo {title} {Direct imaging of a digital-micromirror device
  for configurable microscopic optical potentials},}\ }\href {\doibase
  10.1364/OPTICA.3.001136} {\bibfield  {journal} {\bibinfo  {journal} {Optica}\
  }\textbf {\bibinfo {volume} {3}},\ \bibinfo {pages} {1136--1143} (\bibinfo
  {year} {2016})}\BibitemShut {NoStop}%
\bibitem [{\citenamefont {Mukherjee}\ \emph {et~al.}(2017)\citenamefont
  {Mukherjee}, \citenamefont {Yan}, \citenamefont {Patel}, \citenamefont
  {Hadzibabic}, \citenamefont {Yefsah}, \citenamefont {Struck},\ and\
  \citenamefont {Zwierlein}}]{mukherjee2017homogeneous}%
  \BibitemOpen
  \bibfield  {author} {\bibinfo {author} {\bibfnamefont {Biswaroop}\
  \bibnamefont {Mukherjee}}, \bibinfo {author} {\bibfnamefont {Zhenjie}\
  \bibnamefont {Yan}}, \bibinfo {author} {\bibfnamefont {Parth~B}\ \bibnamefont
  {Patel}}, \bibinfo {author} {\bibfnamefont {Zoran}\ \bibnamefont
  {Hadzibabic}}, \bibinfo {author} {\bibfnamefont {Tarik}\ \bibnamefont
  {Yefsah}}, \bibinfo {author} {\bibfnamefont {Julian}\ \bibnamefont {Struck}},
  \ and\ \bibinfo {author} {\bibfnamefont {Martin~W}\ \bibnamefont
  {Zwierlein}},\ }\bibfield  {title} {\enquote {\bibinfo {title} {Homogeneous
  atomic {F}ermi gases},}\ }\href {\doibase
  https://doi.org/10.1103/PhysRevLett.118.123401} {\bibfield  {journal}
  {\bibinfo  {journal} {Physical Review Letters}\ }\textbf {\bibinfo {volume}
  {118}},\ \bibinfo {pages} {123401} (\bibinfo {year} {2017})}\BibitemShut
  {NoStop}%
\bibitem [{\citenamefont {Tajik}\ \emph {et~al.}(2019)\citenamefont {Tajik},
  \citenamefont {Rauer}, \citenamefont {Schweigler}, \citenamefont {Cataldini},
  \citenamefont {Sabino}, \citenamefont {M{\o}ller}, \citenamefont {Ji},
  \citenamefont {Mazets},\ and\ \citenamefont
  {Schmiedmayer}}]{tajik2019designing}%
  \BibitemOpen
  \bibfield  {author} {\bibinfo {author} {\bibfnamefont {Mohammadamin}\
  \bibnamefont {Tajik}}, \bibinfo {author} {\bibfnamefont {Bernhard}\
  \bibnamefont {Rauer}}, \bibinfo {author} {\bibfnamefont {Thomas}\
  \bibnamefont {Schweigler}}, \bibinfo {author} {\bibfnamefont {Federica}\
  \bibnamefont {Cataldini}}, \bibinfo {author} {\bibfnamefont {Jo{\~a}o}\
  \bibnamefont {Sabino}}, \bibinfo {author} {\bibfnamefont {Frederik~S}\
  \bibnamefont {M{\o}ller}}, \bibinfo {author} {\bibfnamefont {Si-Cong}\
  \bibnamefont {Ji}}, \bibinfo {author} {\bibfnamefont {Igor~E}\ \bibnamefont
  {Mazets}}, \ and\ \bibinfo {author} {\bibfnamefont {J{\"o}rg}\ \bibnamefont
  {Schmiedmayer}},\ }\bibfield  {title} {\enquote {\bibinfo {title} {Designing
  arbitrary one-dimensional potentials on an atom chip},}\ }\href {\doibase
  https://doi.org/10.1364/OE.27.033474} {\bibfield  {journal} {\bibinfo
  {journal} {Optics Express}\ }\textbf {\bibinfo {volume} {27}},\ \bibinfo
  {pages} {33474--33487} (\bibinfo {year} {2019})}\BibitemShut {NoStop}%
\bibitem [{\citenamefont {Navon}\ \emph {et~al.}(2021)\citenamefont {Navon},
  \citenamefont {Smith},\ and\ \citenamefont {Hadzibabic}}]{navon2021quantum}%
  \BibitemOpen
  \bibfield  {author} {\bibinfo {author} {\bibfnamefont {Nir}\ \bibnamefont
  {Navon}}, \bibinfo {author} {\bibfnamefont {Robert~P.}\ \bibnamefont
  {Smith}}, \ and\ \bibinfo {author} {\bibfnamefont {Zoran}\ \bibnamefont
  {Hadzibabic}},\ }\bibfield  {title} {\enquote {\bibinfo {title} {Quantum
  gases in optical boxes},}\ }\href {\doibase
  https://doi.org/10.1038/s41567-021-01403-z} {\bibfield  {journal} {\bibinfo
  {journal} {Nature Physics}\ }\textbf {\bibinfo {volume} {17}},\ \bibinfo
  {pages} {1334--1341} (\bibinfo {year} {2021})}\BibitemShut {NoStop}%
\bibitem [{\citenamefont {Navon}\ \emph {et~al.}(2016)\citenamefont {Navon},
  \citenamefont {Gaunt}, \citenamefont {Smith},\ and\ \citenamefont
  {Hadzibabic}}]{Navon:2016cb}%
  \BibitemOpen
  \bibfield  {author} {\bibinfo {author} {\bibfnamefont {Nir}\ \bibnamefont
  {Navon}}, \bibinfo {author} {\bibfnamefont {Alexander~L.}\ \bibnamefont
  {Gaunt}}, \bibinfo {author} {\bibfnamefont {Robert~P.}\ \bibnamefont
  {Smith}}, \ and\ \bibinfo {author} {\bibfnamefont {Zoran}\ \bibnamefont
  {Hadzibabic}},\ }\bibfield  {title} {\enquote {\bibinfo {title} {{Emergence
  of a Turbulent Cascade in a Quantum Gas}},}\ }\href {\doibase
  https://doi.org/10.1038/nature20114} {\bibfield  {journal} {\bibinfo
  {journal} {Nature}\ }\textbf {\bibinfo {volume} {539}},\ \bibinfo {pages}
  {72--75} (\bibinfo {year} {2016})}\BibitemShut {NoStop}%
\bibitem [{\citenamefont {Navon}\ \emph {et~al.}(2019)\citenamefont {Navon},
  \citenamefont {Eigen}, \citenamefont {Zhang}, \citenamefont {Lopes},
  \citenamefont {Gaunt}, \citenamefont {Fujimoto}, \citenamefont {Tsubota},
  \citenamefont {Smith},\ and\ \citenamefont
  {Hadzibabic}}]{navon2019synthetic}%
  \BibitemOpen
  \bibfield  {author} {\bibinfo {author} {\bibfnamefont {Nir}\ \bibnamefont
  {Navon}}, \bibinfo {author} {\bibfnamefont {Christoph}\ \bibnamefont
  {Eigen}}, \bibinfo {author} {\bibfnamefont {Jinyi}\ \bibnamefont {Zhang}},
  \bibinfo {author} {\bibfnamefont {Raphael}\ \bibnamefont {Lopes}}, \bibinfo
  {author} {\bibfnamefont {Alexander~L.}\ \bibnamefont {Gaunt}}, \bibinfo
  {author} {\bibfnamefont {Kazuya}\ \bibnamefont {Fujimoto}}, \bibinfo {author}
  {\bibfnamefont {Makoto}\ \bibnamefont {Tsubota}}, \bibinfo {author}
  {\bibfnamefont {Robert~P.}\ \bibnamefont {Smith}}, \ and\ \bibinfo {author}
  {\bibfnamefont {Zoran}\ \bibnamefont {Hadzibabic}},\ }\bibfield  {title}
  {\enquote {\bibinfo {title} {Synthetic dissipation and cascade fluxes in a
  turbulent quantum gas},}\ }\href {\doibase 10.1126/science.aau6103}
  {\bibfield  {journal} {\bibinfo  {journal} {Science}\ }\textbf {\bibinfo
  {volume} {366}},\ \bibinfo {pages} {382--385} (\bibinfo {year}
  {2019})}\BibitemShut {NoStop}%
\bibitem [{\citenamefont {Glidden}\ \emph {et~al.}(2021)\citenamefont
  {Glidden}, \citenamefont {Eigen}, \citenamefont {Dogra}, \citenamefont
  {Hilker}, \citenamefont {Smith},\ and\ \citenamefont
  {Hadzibabic}}]{glidden2021bidirectional}%
  \BibitemOpen
  \bibfield  {author} {\bibinfo {author} {\bibfnamefont {Jake~AP}\ \bibnamefont
  {Glidden}}, \bibinfo {author} {\bibfnamefont {Christoph}\ \bibnamefont
  {Eigen}}, \bibinfo {author} {\bibfnamefont {Lena~H}\ \bibnamefont {Dogra}},
  \bibinfo {author} {\bibfnamefont {Timon~A}\ \bibnamefont {Hilker}}, \bibinfo
  {author} {\bibfnamefont {Robert~P}\ \bibnamefont {Smith}}, \ and\ \bibinfo
  {author} {\bibfnamefont {Zoran}\ \bibnamefont {Hadzibabic}},\ }\bibfield
  {title} {\enquote {\bibinfo {title} {Bidirectional dynamic scaling in an
  isolated bose gas far from equilibrium},}\ }\href {\doibase
  https://doi.org/10.1038/s41567-020-01114-x} {\bibfield  {journal} {\bibinfo
  {journal} {Nature Physics}\ }\textbf {\bibinfo {volume} {17}},\ \bibinfo
  {pages} {457--461} (\bibinfo {year} {2021})}\BibitemShut {NoStop}%
\bibitem [{\citenamefont {Gauthier}\ \emph
  {et~al.}(2019{\natexlab{a}})\citenamefont {Gauthier}, \citenamefont {Reeves},
  \citenamefont {Yu}, \citenamefont {Bradley}, \citenamefont {Baker},
  \citenamefont {Bell}, \citenamefont {Rubinsztein-Dunlop}, \citenamefont
  {Davis},\ and\ \citenamefont {Neely}}]{gauthier2019giant}%
  \BibitemOpen
  \bibfield  {author} {\bibinfo {author} {\bibfnamefont {Guillaume}\
  \bibnamefont {Gauthier}}, \bibinfo {author} {\bibfnamefont {Matthew~T}\
  \bibnamefont {Reeves}}, \bibinfo {author} {\bibfnamefont {Xiaoquan}\
  \bibnamefont {Yu}}, \bibinfo {author} {\bibfnamefont {Ashton~S}\ \bibnamefont
  {Bradley}}, \bibinfo {author} {\bibfnamefont {Mark~A}\ \bibnamefont {Baker}},
  \bibinfo {author} {\bibfnamefont {Thomas~A}\ \bibnamefont {Bell}}, \bibinfo
  {author} {\bibfnamefont {Halina}\ \bibnamefont {Rubinsztein-Dunlop}},
  \bibinfo {author} {\bibfnamefont {Matthew~J}\ \bibnamefont {Davis}}, \ and\
  \bibinfo {author} {\bibfnamefont {Tyler~W}\ \bibnamefont {Neely}},\
  }\bibfield  {title} {\enquote {\bibinfo {title} {Giant vortex clusters in a
  two-dimensional quantum fluid},}\ }\href {\doibase 10.1126/science.aat5718}
  {\bibfield  {journal} {\bibinfo  {journal} {Science}\ }\textbf {\bibinfo
  {volume} {364}},\ \bibinfo {pages} {1264--1267} (\bibinfo {year}
  {2019}{\natexlab{a}})}\BibitemShut {NoStop}%
\bibitem [{\citenamefont {Johnstone}\ \emph {et~al.}(2019)\citenamefont
  {Johnstone}, \citenamefont {Groszek}, \citenamefont {Starkey}, \citenamefont
  {Billington}, \citenamefont {Simula},\ and\ \citenamefont
  {Helmerson}}]{johnstone2019evolution}%
  \BibitemOpen
  \bibfield  {author} {\bibinfo {author} {\bibfnamefont {Shaun~P}\ \bibnamefont
  {Johnstone}}, \bibinfo {author} {\bibfnamefont {Andrew~J}\ \bibnamefont
  {Groszek}}, \bibinfo {author} {\bibfnamefont {Philip~T}\ \bibnamefont
  {Starkey}}, \bibinfo {author} {\bibfnamefont {Christopher~J}\ \bibnamefont
  {Billington}}, \bibinfo {author} {\bibfnamefont {Tapio~P}\ \bibnamefont
  {Simula}}, \ and\ \bibinfo {author} {\bibfnamefont {Kristian}\ \bibnamefont
  {Helmerson}},\ }\bibfield  {title} {\enquote {\bibinfo {title} {Evolution of
  large-scale flow from turbulence in a two-dimensional superfluid},}\ }\href
  {\doibase 10.1126/science.aat5793} {\bibfield  {journal} {\bibinfo  {journal}
  {Science}\ }\textbf {\bibinfo {volume} {364}},\ \bibinfo {pages} {1267--1271}
  (\bibinfo {year} {2019})}\BibitemShut {NoStop}%
\bibitem [{\citenamefont {Stockdale}\ \emph {et~al.}(2020)\citenamefont
  {Stockdale}, \citenamefont {Reeves}, \citenamefont {Yu}, \citenamefont
  {Gauthier}, \citenamefont {Goddard-Lee}, \citenamefont {Bowen}, \citenamefont
  {Neely},\ and\ \citenamefont {Davis}}]{stockdale2019universal}%
  \BibitemOpen
  \bibfield  {author} {\bibinfo {author} {\bibfnamefont {Oliver~R.}\
  \bibnamefont {Stockdale}}, \bibinfo {author} {\bibfnamefont {Matthew~T.}\
  \bibnamefont {Reeves}}, \bibinfo {author} {\bibfnamefont {Xiaoquan}\
  \bibnamefont {Yu}}, \bibinfo {author} {\bibfnamefont {Guillaume}\
  \bibnamefont {Gauthier}}, \bibinfo {author} {\bibfnamefont {Kwan}\
  \bibnamefont {Goddard-Lee}}, \bibinfo {author} {\bibfnamefont {Warwick~P.}\
  \bibnamefont {Bowen}}, \bibinfo {author} {\bibfnamefont {Tyler~W.}\
  \bibnamefont {Neely}}, \ and\ \bibinfo {author} {\bibfnamefont {Matthew~J.}\
  \bibnamefont {Davis}},\ }\bibfield  {title} {\enquote {\bibinfo {title}
  {Universal dynamics in the expansion of vortex clusters in a dissipative
  two-dimensional superfluid},}\ }\href
  {https://journals.aps.org/prresearch/abstract/10.1103/PhysRevResearch.2.033138}
  {\bibfield  {journal} {\bibinfo  {journal} {Physical Review Research}\
  }\textbf {\bibinfo {volume} {2}},\ \bibinfo {pages} {033138} (\bibinfo {year}
  {2020})}\BibitemShut {NoStop}%
\bibitem [{\citenamefont {Kwon}\ \emph {et~al.}(2021)\citenamefont {Kwon},
  \citenamefont {Pace}, \citenamefont {Xhani}, \citenamefont {Galantucci},
  \citenamefont {Falconi}, \citenamefont {Inguscio}, \citenamefont {Scazza},\
  and\ \citenamefont {Roati}}]{kwon2021sound}%
  \BibitemOpen
  \bibfield  {author} {\bibinfo {author} {\bibfnamefont {W.~J.}\ \bibnamefont
  {Kwon}}, \bibinfo {author} {\bibfnamefont {G.~Del}\ \bibnamefont {Pace}},
  \bibinfo {author} {\bibfnamefont {K.}~\bibnamefont {Xhani}}, \bibinfo
  {author} {\bibfnamefont {L.}~\bibnamefont {Galantucci}}, \bibinfo {author}
  {\bibfnamefont {A.~Muzi}\ \bibnamefont {Falconi}}, \bibinfo {author}
  {\bibfnamefont {M.}~\bibnamefont {Inguscio}}, \bibinfo {author}
  {\bibfnamefont {F.}~\bibnamefont {Scazza}}, \ and\ \bibinfo {author}
  {\bibfnamefont {G.}~\bibnamefont {Roati}},\ }\bibfield  {title} {\enquote
  {\bibinfo {title} {Sound emission and annihilations in a programmable quantum
  vortex collider},}\ }\href {\doibase
  https://doi.org/10.1038/s41586-021-04047-4} {\bibfield  {journal} {\bibinfo
  {journal} {Nature}\ }\textbf {\bibinfo {volume} {600}},\ \bibinfo {pages}
  {64--69} (\bibinfo {year} {2021})}\BibitemShut {NoStop}%
\bibitem [{\citenamefont {Fetter}(1966)}]{fetter1966vortices}%
  \BibitemOpen
  \bibfield  {author} {\bibinfo {author} {\bibfnamefont {Alexander~L.}\
  \bibnamefont {Fetter}},\ }\bibfield  {title} {\enquote {\bibinfo {title}
  {{V}ortices in an {I}mperfect {B}ose {G}as. {IV}. {T}ranslational
  {V}elocity},}\ }\href {\doibase 10.1103/PhysRev.151.100} {\bibfield
  {journal} {\bibinfo  {journal} {Physical Review}\ }\textbf {\bibinfo {volume}
  {151}},\ \bibinfo {pages} {100--104} (\bibinfo {year} {1966})}\BibitemShut
  {NoStop}%
\bibitem [{\citenamefont {Groszek}\ \emph {et~al.}(2018)\citenamefont
  {Groszek}, \citenamefont {Paganin}, \citenamefont {Helmerson},\ and\
  \citenamefont {Simula}}]{groszek2018motion}%
  \BibitemOpen
  \bibfield  {author} {\bibinfo {author} {\bibfnamefont {Andrew~J}\
  \bibnamefont {Groszek}}, \bibinfo {author} {\bibfnamefont {David~M}\
  \bibnamefont {Paganin}}, \bibinfo {author} {\bibfnamefont {Kristian}\
  \bibnamefont {Helmerson}}, \ and\ \bibinfo {author} {\bibfnamefont {Tapio~P}\
  \bibnamefont {Simula}},\ }\bibfield  {title} {\enquote {\bibinfo {title}
  {Motion of vortices in inhomogeneous {B}ose-{E}instein condensates},}\ }\href
  {\doibase 10.1103/PhysRevA.97.023617} {\bibfield  {journal} {\bibinfo
  {journal} {Physical Review A}\ }\textbf {\bibinfo {volume} {97}},\ \bibinfo
  {pages} {023617} (\bibinfo {year} {2018})}\BibitemShut {NoStop}%
\bibitem [{\citenamefont {Dritschel}\ \emph {et~al.}(2015)\citenamefont
  {Dritschel}, \citenamefont {Lucia},\ and\ \citenamefont
  {Poje}}]{dritschel2015ergodicity}%
  \BibitemOpen
  \bibfield  {author} {\bibinfo {author} {\bibfnamefont {David~G.}\
  \bibnamefont {Dritschel}}, \bibinfo {author} {\bibfnamefont {Marcello}\
  \bibnamefont {Lucia}}, \ and\ \bibinfo {author} {\bibfnamefont {Andrew~C.}\
  \bibnamefont {Poje}},\ }\bibfield  {title} {\enquote {\bibinfo {title}
  {Ergodicity and spectral cascades in point vortex flows on the sphere},}\
  }\href {\doibase 10.1103/PhysRevE.91.063014} {\bibfield  {journal} {\bibinfo
  {journal} {Phys. Rev. E}\ }\textbf {\bibinfo {volume} {91}},\ \bibinfo
  {pages} {063014} (\bibinfo {year} {2015})}\BibitemShut {NoStop}%
\bibitem [{\citenamefont {Esler}\ and\ \citenamefont
  {Ashbee}(2015)}]{esler2015universal}%
  \BibitemOpen
  \bibfield  {author} {\bibinfo {author} {\bibfnamefont {J.~G.}\ \bibnamefont
  {Esler}}\ and\ \bibinfo {author} {\bibfnamefont {T.~L.}\ \bibnamefont
  {Ashbee}},\ }\bibfield  {title} {\enquote {\bibinfo {title} {Universal
  statistics of point vortex turbulence},}\ }\href@noop {} {\bibfield
  {journal} {\bibinfo  {journal} {J. Fluid Mech.}\ }\textbf {\bibinfo {volume}
  {779}},\ \bibinfo {pages} {275--308} (\bibinfo {year} {2015})}\BibitemShut
  {NoStop}%
\bibitem [{\citenamefont {Salman}\ and\ \citenamefont
  {Maestrini}(2016)}]{salman2016long}%
  \BibitemOpen
  \bibfield  {author} {\bibinfo {author} {\bibfnamefont {Hayder}\ \bibnamefont
  {Salman}}\ and\ \bibinfo {author} {\bibfnamefont {Davide}\ \bibnamefont
  {Maestrini}},\ }\bibfield  {title} {\enquote {\bibinfo {title} {Long-range
  ordering of topological excitations in a two-dimensional superfluid far from
  equilibrium},}\ }\href {\doibase 10.1103/PhysRevA.94.043642} {\bibfield
  {journal} {\bibinfo  {journal} {Physical Review A}\ }\textbf {\bibinfo
  {volume} {94}},\ \bibinfo {pages} {043642} (\bibinfo {year}
  {2016})}\BibitemShut {NoStop}%
\bibitem [{\citenamefont {Yu}\ \emph {et~al.}(2016)\citenamefont {Yu},
  \citenamefont {Billam}, \citenamefont {Nian}, \citenamefont {Reeves},\ and\
  \citenamefont {Bradley}}]{yu2016theory}%
  \BibitemOpen
  \bibfield  {author} {\bibinfo {author} {\bibfnamefont {Xiaoquan}\
  \bibnamefont {Yu}}, \bibinfo {author} {\bibfnamefont {Thomas~P.}\
  \bibnamefont {Billam}}, \bibinfo {author} {\bibfnamefont {Jun}\ \bibnamefont
  {Nian}}, \bibinfo {author} {\bibfnamefont {Matthew~T.}\ \bibnamefont
  {Reeves}}, \ and\ \bibinfo {author} {\bibfnamefont {Ashton~S.}\ \bibnamefont
  {Bradley}},\ }\bibfield  {title} {\enquote {\bibinfo {title} {Theory of the
  vortex-clustering transition in a confined two-dimensional quantum fluid},}\
  }\href {\doibase 10.1103/PhysRevA.94.023602} {\bibfield  {journal} {\bibinfo
  {journal} {Physical Review A}\ }\textbf {\bibinfo {volume} {94}},\ \bibinfo
  {pages} {023602} (\bibinfo {year} {2016})}\BibitemShut {NoStop}%
\bibitem [{\citenamefont {Smith}(1989)}]{smith_phase-transition_1989}%
  \BibitemOpen
  \bibfield  {author} {\bibinfo {author} {\bibfnamefont {R~A}\ \bibnamefont
  {Smith}},\ }\bibfield  {title} {\enquote {\bibinfo {title}
  {Phase-{{Transition Behavior}} in a {{Negative}}-{{Temperature
  Guiding}}-{{Center Plasma}}},}\ }\href {\doibase 10.1103/PhysRevLett.63.1479}
  {\bibfield  {journal} {\bibinfo  {journal} {Physical Review Letters}\
  }\textbf {\bibinfo {volume} {63}},\ \bibinfo {pages} {1479--1482} (\bibinfo
  {year} {1989})}\BibitemShut {NoStop}%
\bibitem [{\citenamefont {Newton}(2013)}]{newton2013nvortexproblem}%
  \BibitemOpen
  \bibfield  {author} {\bibinfo {author} {\bibfnamefont {Paul~K.}\ \bibnamefont
  {Newton}},\ }\href {https://link.springer.com/book/10.1007/978-1-4684-9290-3}
  {\emph {\bibinfo {title} {The {N}-Vortex Problem: Analytical Techniques}}},\
  Vol.\ \bibinfo {volume} {145}\ (\bibinfo  {publisher} {Springer Science \&
  Business Media},\ \bibinfo {year} {2013})\BibitemShut {NoStop}%
\bibitem [{\citenamefont {Smith}\ and\ \citenamefont
  {O'Neil}(1990)}]{smith1990nonaxisymmetric}%
  \BibitemOpen
  \bibfield  {author} {\bibinfo {author} {\bibfnamefont {Ralph~A.}\
  \bibnamefont {Smith}}\ and\ \bibinfo {author} {\bibfnamefont {Thomas~M.}\
  \bibnamefont {O'Neil}},\ }\bibfield  {title} {\enquote {\bibinfo {title}
  {Nonaxisymmetric thermal equilibria of a cylindrically bounded guiding-center
  plasma or discrete vortex system},}\ }\href {\doibase 10.1063/1.859362}
  {\bibfield  {journal} {\bibinfo  {journal} {Physics of Fluids B: Plasma
  Physics}\ }\textbf {\bibinfo {volume} {2}},\ \bibinfo {pages} {2961--2975}
  (\bibinfo {year} {1990})}\BibitemShut {NoStop}%
\bibitem [{Note1()}]{Note1}%
  \BibitemOpen
  \bibinfo {note} {$M$ is often referred to as the angular momentum~\cite
  {newton2013nvortexproblem,aref1979motion}. The distinction is not so
  important when applying the point-vortex model to the dynamics of a classical
  Euler fluid (where the vortex number is conserved), but is for a superfluid
  as vortex annihilation can occur at the boundary; here $L$ varies
  continuously as a vortex leaves at the boundary, whereas $M$ changes
  discontinuously.}\BibitemShut {Stop}%
\bibitem [{Note2()}]{Note2}%
  \BibitemOpen
  \bibinfo {note} {Note however that for a given solution $\omega $ does not
  necessarily correspond to any particular physical rotation rate in the
  system; see Ref.~\cite {smith1990nonaxisymmetric}.}\BibitemShut {Stop}%
\bibitem [{Note3()}]{Note3}%
  \BibitemOpen
  \bibinfo {note} {This behaviour follows from the fact that below the off-axis
  transition the vortex positions are approximately normally distributed about
  the origin, with the width of the normal distribution determined by the
  angular momentum.}\BibitemShut {Stop}%
\bibitem [{\citenamefont {Samson}(2012)}]{samson2012generating}%
  \BibitemOpen
  \bibfield  {author} {\bibinfo {author} {\bibfnamefont {Edward Carlo~Copon}\
  \bibnamefont {Samson}},\ }\emph {\bibinfo {title} {Generating and
  manipulating quantized vortices in highly oblate {B}ose-{E}instein
  condensates}},\ \href@noop {} {Ph.D. thesis},\ \bibinfo  {school} {The
  University of Arizona} (\bibinfo {year} {2012})\BibitemShut {NoStop}%
\bibitem [{\citenamefont {Bradley}\ \emph {et~al.}(1997)\citenamefont
  {Bradley}, \citenamefont {Sackett},\ and\ \citenamefont
  {Hulet}}]{bradley1997bose}%
  \BibitemOpen
  \bibfield  {author} {\bibinfo {author} {\bibfnamefont {Curtis~Charles}\
  \bibnamefont {Bradley}}, \bibinfo {author} {\bibfnamefont {C.~A.}\
  \bibnamefont {Sackett}}, \ and\ \bibinfo {author} {\bibfnamefont {R.~G.}\
  \bibnamefont {Hulet}},\ }\bibfield  {title} {\enquote {\bibinfo {title}
  {{B}ose-{E}instein condensation of lithium: {O}bservation of limited
  condensate number},}\ }\href {\doibase 10.1103/PhysRevLett.78.985} {\bibfield
   {journal} {\bibinfo  {journal} {Physical Review Letters}\ }\textbf {\bibinfo
  {volume} {78}},\ \bibinfo {pages} {985} (\bibinfo {year} {1997})}\BibitemShut
  {NoStop}%
\bibitem [{\citenamefont {Rakonjac}\ \emph {et~al.}(2016)\citenamefont
  {Rakonjac}, \citenamefont {Marchant}, \citenamefont {Billam}, \citenamefont
  {Helm}, \citenamefont {Yu}, \citenamefont {Gardiner},\ and\ \citenamefont
  {Cornish}}]{rakonjac2016measuring}%
  \BibitemOpen
  \bibfield  {author} {\bibinfo {author} {\bibfnamefont {A.}~\bibnamefont
  {Rakonjac}}, \bibinfo {author} {\bibfnamefont {A.~L.}\ \bibnamefont
  {Marchant}}, \bibinfo {author} {\bibfnamefont {T.~P.}\ \bibnamefont
  {Billam}}, \bibinfo {author} {\bibfnamefont {J.~L.}\ \bibnamefont {Helm}},
  \bibinfo {author} {\bibfnamefont {M.~M.~H.}\ \bibnamefont {Yu}}, \bibinfo
  {author} {\bibfnamefont {S.~A.}\ \bibnamefont {Gardiner}}, \ and\ \bibinfo
  {author} {\bibfnamefont {S.~L.}\ \bibnamefont {Cornish}},\ }\bibfield
  {title} {\enquote {\bibinfo {title} {Measuring the disorder of vortex
  lattices in a {B}ose-{E}instein condensate},}\ }\href {\doibase
  10.1103/PhysRevA.93.013607} {\bibfield  {journal} {\bibinfo  {journal}
  {Physical Review A}\ }\textbf {\bibinfo {volume} {93}},\ \bibinfo {pages}
  {013607} (\bibinfo {year} {2016})}\BibitemShut {NoStop}%
\bibitem [{\citenamefont {Chavanis}\ and\ \citenamefont
  {Sommeria}(1996)}]{chavanis1996classification}%
  \BibitemOpen
  \bibfield  {author} {\bibinfo {author} {\bibfnamefont {Pierre-Henri}\
  \bibnamefont {Chavanis}}\ and\ \bibinfo {author} {\bibfnamefont {Joel}\
  \bibnamefont {Sommeria}},\ }\bibfield  {title} {\enquote {\bibinfo {title}
  {Classification of self-organized vortices in two-dimensional turbulence: the
  case of a bounded domain},}\ }\href {\doibase
  https://doi.org/10.1017/S0022112096000316} {\bibfield  {journal} {\bibinfo
  {journal} {Journal of Fluid Mechanics}\ }\textbf {\bibinfo {volume} {314}},\
  \bibinfo {pages} {267--297} (\bibinfo {year} {1996})}\BibitemShut {NoStop}%
\bibitem [{\citenamefont {Weiss}\ and\ \citenamefont
  {McWilliams}(1993)}]{weiss1993temporal}%
  \BibitemOpen
  \bibfield  {author} {\bibinfo {author} {\bibfnamefont {Jeffrey~B.}\
  \bibnamefont {Weiss}}\ and\ \bibinfo {author} {\bibfnamefont {James~C.}\
  \bibnamefont {McWilliams}},\ }\bibfield  {title} {\enquote {\bibinfo {title}
  {Temporal scaling behavior of decaying two‐dimensional turbulence},}\
  }\href {\doibase 10.1063/1.858647} {\bibfield  {journal} {\bibinfo  {journal}
  {Physics of Fluids A: Fluid Dynamics}\ }\textbf {\bibinfo {volume} {5}},\
  \bibinfo {pages} {608--621} (\bibinfo {year} {1993})}\BibitemShut {NoStop}%
\bibitem [{\citenamefont {Simula}\ \emph {et~al.}(2014)\citenamefont {Simula},
  \citenamefont {Davis},\ and\ \citenamefont
  {Helmerson}}]{simula2014emergence}%
  \BibitemOpen
  \bibfield  {author} {\bibinfo {author} {\bibfnamefont {Tapio}\ \bibnamefont
  {Simula}}, \bibinfo {author} {\bibfnamefont {Matthew~J.}\ \bibnamefont
  {Davis}}, \ and\ \bibinfo {author} {\bibfnamefont {Kristian}\ \bibnamefont
  {Helmerson}},\ }\bibfield  {title} {\enquote {\bibinfo {title} {Emergence of
  order from turbulence in an isolated planar superfluid},}\ }\href {\doibase
  10.1103/PhysRevLett.113.165302} {\bibfield  {journal} {\bibinfo  {journal}
  {Physical Review Letters}\ }\textbf {\bibinfo {volume} {113}},\ \bibinfo
  {pages} {165302} (\bibinfo {year} {2014})}\BibitemShut {NoStop}%
\bibitem [{\citenamefont {Billam}\ \emph {et~al.}(2015)\citenamefont {Billam},
  \citenamefont {Reeves},\ and\ \citenamefont {Bradley}}]{billam2015spectral}%
  \BibitemOpen
  \bibfield  {author} {\bibinfo {author} {\bibfnamefont {T.~P.}\ \bibnamefont
  {Billam}}, \bibinfo {author} {\bibfnamefont {M.~T.}\ \bibnamefont {Reeves}},
  \ and\ \bibinfo {author} {\bibfnamefont {A.~S.}\ \bibnamefont {Bradley}},\
  }\bibfield  {title} {\enquote {\bibinfo {title} {Spectral energy transport in
  two-dimensional quantum vortex dynamics},}\ }\href {\doibase
  10.1103/PhysRevA.91.023615} {\bibfield  {journal} {\bibinfo  {journal} {Phys.
  Rev. A}\ }\textbf {\bibinfo {volume} {91}},\ \bibinfo {pages} {023615}
  (\bibinfo {year} {2015})}\BibitemShut {NoStop}%
\bibitem [{Note4()}]{Note4}%
  \BibitemOpen
  \bibinfo {note} {The positions of the additional vortices were determined by
  using the experimental histograms [cf.~Fig.~\ref {fig:StirringSchematic}] as
  probability distributions for rejection sampling. For experiments I and V,
  one additional vortex was added to each experimental run. For experiment II,
  between 2 and 3 vortices were added with equal probability.}\BibitemShut
  {Stop}%
\bibitem [{\citenamefont {T{\"o}rnkvist}\ and\ \citenamefont
  {Schr{\"o}der}(1997)}]{tornkvist1997vortex}%
  \BibitemOpen
  \bibfield  {author} {\bibinfo {author} {\bibfnamefont {Ola}\ \bibnamefont
  {T{\"o}rnkvist}}\ and\ \bibinfo {author} {\bibfnamefont {Elsebeth}\
  \bibnamefont {Schr{\"o}der}},\ }\bibfield  {title} {\enquote {\bibinfo
  {title} {Vortex dynamics in dissipative systems},}\ }\href {\doibase
  10.1103/PhysRevLett.78.1908} {\bibfield  {journal} {\bibinfo  {journal}
  {Physical Review Letters}\ }\textbf {\bibinfo {volume} {78}},\ \bibinfo
  {pages} {1908} (\bibinfo {year} {1997})}\BibitemShut {NoStop}%
\bibitem [{\citenamefont {Blakie}\ and\ \citenamefont
  {Davis}(2005)}]{blakie2005projected}%
  \BibitemOpen
  \bibfield  {author} {\bibinfo {author} {\bibfnamefont {P~Blair}\ \bibnamefont
  {Blakie}}\ and\ \bibinfo {author} {\bibfnamefont {Matthew~J}\ \bibnamefont
  {Davis}},\ }\bibfield  {title} {\enquote {\bibinfo {title} {Projected
  {G}ross-{P}itaevskii equation for harmonically confined {B}ose gases at
  finite temperature},}\ }\href
  {https://journals.aps.org/pra/abstract/10.1103/PhysRevA.72.063608} {\bibfield
   {journal} {\bibinfo  {journal} {Physical Review A}\ }\textbf {\bibinfo
  {volume} {72}},\ \bibinfo {pages} {063608} (\bibinfo {year}
  {2005})}\BibitemShut {NoStop}%
\bibitem [{\citenamefont {Gardiner}\ \emph {et~al.}(2002)\citenamefont
  {Gardiner}, \citenamefont {Anglin},\ and\ \citenamefont
  {Fudge}}]{Gardiner_2002}%
  \BibitemOpen
  \bibfield  {author} {\bibinfo {author} {\bibfnamefont {C~W}\ \bibnamefont
  {Gardiner}}, \bibinfo {author} {\bibfnamefont {J~R}\ \bibnamefont {Anglin}},
  \ and\ \bibinfo {author} {\bibfnamefont {T~I~A}\ \bibnamefont {Fudge}},\
  }\bibfield  {title} {\enquote {\bibinfo {title} {The {S}tochastic
  {G}ross-{P}itaevskii equation},}\ }\href {\doibase
  10.1088/0953-4075/35/6/310} {\bibfield  {journal} {\bibinfo  {journal}
  {Journal of Physics B: Atomic, Molecular and Optical Physics}\ }\textbf
  {\bibinfo {volume} {35}},\ \bibinfo {pages} {1555--1582} (\bibinfo {year}
  {2002})}\BibitemShut {NoStop}%
\bibitem [{\citenamefont {Gardiner}\ and\ \citenamefont
  {Davis}(2003)}]{gardiner2003stochastic}%
  \BibitemOpen
  \bibfield  {author} {\bibinfo {author} {\bibfnamefont {C.~W.}\ \bibnamefont
  {Gardiner}}\ and\ \bibinfo {author} {\bibfnamefont {M.~J.}\ \bibnamefont
  {Davis}},\ }\bibfield  {title} {\enquote {\bibinfo {title} {The stochastic
  {G}ross--{P}itaevskii equation: Ii},}\ }\href
  {https://iopscience.iop.org/article/10.1088/0953-4075/36/23/010} {\bibfield
  {journal} {\bibinfo  {journal} {Journal of Physics B: Atomic, Molecular and
  Optical Physics}\ }\textbf {\bibinfo {volume} {36}},\ \bibinfo {pages} {4731}
  (\bibinfo {year} {2003})}\BibitemShut {NoStop}%
\bibitem [{\citenamefont {Rooney}\ \emph {et~al.}(2012)\citenamefont {Rooney},
  \citenamefont {Blakie},\ and\ \citenamefont
  {Bradley}}]{rooney2012stochastic}%
  \BibitemOpen
  \bibfield  {author} {\bibinfo {author} {\bibfnamefont {S.~J.}\ \bibnamefont
  {Rooney}}, \bibinfo {author} {\bibfnamefont {P.~B.}\ \bibnamefont {Blakie}},
  \ and\ \bibinfo {author} {\bibfnamefont {A.~S.}\ \bibnamefont {Bradley}},\
  }\bibfield  {title} {\enquote {\bibinfo {title} {Stochastic projected
  {G}ross-{P}itaevskii equation},}\ }\href
  {https://journals.aps.org/pra/abstract/10.1103/PhysRevA.90.023631} {\bibfield
   {journal} {\bibinfo  {journal} {Physical Review A}\ }\textbf {\bibinfo
  {volume} {86}},\ \bibinfo {pages} {053634} (\bibinfo {year}
  {2012})}\BibitemShut {NoStop}%
\bibitem [{\citenamefont {Blakie}\ \emph {et~al.}(2008)\citenamefont {Blakie},
  \citenamefont {Bradley}, \citenamefont {Davis}, \citenamefont {Ballagh},\
  and\ \citenamefont {Gardiner}}]{blakie2008dynamics}%
  \BibitemOpen
  \bibfield  {author} {\bibinfo {author} {\bibfnamefont {P.~B.}\ \bibnamefont
  {Blakie}}, \bibinfo {author} {\bibfnamefont {A.~S.}\ \bibnamefont {Bradley}},
  \bibinfo {author} {\bibfnamefont {M.~J.}\ \bibnamefont {Davis}}, \bibinfo
  {author} {\bibfnamefont {R.~J.}\ \bibnamefont {Ballagh}}, \ and\ \bibinfo
  {author} {\bibfnamefont {C.~W.}\ \bibnamefont {Gardiner}},\ }\bibfield
  {title} {\enquote {\bibinfo {title} {Dynamics and statistical mechanics of
  ultra-cold bose gases using c-field techniques},}\ }\href
  {https://www.tandfonline.com/doi/full/10.1080/00018730802564254} {\bibfield
  {journal} {\bibinfo  {journal} {Advances in Physics}\ }\textbf {\bibinfo
  {volume} {57}},\ \bibinfo {pages} {363--455} (\bibinfo {year}
  {2008})}\BibitemShut {NoStop}%
\bibitem [{\citenamefont {Kim}\ \emph {et~al.}(2016)\citenamefont {Kim},
  \citenamefont {Kwon},\ and\ \citenamefont {Shin}}]{kim2016role}%
  \BibitemOpen
  \bibfield  {author} {\bibinfo {author} {\bibfnamefont {Joon~Hyun}\
  \bibnamefont {Kim}}, \bibinfo {author} {\bibfnamefont {Woo~Jin}\ \bibnamefont
  {Kwon}}, \ and\ \bibinfo {author} {\bibfnamefont {Y.}~\bibnamefont {Shin}},\
  }\bibfield  {title} {\enquote {\bibinfo {title} {Role of thermal friction in
  relaxation of turbulent {B}ose-{E}instein condensates},}\ }\href {\doibase
  10.1103/PhysRevA.94.033612} {\bibfield  {journal} {\bibinfo  {journal}
  {Physical Review A}\ }\textbf {\bibinfo {volume} {94}},\ \bibinfo {pages}
  {033612} (\bibinfo {year} {2016})}\BibitemShut {NoStop}%
\bibitem [{\citenamefont {Sachkou}\ \emph {et~al.}(2019)\citenamefont
  {Sachkou}, \citenamefont {Baker}, \citenamefont {Harris}, \citenamefont
  {Stockdale}, \citenamefont {Forstner}, \citenamefont {Reeves}, \citenamefont
  {He}, \citenamefont {McAuslan}, \citenamefont {Bradley}, \citenamefont
  {Davis},\ and\ \citenamefont {Bowen}}]{sachkou2019coherent}%
  \BibitemOpen
  \bibfield  {author} {\bibinfo {author} {\bibfnamefont {Yauhen~P.}\
  \bibnamefont {Sachkou}}, \bibinfo {author} {\bibfnamefont {Christopher~G.}\
  \bibnamefont {Baker}}, \bibinfo {author} {\bibfnamefont {Glen~I.}\
  \bibnamefont {Harris}}, \bibinfo {author} {\bibfnamefont {Oliver~R.}\
  \bibnamefont {Stockdale}}, \bibinfo {author} {\bibfnamefont {Stefan}\
  \bibnamefont {Forstner}}, \bibinfo {author} {\bibfnamefont {Matthew~T.}\
  \bibnamefont {Reeves}}, \bibinfo {author} {\bibfnamefont {Xin}\ \bibnamefont
  {He}}, \bibinfo {author} {\bibfnamefont {David~L.}\ \bibnamefont {McAuslan}},
  \bibinfo {author} {\bibfnamefont {Ashton~S.}\ \bibnamefont {Bradley}},
  \bibinfo {author} {\bibfnamefont {Matthew~J.}\ \bibnamefont {Davis}}, \ and\
  \bibinfo {author} {\bibfnamefont {Warwick~P.}\ \bibnamefont {Bowen}},\
  }\bibfield  {title} {\enquote {\bibinfo {title} {Coherent vortex dynamics in
  a strongly interacting superfluid on a silicon chip},}\ }\href {\doibase
  10.1126/science.aat5793} {\bibfield  {journal} {\bibinfo  {journal}
  {Science}\ }\textbf {\bibinfo {volume} {366}},\ \bibinfo {pages} {1480--1485}
  (\bibinfo {year} {2019})}\BibitemShut {NoStop}%
\bibitem [{\citenamefont {Chorin}(2013)}]{chorin2013vorticity}%
  \BibitemOpen
  \bibfield  {author} {\bibinfo {author} {\bibfnamefont {Alexandre~J}\
  \bibnamefont {Chorin}},\ }\href
  {https://www.springer.com/gp/book/9780387941974} {\emph {\bibinfo {title}
  {Vorticity and turbulence}}},\ Vol.\ \bibinfo {volume} {103}\ (\bibinfo
  {publisher} {Springer Science \& Business Media},\ \bibinfo {year}
  {2013})\BibitemShut {NoStop}%
\bibitem [{\citenamefont {Bogatskiy}\ and\ \citenamefont
  {Wiegmann}(2019)}]{bogatskiy2019edge}%
  \BibitemOpen
  \bibfield  {author} {\bibinfo {author} {\bibfnamefont {A.}~\bibnamefont
  {Bogatskiy}}\ and\ \bibinfo {author} {\bibfnamefont {P.}~\bibnamefont
  {Wiegmann}},\ }\bibfield  {title} {\enquote {\bibinfo {title} {Edge wave and
  boundary layer of vortex matter},}\ }\href {\doibase
  10.1103/PhysRevLett.122.214505} {\bibfield  {journal} {\bibinfo  {journal}
  {Physical Review Letters}\ }\textbf {\bibinfo {volume} {122}},\ \bibinfo
  {pages} {214505} (\bibinfo {year} {2019})}\BibitemShut {NoStop}%
\bibitem [{\citenamefont {Tabeling}\ and\ \citenamefont
  {Cardoso}(2012)}]{tabeling2012turbulence}%
  \BibitemOpen
  \bibfield  {author} {\bibinfo {author} {\bibfnamefont {Patrick}\ \bibnamefont
  {Tabeling}}\ and\ \bibinfo {author} {\bibfnamefont {O}~\bibnamefont
  {Cardoso}},\ }\href@noop {} {\emph {\bibinfo {title} {Turbulence: a tentative
  dictionary}}},\ Vol.\ \bibinfo {volume} {341}\ (\bibinfo  {publisher}
  {Springer Science \& Business Media},\ \bibinfo {year} {2012})\BibitemShut
  {NoStop}%
\bibitem [{\citenamefont {Hadzibabic}\ \emph {et~al.}(2006)\citenamefont
  {Hadzibabic}, \citenamefont {Kr{\"u}ger}, \citenamefont {Cheneau},
  \citenamefont {Battelier},\ and\ \citenamefont
  {Dalibard}}]{hadzibabic2006berezinskii}%
  \BibitemOpen
  \bibfield  {author} {\bibinfo {author} {\bibfnamefont {Zoran}\ \bibnamefont
  {Hadzibabic}}, \bibinfo {author} {\bibfnamefont {Peter}\ \bibnamefont
  {Kr{\"u}ger}}, \bibinfo {author} {\bibfnamefont {Marc}\ \bibnamefont
  {Cheneau}}, \bibinfo {author} {\bibfnamefont {Baptiste}\ \bibnamefont
  {Battelier}}, \ and\ \bibinfo {author} {\bibfnamefont {Jean}\ \bibnamefont
  {Dalibard}},\ }\bibfield  {title} {\enquote {\bibinfo {title}
  {Berezinskii--{K}osterlitz--{T}houless crossover in a trapped atomic gas},}\
  }\href {https://www.nature.com/articles/nature04851} {\bibfield  {journal}
  {\bibinfo  {journal} {Nature}\ }\textbf {\bibinfo {volume} {441}},\ \bibinfo
  {pages} {1118--1121} (\bibinfo {year} {2006})}\BibitemShut {NoStop}%
\bibitem [{\citenamefont {Fine}\ \emph {et~al.}(1995)\citenamefont {Fine},
  \citenamefont {Cass}, \citenamefont {Flynn},\ and\ \citenamefont
  {Driscoll}}]{fine1995}%
  \BibitemOpen
  \bibfield  {author} {\bibinfo {author} {\bibfnamefont {K.~S.}\ \bibnamefont
  {Fine}}, \bibinfo {author} {\bibfnamefont {A.~C.}\ \bibnamefont {Cass}},
  \bibinfo {author} {\bibfnamefont {W.~G.}\ \bibnamefont {Flynn}}, \ and\
  \bibinfo {author} {\bibfnamefont {C.~F.}\ \bibnamefont {Driscoll}},\
  }\bibfield  {title} {\enquote {\bibinfo {title} {Relaxation of {2D}
  {T}urbulence to {V}ortex {C}rystals},}\ }\href {\doibase
  10.1103/PhysRevLett.75.3277} {\bibfield  {journal} {\bibinfo  {journal}
  {Physical Review Letters}\ }\textbf {\bibinfo {volume} {75}},\ \bibinfo
  {pages} {3277--3280} (\bibinfo {year} {1995})}\BibitemShut {NoStop}%
\bibitem [{\citenamefont {Jin}\ and\ \citenamefont {Dubin}(1998)}]{jin1998}%
  \BibitemOpen
  \bibfield  {author} {\bibinfo {author} {\bibfnamefont {D.~Z.}\ \bibnamefont
  {Jin}}\ and\ \bibinfo {author} {\bibfnamefont {Daniel H.~E.}\ \bibnamefont
  {Dubin}},\ }\bibfield  {title} {\enquote {\bibinfo {title} {Regional maximum
  entropy theory of vortex crystal formation},}\ }\href {\doibase
  10.1103/PhysRevLett.80.4434} {\bibfield  {journal} {\bibinfo  {journal}
  {Physical Review Letters}\ }\textbf {\bibinfo {volume} {80}},\ \bibinfo
  {pages} {4434--4437} (\bibinfo {year} {1998})}\BibitemShut {NoStop}%
\bibitem [{\citenamefont {Simula}(2018)}]{simula2018vortex}%
  \BibitemOpen
  \bibfield  {author} {\bibinfo {author} {\bibfnamefont {Tapio}\ \bibnamefont
  {Simula}},\ }\bibfield  {title} {\enquote {\bibinfo {title} {Vortex mass in a
  superfluid},}\ }\href {\doibase 10.1103/PhysRevA.97.023609} {\bibfield
  {journal} {\bibinfo  {journal} {Physical Review A}\ }\textbf {\bibinfo
  {volume} {97}},\ \bibinfo {pages} {023609} (\bibinfo {year}
  {2018})}\BibitemShut {NoStop}%
\bibitem [{\citenamefont {Thouless}\ and\ \citenamefont
  {Anglin}(2007)}]{thouless2007vortex}%
  \BibitemOpen
  \bibfield  {author} {\bibinfo {author} {\bibfnamefont {D.~J.}\ \bibnamefont
  {Thouless}}\ and\ \bibinfo {author} {\bibfnamefont {J.~R.}\ \bibnamefont
  {Anglin}},\ }\bibfield  {title} {\enquote {\bibinfo {title} {Vortex mass in a
  superfluid at low frequencies},}\ }\href {\doibase
  10.1103/PhysRevLett.99.105301} {\bibfield  {journal} {\bibinfo  {journal}
  {Physical Review Letters}\ }\textbf {\bibinfo {volume} {99}},\ \bibinfo
  {pages} {105301} (\bibinfo {year} {2007})}\BibitemShut {NoStop}%
\bibitem [{\citenamefont {Lucas}\ and\ \citenamefont
  {Sur\'owka}(2014)}]{lucas2014sound}%
  \BibitemOpen
  \bibfield  {author} {\bibinfo {author} {\bibfnamefont {Andrew}\ \bibnamefont
  {Lucas}}\ and\ \bibinfo {author} {\bibfnamefont {Piotr}\ \bibnamefont
  {Sur\'owka}},\ }\bibfield  {title} {\enquote {\bibinfo {title} {Sound-induced
  vortex interactions in a zero-temperature two-dimensional superfluid},}\
  }\href@noop {} {\bibfield  {journal} {\bibinfo  {journal} {Physical Review
  A}\ }\textbf {\bibinfo {volume} {90}},\ \bibinfo {pages} {053617} (\bibinfo
  {year} {2014})}\BibitemShut {NoStop}%
\bibitem [{\citenamefont {Griffin}\ \emph {et~al.}(2020)\citenamefont
  {Griffin}, \citenamefont {Shukla}, \citenamefont {Brachet},\ and\
  \citenamefont {Nazarenko}}]{shukla2020Magnus}%
  \BibitemOpen
  \bibfield  {author} {\bibinfo {author} {\bibfnamefont {Adam}\ \bibnamefont
  {Griffin}}, \bibinfo {author} {\bibfnamefont {Vishwanath}\ \bibnamefont
  {Shukla}}, \bibinfo {author} {\bibfnamefont {Marc-Etienne}\ \bibnamefont
  {Brachet}}, \ and\ \bibinfo {author} {\bibfnamefont {Sergey}\ \bibnamefont
  {Nazarenko}},\ }\bibfield  {title} {\enquote {\bibinfo {title} {Magnus-force
  model for active particles trapped on superfluid vortices},}\ }\href
  {\doibase 10.1103/PhysRevA.101.053601} {\bibfield  {journal} {\bibinfo
  {journal} {Physical Review A}\ }\textbf {\bibinfo {volume} {101}},\ \bibinfo
  {pages} {053601} (\bibinfo {year} {2020})}\BibitemShut {NoStop}%
\bibitem [{\citenamefont {Richaud}\ \emph {et~al.}(2020)\citenamefont
  {Richaud}, \citenamefont {Penna}, \citenamefont {Mayol},\ and\ \citenamefont
  {Guilleumas}}]{richaud2020vortices}%
  \BibitemOpen
  \bibfield  {author} {\bibinfo {author} {\bibfnamefont {Andrea}\ \bibnamefont
  {Richaud}}, \bibinfo {author} {\bibfnamefont {Vittorio}\ \bibnamefont
  {Penna}}, \bibinfo {author} {\bibfnamefont {Ricardo}\ \bibnamefont {Mayol}},
  \ and\ \bibinfo {author} {\bibfnamefont {Montserrat}\ \bibnamefont
  {Guilleumas}},\ }\bibfield  {title} {\enquote {\bibinfo {title} {Vortices
  with massive cores in a binary mixture of {B}ose-{E}instein condensates},}\
  }\href {https://journals.aps.org/pra/abstract/10.1103/PhysRevA.101.013630}
  {\bibfield  {journal} {\bibinfo  {journal} {Physical Review A}\ }\textbf
  {\bibinfo {volume} {101}},\ \bibinfo {pages} {013630} (\bibinfo {year}
  {2020})}\BibitemShut {NoStop}%
\bibitem [{\citenamefont {Eckel}\ \emph {et~al.}(2016)\citenamefont {Eckel},
  \citenamefont {Lee}, \citenamefont {Jendrzejewski}, \citenamefont {Lobb},
  \citenamefont {Campbell},\ and\ \citenamefont {Hill~III}}]{eckel2016contact}%
  \BibitemOpen
  \bibfield  {author} {\bibinfo {author} {\bibfnamefont {S}~\bibnamefont
  {Eckel}}, \bibinfo {author} {\bibfnamefont {Jeffrey~G}\ \bibnamefont {Lee}},
  \bibinfo {author} {\bibfnamefont {F}~\bibnamefont {Jendrzejewski}}, \bibinfo
  {author} {\bibfnamefont {C.~J.}\ \bibnamefont {Lobb}}, \bibinfo {author}
  {\bibfnamefont {G.~K.}\ \bibnamefont {Campbell}}, \ and\ \bibinfo {author}
  {\bibfnamefont {W.~T.}\ \bibnamefont {Hill~III}},\ }\bibfield  {title}
  {\enquote {\bibinfo {title} {Contact resistance and phase slips in mesoscopic
  superfluid-atom transport},}\ }\href
  {https://journals.aps.org/pra/abstract/10.1103/PhysRevA.93.063619} {\bibfield
   {journal} {\bibinfo  {journal} {Physical Review A}\ }\textbf {\bibinfo
  {volume} {93}},\ \bibinfo {pages} {063619} (\bibinfo {year}
  {2016})}\BibitemShut {NoStop}%
\bibitem [{\citenamefont {Burchianti}\ \emph {et~al.}(2018)\citenamefont
  {Burchianti}, \citenamefont {Scazza}, \citenamefont {Amico}, \citenamefont
  {Valtolina}, \citenamefont {Seman}, \citenamefont {Fort}, \citenamefont
  {Zaccanti}, \citenamefont {Inguscio},\ and\ \citenamefont
  {Roati}}]{burchianti2018connecting}%
  \BibitemOpen
  \bibfield  {author} {\bibinfo {author} {\bibfnamefont {A}~\bibnamefont
  {Burchianti}}, \bibinfo {author} {\bibfnamefont {F}~\bibnamefont {Scazza}},
  \bibinfo {author} {\bibfnamefont {A}~\bibnamefont {Amico}}, \bibinfo {author}
  {\bibfnamefont {G}~\bibnamefont {Valtolina}}, \bibinfo {author}
  {\bibfnamefont {J.~A.}\ \bibnamefont {Seman}}, \bibinfo {author}
  {\bibfnamefont {C}~\bibnamefont {Fort}}, \bibinfo {author} {\bibfnamefont
  {M}~\bibnamefont {Zaccanti}}, \bibinfo {author} {\bibfnamefont
  {M}~\bibnamefont {Inguscio}}, \ and\ \bibinfo {author} {\bibfnamefont
  {G}~\bibnamefont {Roati}},\ }\bibfield  {title} {\enquote {\bibinfo {title}
  {Connecting dissipation and phase slips in a {J}osephson junction between
  fermionic superfluids},}\ }\href
  {https://journals.aps.org/prl/abstract/10.1103/PhysRevLett.120.025302}
  {\bibfield  {journal} {\bibinfo  {journal} {Physical Review Letters}\
  }\textbf {\bibinfo {volume} {120}},\ \bibinfo {pages} {025302} (\bibinfo
  {year} {2018})}\BibitemShut {NoStop}%
\bibitem [{\citenamefont {Gauthier}\ \emph
  {et~al.}(2019{\natexlab{b}})\citenamefont {Gauthier}, \citenamefont
  {Szigeti}, \citenamefont {Reeves}, \citenamefont {Baker}, \citenamefont
  {Bell}, \citenamefont {Rubinsztein-Dunlop}, \citenamefont {Davis},\ and\
  \citenamefont {Neely}}]{gauthier2019quantitative}%
  \BibitemOpen
  \bibfield  {author} {\bibinfo {author} {\bibfnamefont {Guillaume}\
  \bibnamefont {Gauthier}}, \bibinfo {author} {\bibfnamefont {Stuart~S}\
  \bibnamefont {Szigeti}}, \bibinfo {author} {\bibfnamefont {Matthew~T}\
  \bibnamefont {Reeves}}, \bibinfo {author} {\bibfnamefont {Mark}\ \bibnamefont
  {Baker}}, \bibinfo {author} {\bibfnamefont {Thomas~A}\ \bibnamefont {Bell}},
  \bibinfo {author} {\bibfnamefont {Halina}\ \bibnamefont
  {Rubinsztein-Dunlop}}, \bibinfo {author} {\bibfnamefont {Matthew~J}\
  \bibnamefont {Davis}}, \ and\ \bibinfo {author} {\bibfnamefont {Tyler~W}\
  \bibnamefont {Neely}},\ }\bibfield  {title} {\enquote {\bibinfo {title}
  {Quantitative {A}coustic {M}odels for {S}uperfluid {C}ircuits},}\ }\href
  {\doibase 10.1103/PhysRevLett.123.260402} {\bibfield  {journal} {\bibinfo
  {journal} {Physical Review Letters}\ }\textbf {\bibinfo {volume} {123}},\
  \bibinfo {pages} {260402} (\bibinfo {year} {2019}{\natexlab{b}})}\BibitemShut
  {NoStop}%
\bibitem [{\citenamefont {Harris}\ \emph {et~al.}(2016)\citenamefont {Harris},
  \citenamefont {McAuslan}, \citenamefont {Sheridan}, \citenamefont {Sachkou},
  \citenamefont {Baker},\ and\ \citenamefont {Bowen}}]{harris2016laser}%
  \BibitemOpen
  \bibfield  {author} {\bibinfo {author} {\bibfnamefont {GI}~\bibnamefont
  {Harris}}, \bibinfo {author} {\bibfnamefont {DL}~\bibnamefont {McAuslan}},
  \bibinfo {author} {\bibfnamefont {E}~\bibnamefont {Sheridan}}, \bibinfo
  {author} {\bibfnamefont {Y}~\bibnamefont {Sachkou}}, \bibinfo {author}
  {\bibfnamefont {C}~\bibnamefont {Baker}}, \ and\ \bibinfo {author}
  {\bibfnamefont {WP}~\bibnamefont {Bowen}},\ }\bibfield  {title} {\enquote
  {\bibinfo {title} {Laser cooling and control of excitations in superfluid
  helium},}\ }\href {https://www.nature.com/articles/nphys3714} {\bibfield
  {journal} {\bibinfo  {journal} {Nature Physics}\ }\textbf {\bibinfo {volume}
  {12}},\ \bibinfo {pages} {788--793} (\bibinfo {year} {2016})}\BibitemShut
  {NoStop}%
\bibitem [{\citenamefont {Souris}\ \emph {et~al.}(2017)\citenamefont {Souris},
  \citenamefont {Rojas}, \citenamefont {Kim},\ and\ \citenamefont
  {Davis}}]{souris2017ultralow}%
  \BibitemOpen
  \bibfield  {author} {\bibinfo {author} {\bibfnamefont {F.}~\bibnamefont
  {Souris}}, \bibinfo {author} {\bibfnamefont {X.}~\bibnamefont {Rojas}},
  \bibinfo {author} {\bibfnamefont {P.~H.}\ \bibnamefont {Kim}}, \ and\
  \bibinfo {author} {\bibfnamefont {J.~P.}\ \bibnamefont {Davis}},\ }\bibfield
  {title} {\enquote {\bibinfo {title} {Ultralow-dissipation superfluid
  micromechanical resonator},}\ }\href {\doibase
  10.1103/PhysRevApplied.7.044008} {\bibfield  {journal} {\bibinfo  {journal}
  {Physical Review Applied}\ }\textbf {\bibinfo {volume} {7}},\ \bibinfo
  {pages} {044008} (\bibinfo {year} {2017})}\BibitemShut {NoStop}%
\bibitem [{\citenamefont {He}\ \emph {et~al.}(2020)\citenamefont {He},
  \citenamefont {Harris}, \citenamefont {Baker}, \citenamefont {Sawadsky},
  \citenamefont {Sfendla}, \citenamefont {Sachkou}, \citenamefont {Forstner},\
  and\ \citenamefont {Bowen}}]{he2020strong}%
  \BibitemOpen
  \bibfield  {author} {\bibinfo {author} {\bibfnamefont {Xin}\ \bibnamefont
  {He}}, \bibinfo {author} {\bibfnamefont {Glen~I}\ \bibnamefont {Harris}},
  \bibinfo {author} {\bibfnamefont {Christopher~G}\ \bibnamefont {Baker}},
  \bibinfo {author} {\bibfnamefont {Andreas}\ \bibnamefont {Sawadsky}},
  \bibinfo {author} {\bibfnamefont {Yasmine~L}\ \bibnamefont {Sfendla}},
  \bibinfo {author} {\bibfnamefont {Yauhen~P}\ \bibnamefont {Sachkou}},
  \bibinfo {author} {\bibfnamefont {Stefan}\ \bibnamefont {Forstner}}, \ and\
  \bibinfo {author} {\bibfnamefont {Warwick~P}\ \bibnamefont {Bowen}},\
  }\bibfield  {title} {\enquote {\bibinfo {title} {Strong optical coupling
  through superfluid brillouin lasing},}\ }\href
  {https://www.nature.com/articles/s41567-020-0785-0} {\bibfield  {journal}
  {\bibinfo  {journal} {Nature Physics}\ }\textbf {\bibinfo {volume} {16}},\
  \bibinfo {pages} {417--421} (\bibinfo {year} {2020})}\BibitemShut {NoStop}%
\bibitem [{\citenamefont {Varga}\ \emph {et~al.}(2020)\citenamefont {Varga},
  \citenamefont {Vadakkumbatt}, \citenamefont {Shook}, \citenamefont {Kim},\
  and\ \citenamefont {Davis}}]{varga2020observation}%
  \BibitemOpen
  \bibfield  {author} {\bibinfo {author} {\bibfnamefont {E.}~\bibnamefont
  {Varga}}, \bibinfo {author} {\bibfnamefont {V.}~\bibnamefont {Vadakkumbatt}},
  \bibinfo {author} {\bibfnamefont {A.~J.}\ \bibnamefont {Shook}}, \bibinfo
  {author} {\bibfnamefont {P.~H.}\ \bibnamefont {Kim}}, \ and\ \bibinfo
  {author} {\bibfnamefont {J.~P.}\ \bibnamefont {Davis}},\ }\bibfield  {title}
  {\enquote {\bibinfo {title} {Observation of bistable turbulence in
  quasi-two-dimensional superflow},}\ }\href {\doibase
  10.1103/PhysRevLett.125.025301} {\bibfield  {journal} {\bibinfo  {journal}
  {Phys. Rev. Lett.}\ }\textbf {\bibinfo {volume} {125}},\ \bibinfo {pages}
  {025301} (\bibinfo {year} {2020})}\BibitemShut {NoStop}%
\bibitem [{\citenamefont {Foss-Feig}\ \emph {et~al.}(2017)\citenamefont
  {Foss-Feig}, \citenamefont {Niroula}, \citenamefont {Young}, \citenamefont
  {Hafezi}, \citenamefont {Gorshkov}, \citenamefont {Wilson},\ and\
  \citenamefont {Maghrebi}}]{FossFeig2017}%
  \BibitemOpen
  \bibfield  {author} {\bibinfo {author} {\bibfnamefont {M.}~\bibnamefont
  {Foss-Feig}}, \bibinfo {author} {\bibfnamefont {P.}~\bibnamefont {Niroula}},
  \bibinfo {author} {\bibfnamefont {J.~T.}\ \bibnamefont {Young}}, \bibinfo
  {author} {\bibfnamefont {M.}~\bibnamefont {Hafezi}}, \bibinfo {author}
  {\bibfnamefont {A.~V.}\ \bibnamefont {Gorshkov}}, \bibinfo {author}
  {\bibfnamefont {R.~M.}\ \bibnamefont {Wilson}}, \ and\ \bibinfo {author}
  {\bibfnamefont {M.~F.}\ \bibnamefont {Maghrebi}},\ }\bibfield  {title}
  {\enquote {\bibinfo {title} {Emergent equilibrium in many-body optical
  bistability},}\ }\href {\doibase 10.1103/PhysRevA.95.043826} {\bibfield
  {journal} {\bibinfo  {journal} {Physical Review A}\ }\textbf {\bibinfo
  {volume} {95}},\ \bibinfo {pages} {043826} (\bibinfo {year}
  {2017})}\BibitemShut {NoStop}%
\bibitem [{\citenamefont {Eckel}\ \emph {et~al.}(2014)\citenamefont {Eckel},
  \citenamefont {Lee}, \citenamefont {Jendrzejewski}, \citenamefont {Murray},
  \citenamefont {Clark}, \citenamefont {Lobb}, \citenamefont {Phillips},
  \citenamefont {Edwards},\ and\ \citenamefont
  {Campbell}}]{eckel2014hysteresis}%
  \BibitemOpen
  \bibfield  {author} {\bibinfo {author} {\bibfnamefont {Stephen}\ \bibnamefont
  {Eckel}}, \bibinfo {author} {\bibfnamefont {Jeffrey~G}\ \bibnamefont {Lee}},
  \bibinfo {author} {\bibfnamefont {Fred}\ \bibnamefont {Jendrzejewski}},
  \bibinfo {author} {\bibfnamefont {Noel}\ \bibnamefont {Murray}}, \bibinfo
  {author} {\bibfnamefont {Charles~W}\ \bibnamefont {Clark}}, \bibinfo {author}
  {\bibfnamefont {Christopher~J}\ \bibnamefont {Lobb}}, \bibinfo {author}
  {\bibfnamefont {William~D}\ \bibnamefont {Phillips}}, \bibinfo {author}
  {\bibfnamefont {Mark}\ \bibnamefont {Edwards}}, \ and\ \bibinfo {author}
  {\bibfnamefont {Gretchen~K}\ \bibnamefont {Campbell}},\ }\bibfield  {title}
  {\enquote {\bibinfo {title} {Hysteresis in a quantized superfluid
  `atomtronic' circuit},}\ }\href {\doibase 10.1038/nature12958} {\bibfield
  {journal} {\bibinfo  {journal} {Nature}\ }\textbf {\bibinfo {volume} {506}},\
  \bibinfo {pages} {200} (\bibinfo {year} {2014})}\BibitemShut {NoStop}%
\bibitem [{\citenamefont {Wright}\ \emph {et~al.}(2013)\citenamefont {Wright},
  \citenamefont {Blakestad}, \citenamefont {Lobb}, \citenamefont {Phillips},\
  and\ \citenamefont {Campbell}}]{wright2013driving}%
  \BibitemOpen
  \bibfield  {author} {\bibinfo {author} {\bibfnamefont {Kevin~C}\ \bibnamefont
  {Wright}}, \bibinfo {author} {\bibfnamefont {RB}~\bibnamefont {Blakestad}},
  \bibinfo {author} {\bibfnamefont {C.~J.}\ \bibnamefont {Lobb}}, \bibinfo
  {author} {\bibfnamefont {W.~D.}\ \bibnamefont {Phillips}}, \ and\ \bibinfo
  {author} {\bibfnamefont {G.~K.}\ \bibnamefont {Campbell}},\ }\bibfield
  {title} {\enquote {\bibinfo {title} {Driving phase slips in a superfluid atom
  circuit with a rotating weak link},}\ }\href {\doibase
  10.1103/PhysRevLett.110.025302} {\bibfield  {journal} {\bibinfo  {journal}
  {Physical Review Letters}\ }\textbf {\bibinfo {volume} {110}},\ \bibinfo
  {pages} {025302} (\bibinfo {year} {2013})}\BibitemShut {NoStop}%
\bibitem [{\citenamefont {Wilson}\ \emph {et~al.}(2021)\citenamefont {Wilson},
  \citenamefont {Samson}, \citenamefont {Newman},\ and\ \citenamefont
  {Anderson}}]{wilson2021generation}%
  \BibitemOpen
  \bibfield  {author} {\bibinfo {author} {\bibfnamefont {Kali~E}\ \bibnamefont
  {Wilson}}, \bibinfo {author} {\bibfnamefont {E}~\bibnamefont {Samson}},
  \bibinfo {author} {\bibfnamefont {Zachary~L}\ \bibnamefont {Newman}}, \ and\
  \bibinfo {author} {\bibfnamefont {Brian~P}\ \bibnamefont {Anderson}},\
  }\bibfield  {title} {\enquote {\bibinfo {title} {Generation of high
  winding-number superfluid circulation in {B}ose-{E}instein condensates},}\
  }\href@noop {} {\bibfield  {journal} {\bibinfo  {journal} {arXiv preprint
  arXiv:2109.12945}\ } (\bibinfo {year} {2021})}\BibitemShut {NoStop}%
\bibitem [{SM()}]{SM}%
  \BibitemOpen
  \href@noop {} {}\bibinfo {note} {See Supplemental Material at [URL will be
  inserted by publisher] for movies of damped GPE simulations of the three
  stirring protocols.}\BibitemShut {Stop}%
\bibitem [{\citenamefont {Turkington}\ and\ \citenamefont
  {Whitaker}(1996)}]{turkington1996statistical}%
  \BibitemOpen
  \bibfield  {author} {\bibinfo {author} {\bibfnamefont {Bruce}\ \bibnamefont
  {Turkington}}\ and\ \bibinfo {author} {\bibfnamefont {Nathaniel}\
  \bibnamefont {Whitaker}},\ }\bibfield  {title} {\enquote {\bibinfo {title}
  {Statistical equilibrium computations of coherent structures in turbulent
  shear layers},}\ }\href {\doibase 10.1137/S1064827593251708} {\bibfield
  {journal} {\bibinfo  {journal} {SIAM Journal on Scientific Computing}\
  }\textbf {\bibinfo {volume} {17}},\ \bibinfo {pages} {1414--1433} (\bibinfo
  {year} {1996})}\BibitemShut {NoStop}%
\bibitem [{\citenamefont {Aref}(1979)}]{aref1979motion}%
  \BibitemOpen
  \bibfield  {author} {\bibinfo {author} {\bibfnamefont {Hassan}\ \bibnamefont
  {Aref}},\ }\bibfield  {title} {\enquote {\bibinfo {title} {Motion of three
  vortices},}\ }\href {\doibase 10.1063/1.862605} {\bibfield  {journal}
  {\bibinfo  {journal} {The Physics of Fluids}\ }\textbf {\bibinfo {volume}
  {22}},\ \bibinfo {pages} {393--400} (\bibinfo {year} {1979})}\BibitemShut
  {NoStop}%
\end{thebibliography}
%

\end{document}